\documentclass[5p]{elsarticle}

\makeatletter
\def\ps@pprintTitle{%
  \let\@oddhead\@empty
  \let\@evenhead\@empty
  \let\@oddfoot\@empty
  \let\@evenfoot\@empty
}
\makeatother

\usepackage[title]{appendix}

\biboptions{numbers,sort&compress}

\usepackage{amssymb}
\usepackage{pifont}

\usepackage{libertine}
\usepackage{libertinust1math}
\usepackage{geometry}
\usepackage{amsmath}
\usepackage{bbold}
\usepackage{graphicx}
\usepackage{eurosym}
\usepackage{mathtools}
\usepackage{url}
\usepackage{booktabs}
\usepackage{epstopdf}
\usepackage{xfrac}
\usepackage{tabularx}
\usepackage{bm}
\usepackage{subcaption}
\usepackage{longtable}
\usepackage{multirow}
\usepackage{threeparttable}
\usepackage[export]{adjustbox}
\usepackage[version=4]{mhchem}
\usepackage[colorlinks]{hyperref}
\usepackage[nameinlink,sort&compress,capitalise]{cleveref}
\usepackage[leftcaption,raggedright]{sidecap}
\usepackage[prependcaption,textsize=footnotesize]{todonotes}
\usepackage{blindtext}
\usepackage{url}

\usepackage{xr}
\usepackage{lipsum}

\graphicspath{
    {../workflow/notebooks/},
    {../workflow/pypsa-eur/resources/},
    {../workflow/pypsa-eur/results/},
}

\usepackage{siunitx}
\DeclareSIUnit\year{a}
\DeclareSIUnit{\tco}{t_{\ce{CO2}}}
\DeclareSIUnit{\sieuro}{\mbox{\euro}}
\DeclareSIUnit{\twh}{{\tera\watt\hour}}
\DeclareSIUnit{\mwh}{{\mega\watt\hour}}
\DeclareSIUnit{\kwh}{{\kilo\watt\hour}}

\renewcommand{\ttdefault}{\sfdefault}

\begin{document}

\begin{frontmatter}

    \title{Managing the Mismatch:\\ The Role of Flexibility on the Path to a Carbon-Neutral Energy System}

	\author[tub]{Julian Geis\corref{correspondingauthor}}
	\ead{j.geis@tu-berlin.de}
    \author[tub]{Michael Lindner}
	\author[tub]{Tom Brown}

	\address[tub]{Department of Digital Transformation in Energy Systems, Institute of Energy Technology,\\Technische Universität Berlin, Fakultät III, Einsteinufer 25 (TA 8), 10587 Berlin, Germany}
		
    \begin{abstract}
A rapid expansion of system flexibility is essential to integrate increasing shares of renewable energy into future energy systems.
However, flexibility needs and technology-specific contributions to flexibility remain poorly quantified in energy system modelling. Existing methods are not widely applied, leaving key questions unanswered: which flexibility technologies are critical for climate neutrality, and what are the cost implications of alternative deployment strategies?
To address this gap, we apply a correlation-based flexibility metric to a high-resolution, sector-coupled model of the German energy system, covering its transformation towards climate neutrality.
For our default scenario, we find that daily flexibility needs increase by a factor of 3.7 between 2025 and 2045, driven primarily by the expansion of solar PV. By 2045, stationary batteries provide 38\% of daily flexibility, while flexible electric vehicle charging contributes 30\%. Systems with constrained flexibility increase system costs by 6.9\%, electricity prices by 14~\euro/MWh and trigger 47\% higher hydrogen and e-fuel imports compared to an unconstrained system in 2045. In contrast, scenarios with high shares of flexible electric vehicle charging, vehicle-to-grid, and industrial demand-side management achieve system cost reductions of 3.3\%, while also reducing import dependence. Higher flexibility also reduces electricity price ranges, decreases average electricity prices by 3~\euro/MWh, and reduces backup capacity by 22\% (22 GW).
Overall, our results highlight the decisive role of specific flexibility technologies in achieving cost-efficient and energy-secure climate-neutral energy systems, providing quantitative guidance for policy and investment decisions.

 	\end{abstract}


\end{frontmatter}

\section{Introduction}
\label{sec:intro}

\subsection{Motivation}

The targets are set, and key elements of the path are clear: achieving climate-neutral energy systems within the coming decades requires a massive expansion of variable renewable energy sources (VRES), most importantly, wind and solar power. These technologies offer the most cost-effective route to decarbonisation and reduce dependence on fossil fuel imports in increasingly volatile geopolitical contexts. At the same time, their inherent variability and intermittency pose significant challenges for grid stability and security of supply.

As the share of VRES increases, so does the need for flexibility to balance supply and demand efficiently. The nature of this flexibility need, and the question of how it should be provided, has become one of the central yet least well-defined aspects of energy system transformation. Despite its importance, coherent strategies for achieving adequate flexibility remain largely absent across many energy systems.

The reasons are manifold and lie in the complexity of flexibility itself. There is substantial uncertainty regarding the future cost of flexibility, including potential cost reductions and performance improvements driven by technological learning, market design, and policy support, which complicates investment decisions. Moreover, flexibility needs are difficult to quantify precisely, as they depend on weather variability, demand patterns, and the pace of technological progress. 

Additionally, while sector coupling offers new opportunities to provide flexibility, it requires careful coordination of flexibility across sectors to ensure optimal performance and cost-effectiveness. The wide range of technologies that can provide flexibility across different timescales adds complexity to the task of identifying the most suitable options for a given energy system and quantifying their contributions to overall flexibility needs.

\subsection{Literature}

The challenge begins with the definition of flexibility itself. The literature presents diverse conceptual approaches, ranging from flexibility as the capacity to integrate high shares of renewable energy \cite{papaefthymiou_power_2018, IRENA_2018} to the ``ability of the aggregated set of generators to respond to variations and uncertainty in net load'' \cite{denholm_grid_2011}. More general definitions characterize flexibility as ``the power system's ability to cope with variability and uncertainty'' \cite{heggarty_quantifying_2020}. This conceptual diversity extends to measurement approaches, as metrics for quantifying flexibility ``may be ambiguous to different definitions'' \cite{lund_review_2015}, leading some researchers to conclude that a single indicator is insufficient to capture flexibility's multi-dimensional nature, spanning temporal (short-term vs long-term), spatial (local vs system-wide), and technological (supply-side vs demand-side) aspects \cite{emmanuel_review_2020}.\newline

Several methodologies have emerged to quantify flexibility requirements and provision, typically through analysis of the residual load which is the difference between inflexible electricity demand and non-dispatchable generation. Flexibility metrics can be distinguished by four facets: (1) how much flexibility is required, (2) how flexible is the flexibility solution, (3) how flexible is the power system, and (4) who is providing the flexibility \cite{heggarty_quantifying_2020}. This study primarily addresses questions (1) and (4), while also providing insights into (2) and (3).

Statistical approaches define flexibility requirements through residual load characteristics. Brunner et al. \cite{brunner_future_2020} use three metrics - residual load range, surplus energy, and surplus time - to optimize renewable energy mixes. Other approaches focus on ramping capabilities such as ramping times \cite{huber_integration_2014, deetjen_impacts_2017}, on the volatility of residual load or VRES output \cite{lannoye_power_2012, epri_metrics_2014}, or on a combination of both aspects \cite{dvorkin_assessing_2014}. Storage requirements can also serve as a proxy for flexibility needs \cite{zerrahn_long-run_2017, oh_energy-storage_2018}. %

Frequency-based methods decompose residual load into temporal components. Heggarty et al. \cite{heggarty_multi-temporal_2019} isolate daily, weekly and annual signals with discrete Fourier transform (DFT) to quantify flexibility requirements. This methodology is later used in \cite{heggarty_quantifying_2020} to decompose flexibility provision per technology into different timescales. Their approach yields two metrics: a stacked area chart showing each technology's generation deviations from average, and percentage flexibility contributions per technology. While powerful, these methods rely on band-pass filters to assign variability to predefined frequency ranges. The choice of filter specifications (e.g., cutoff frequencies and bandwidths) is not unique, and results can be sensitive to these assumptions, leading to shifts in attributed flexibility across timescales and technologies \cite{huclin_methodological_2023}. In addition, their complexity reduces transparency and makes the results less intuitive and harder to interpret. %

A simpler, correlation-based approach, adopted by Artelys, IEA, and others \cite{artelys_mainstreaming_2017, iea_managing_2023, trinomics_power_2023, solarpower_europe_mission_2024}, defines flexibility needs as the energy that must be shifted to achieve a flat residual load. Originally proposed by the French TSO RTE \cite{rte_bilan_2015}, Artelys extended this framework to assess technology contributions by subtracting individual generation profiles from the residual load and recalculating flexibility needs \cite{artelys_mainstreaming_2017}. However, the methodology used to attribute flexibility provision to individual technologies is only sparsely documented in the literature and relies primarily on project reports rather than peer-reviewed methodological descriptions. Moreover, the subtraction-based attribution approach is not mathematically additive and can therefore lead to inconsistent allocation of total flexibility needs across technologies. A detailed methodological discussion is provided in Section~\ref{sec:methods:flexibility-correlation-method}.\newline

At the European level, several studies apply flexibility quantification to energy system models. The METIS project \cite{european_commission_metis_2015} finds that EU flexibility needs increase by 26\% daily, 27\% weekly, and 14\% annually between 2020 and 2030 \cite{artelys_mainstreaming_2017}, intensifying to 80\%, 60\%, and 50\% increases respectively between 2030 and 2050 \cite{artelys_optimal_2018}. Interconnectors emerge as dominant flexibility providers across all timescales, while batteries handle daily variations, pumped hydro storage (PHS) balances weekly patterns, and thermal units manage annual fluctuations. A Penta region study \cite{trinomics_power_2023} estimates that in Germany's 2050 climate-neutral system, daily flexibility is provided by electrolyzers (27\%), interconnectors (20\%), electric vehicles (20\%), and PHS (10\%), while weekly and annual flexibility is dominated by electrolyzers (35-40\%) and interconnectors (20-30\%). ENTSO-E \cite{entso-e_system_2024} applies Fourier transform and Artelys-style time decomposition, finding that flexibility needs double across all timescales between 2025--2033, with transmission capacity reducing needs by 15-35\%.\newline
The flexibility literature for Germany is fragmented across distinct streams, with no comprehensive system-level quantification.
\textbf{Residual load and storage studies} analyze how increasing VRES affects storage requirements. Schill \cite{schill_residual_2014} estimates at least 10~GW of additional storage capacity needed by 2050 to absorb surplus renewable energy. Schlachtberger et al. \cite{schlachtberger_backup_2016} find that highly flexible backup becomes dominant beyond 50\% renewable generation. Storage analyses \cite{ruhnau_storage_2022, thimet_what-where-when_2023, open_energy_transition_role_2025} consistently show that battery storage and PHS handle short-term fluctuations while hydrogen storage provides seasonal balancing, though capacities vary with weather patterns, flexibility options, and cost assumptions.
\textbf{Near-term analyses} (2030-2035) identify critical flexibility technologies for Germany's transition phase. Gillich et al. \cite{gillich_schlusselrolle_2024} find that long-term hydrogen storage and power-to-heat in district heating networks are critical for the 2030 system, while short-term providers like batteries, PHS, or Demand Side Management (DSM) show lower system benefits and can be more easily substituted. Studies on grid stability and system robustness \cite{buttner_influence_2024, ackermann_cost-benefit_2024} identify flexible BEV charging, thermal storage in district heating, and dynamic line rating as high-impact strategies for cost reduction and renewable integration.
\textbf{Technology-specific analyses} examine individual flexibility providers in detail. Multiple studies \cite{frank_potential_2025, ghaiurane_integrating_2025, gombodshaw-johann_bos_bidirektionales_2025} confirm BEVs' significant potential for daily flexibility through smart charging and vehicle-to-grid (V2G) capabilities. Scenario-based studies \cite{brunner_what_2025, scharf_gas_2024} find that hydrogen and gas backup requirements decrease with greater system flexibility. 
\textbf{Sector-coupling studies} reveal complementary flexibility roles across technologies. Gils et al. \cite{gils_interaction_2021} show that electrolysis enables both short-term demand response and seasonal balancing, heat pumps with thermal storage shift midday solar generation, and battery-based options manage diurnal variations \cite{maruf_open_2021, nebel_role_2022}. Göke et al. \cite{goke_how_2023} quantify that electrolyzers reduce residual peak load most (42\% reduction), followed by district heating (33\%) and BEVs (20\%), while space and process heating show negligible effects. Gaafar et al. \cite{gaafar_system_2024} find that electrolysis and power-to-heat dominate flexibility during surplus periods, while gas turbines provide flexibility in winter and batteries in summer correlated with photovoltaic patterns.

\subsection{Research Gap}

Despite extensive literature on energy system flexibility, a critical gap remains in the systematic quantification of flexibility needs and technology contributions in sector-coupled energy system models with high spatial and temporal resolution. While Heggarty et al. \cite{heggarty_quantifying_2020} and Artelys \cite{artelys_mainstreaming_2017} developed promising quantitative methodologies, these have rarely been applied in fully sector-coupled, climate-neutral case studies such as Germany’s 2045 pathway. They also have methodological drawbacks that do not ensure the decomposition into flexibility providers adding back up to the total flexibility needs.

Existing German flexibility analyses further suffer from three key limitations. First, they predominantly focus on 2030-2035 horizons \cite{buttner_influence_2024, gillich_schlusselrolle_2024, ackermann_cost-benefit_2024, nebel_role_2022}, leaving the 2045 climate-neutral target underexplored. Second, they emphasize supply-side flexibility (storage, gas backup) while inadequately representing sector-coupling options like industrial DSM, power-to-heat, and BEV flexibility in integrated system contexts \cite{schill_residual_2014, schlachtberger_backup_2016, scharf_gas_2024, ruhnau_storage_2022, frank_potential_2025}. Third, they model 
Germany either as a single node \cite{gaafar_system_2024, nebel_role_2022, ruhnau_storage_2022} or sacrifice temporal resolution when including spatial detail \cite{buttner_influence_2024}, preventing accurate assessment of transmission constraints and regional dynamics.

No study combines quantitative flexibility metrics with high spatiotemporal resolution and comprehensive technology representation to assess which technologies provide flexibility in Germany's 2045 energy system.\newline

This study makes three original contributions: it provides the first formal definition and additivity proof of the correlation-based time-decomposition methodology; it is the first application of this framework to a spatially-explicit, hourly-resolution, fully sector-coupled model of a national energy system; and it delivers the first systematic quantification of flexibility needs and technology contributions along Germany's path to climate neutrality (2025–2045).

We answer three research questions: (1) How much flexibility is necessary for the electricity system in a climate-neutral Germany by 2045? (2) Which technologies provide this flexibility under different system configurations? (3) What are the cost, price, and energy security implications of varying flexibility levels?

This approach advances beyond previous applications of the correlation-based methodology (limited to European-scale analysis or lower spatial resolution) and existing German studies (which lack comprehensive DSM and quantitative metrics in their flexibility assessments). The combination of high spatio-temporal resolution, multi-technology representation, and quantitative flexibility metrics enables systematic comparison of flexibility strategies for Germany's path to climate neutrality.

\section{Methods}
\label{sec:methods}

\subsection{Flexibility quantification}

\subsubsection{Method Selection and Rationale}
\label{sec:methods:flexibility-method-selection}

This study applies the correlation-based methodology \cite{artelys_mainstreaming_2017} for several reasons:

\textbf{Simplicity}: The methodology provides clear steps for calculating flexibility needs from residual load time series and technology generation/consumption profiles. Unlike frequency-based methods \cite{heggarty_multi-temporal_2019, heggarty_quantifying_2020}, it avoids sensitivity to filter specification \cite{huclin_methodological_2023} and remains accessible to policymakers and diverse stakeholders.

\textbf{Comprehensiveness}: The method captures flexibility across multiple timescales. This study focuses on daily, weekly, and annual timeframes corresponding to natural system cycles. A single calculation quantifies both flexibility requirements and technology contributions, distinguishing flexibility providers (positive contributions) from flexibility drivers (negative contributions).

\textbf{Interpretability}: Results provide actionable insights for system operators and planners regarding specific flexibility needs and which technologies address them.

\textbf{Policy relevance}: Article 19e(1) of Regulation (EU) 2019/943 requires regulatory authorities to deliver Flexibility Needs Assessment (FNA) reports to support Member States in setting national non-fossil flexibility objectives. ACER's FNA methodology explicitly demands that TSOs use ``established decomposition methodologies (time decomposition, Fourier decomposition, etc.)'' to calculate daily, weekly, and seasonal flexibility indicators \cite{acer_flexibility_2025}, methodologically aligned with the approach presented here. The method is already extensively applied in European policy studies \cite{artelys_effect_2019, artelys_mainstreaming_2017, trinomics_power_2023, entso-e_system_2024, rte_bilan_2015,iea_managing_2023}.

\subsubsection{Correlation-Based Flexibility Contribution Method}
\label{sec:methods:flexibility-correlation-method}

\subsubsection*{Evaluation of flexibility needs}

\begin{figure}[htb]
    \centering
    \includegraphics[width=\columnwidth]{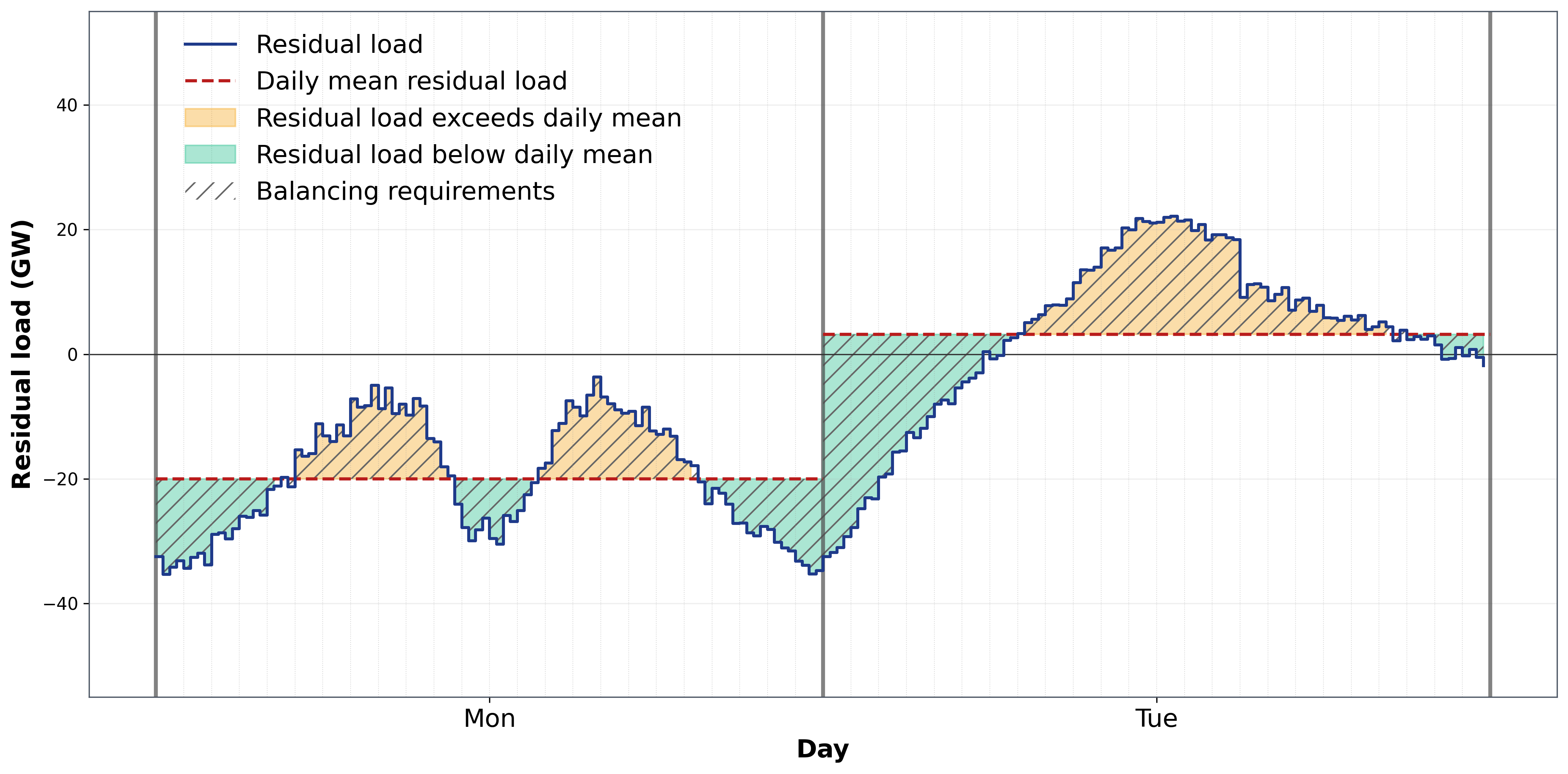} \\    
    \caption{\textbf{Illustration of daily flexibility needs derived from residual load profiles.}
    The residual load fluctuates around its daily mean, with deviations above and below defining the upward and downward balancing requirements for each day. Daily flexibility needs are computed as half the sum of absolute deviations from the daily mean, integrated over time.}
    \label{fig:methods:flex-needs}
\end{figure}

In this study, flexibility needs are quantified by analysing residual load variability. The residual load $RL(t)$ is defined as the net load that must be served by dispatchable technologies. It is calculated as non-dispatchable electricity demand minus non-dispatchable supply, i.e., variable renewable generation and must-run generation. By construction of the model, residual load is fully balanced by dispatchable generation at each time step.
The underlying principle is that a temporally flat residual load implies no need for flexibility from dispatchable units. Flexibility needs therefore represent a metric of the energy that would need to be shifted in time to remove variability in the residual load profile, averaged over a given period \cite{artelys_mainstreaming_2017, rte_bilan_2015}.
In this study, non-dispatchable demand includes inelastic electricity demand from households, industry, agriculture and BEVs. Decentral electrified heat (heat pumps and resistive heaters) is also included, as the inflexible component driven by heat demand outweighs the available flexibility from building thermal inertia. Non-dispatchable supply includes variable renewable generation from wind, solar and run-of-river (ror).

Flexibility needs are defined with respect to two temporal resolutions: a higher resolution $h$ (at which variability is evaluated) and a lower resolution $\ell$ (which defines the balancing timescale), where $h \ll \ell$. Let the study period be $[0,T]$. The lower resolution $\ell$ partitions the time horizon into disjoint intervals $I^l(t)$ of length $\ell$. The period-wise average residual load at resolution $\ell$ ($\ell$-period mean) is defined as
\begin{equation}
\overline{RL}^\ell(t) := \frac{1}{|I^\ell(t)|} \sum_{s \in I^\ell(t)} RL(s).
\end{equation}

If the variability of interest occurs at a higher resolution $h$, the residual load is first averaged to resolution $h$, yielding $\overline{RL}^h(t)$. The flexibility curve at resolution pair $\ell|h$ is then defined as
\begin{equation}
FlexCurve^{\ell|h}(t) := \overline{RL}^h(t) - \overline{RL}^\ell(t).
\end{equation}

The sign of the deviation,
\begin{equation}
FlexSign^{\ell|h}(t) := \mathrm{sign}\big(FlexCurve^{\ell|h}(t)\big),
\end{equation}
indicates whether the system experiences a surplus or shortage relative to the $\ell$-period mean.

Total flexibility needs over the study period for balancing $h$-scale variability around the $\ell$-scale mean are defined as
\begin{align}
FlexNeed^{\ell|h}
&= \frac{1}{2} \sum_{t=0}^{T} \left| FlexCurve^{\ell|h}(t) \right| \\
&= \frac{1}{2} \sum_{t=0}^{T} FlexCurve^{\ell|h}(t)\, FlexSign^{\ell|h}(t).
\end{align}

The factor $\tfrac{1}{2}$ ensures that upward and downward deviations are not double-counted, so that flexibility needs correspond to the total energy that must be shifted in time.

Following this time-decomposition framework, flexibility needs are quantified at three temporal scales:

\begin{itemize}
    \item \textit{Daily flexibility needs}: balancing hourly variability ($h=$ hourly) around the daily mean ($\ell=$ daily), capturing intra-day fluctuations. See Figure~\ref{fig:methods:flex-needs} for an illustration of daily flexibility needs.
    \item \textit{Weekly flexibility needs}: balancing daily variability ($h=$ daily) around the weekly mean ($\ell=$ weekly), capturing intra-week effects while excluding intra-day variability.
    \item \textit{Annual flexibility needs}: balancing weekly variability ($h=$ weekly) around the annual mean ($\ell=$ annual), capturing seasonal fluctuations.
\end{itemize}

This hierarchical structure ensures that variability at shorter timescales is not double-counted at longer timescales and enables a consistent decomposition of flexibility needs across temporal dimensions \cite{artelys_mainstreaming_2017,acer_flexibility_2025,trinomics_study_2022,european_commission_role_2019,rte_bilan_2015}.\newline

\subsubsection*{Evaluation of flexibility provision}

\begin{figure}[htb]
    \centering
    \includegraphics[width=\columnwidth]{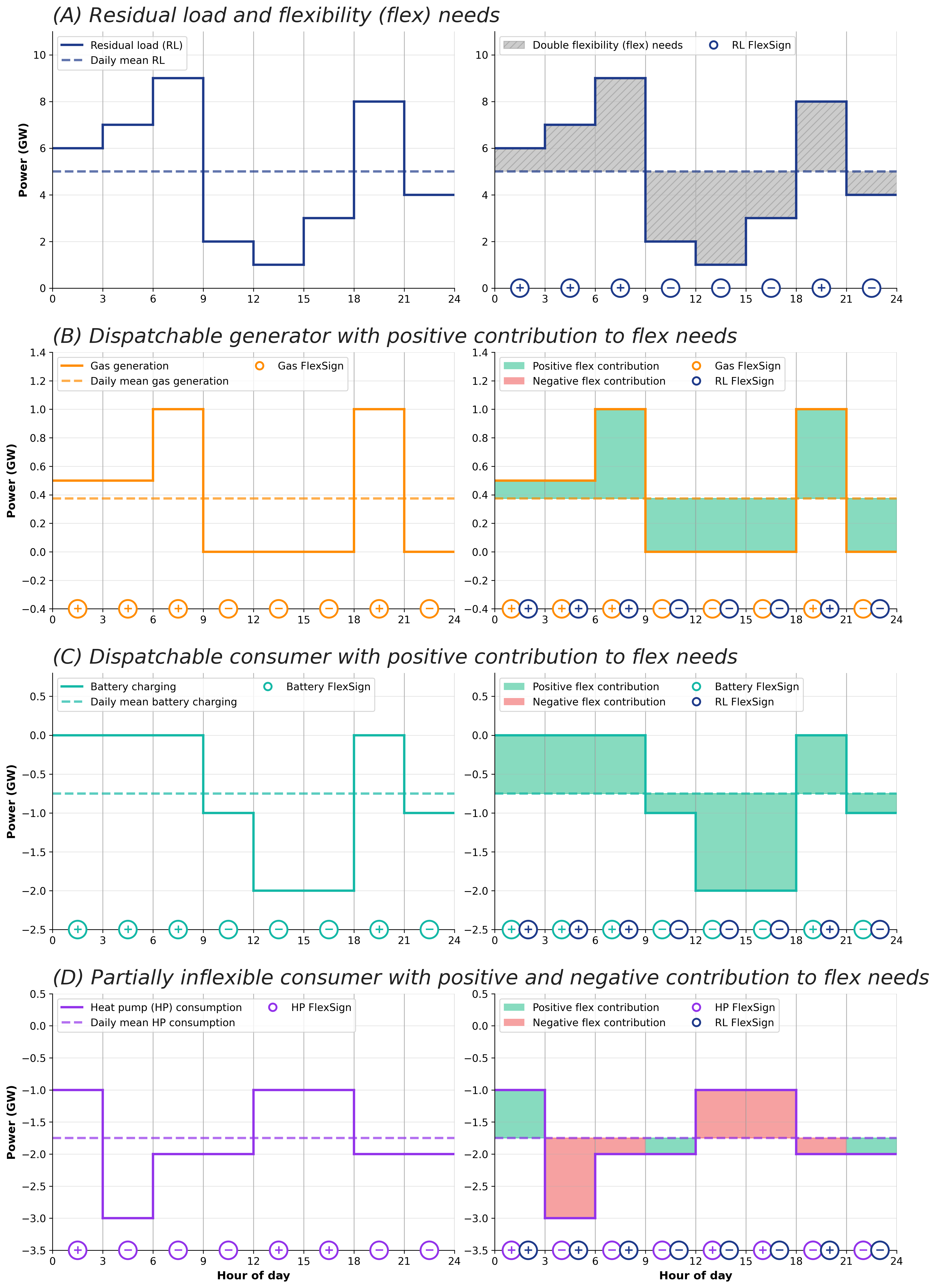} \\    
    \caption{\textbf{Illustration of the flexibility contribution methodology for different technology types.}
    The left column shows the hourly power profile of each technology alongside its daily mean. The right column shows how each timestep contributes positively or negatively to covering flexibility needs, determined by the product of the technology's FlexSign (deviation from its own mean) and the residual load FlexSign (deviation of RL from its daily mean). Panel (A) defines daily flexibility needs as the total area of deviations from the mean RL. Panels (B)–(D) illustrate a dispatchable generator, a dispatchable consumer, and a partially inflexible consumer, respectively, with the latter producing both positive and negative contributions depending on the alignment of its dispatch pattern with flexibility needs.}
    \label{fig:methods:flex-provision}
\end{figure}

Flexibility provision quantifies which technologies help satisfy flexibility needs and which technologies create additional mismatches between supply and demand. A technology contributes positively to flexibility when its generation or consumption pattern aligns with residual load variations, producing more when residual load is above its $\ell$-period mean (positive deviation) and less when it is below (negative deviation). Conversely, technologies whose output varies opposite to system requirements increase flexibility needs. While the flexibility curve of the residual load indicates when the system requires upward or downward balancing, flexibility provision measures how closely each technology follows these variations. For example, a baseload generator with a perfectly flat output profile exhibits zero flexibility provision. Figure~\ref{fig:methods:flex-provision} illustrates this for two example technologies over one day.\footnote{
    The illustrative example in Figure~\ref{fig:methods:flex-provision} uses 3-hour resolution 
    over one day with eight timesteps. Residual load values are $[6,7,9,2,1,3,8,4]$\,GW, giving a daily mean of $\overline{RL}=5$\,GW, $\text{FlexSign}=[+1,+1,+1,-1,-1,-1,+1,-1]$, and daily flexibility needs of $\frac{1}{2}(1+2+4+3+4+2+3+1)\times 3\,\text{h}=30.0$\,GWh. Gas generation (mean $0.375$\,GW) and battery charging (mean $-0.75$\,GW) are both fully 
    aligned with the RL FlexSign, contributing $4.5$\,GWh and $9.0$\,GWh, respectively. The heat pump (mean $-1.75$\,GW) is partially misaligned with the RL FlexSign, yielding $\tfrac{1}{2}\bigl((-1-(-1.75))\cdot(+1)+(-3-(-1.75))\cdot(+1)+(-2-(-1.75))\cdot(+1)+0.25-0.75-0.25+0.25\bigr)\times 3\,\text{h}=\tfrac{1}{2}(-2.0)\times 3\,\text{h}=-3.0$\,GWh. The three contributions sum to $10.5$\,GWh; the remainder is covered by other technologies not shown.
}

The calculation follows the same time-decomposition framework introduced for flexibility needs. For each technology $i$, deviations of its generation or consumption from its $\ell$-period mean are evaluated at resolution $h$ and multiplied by the flexibility sign derived from residual load variations. This captures whether the technology moves in sync with (positive contribution) or against (negative contribution) system requirements. Formally, let $P_i(t)$ denote the generation
(positive) or consumption (negative) of technology $i$ at time step $t$.
At every time step, residual load equals the sum of net outputs of all
dispatchable technologies \textit{i}:
\begin{equation}
    \sum_i P_i(t) = RL(t) \quad \forall\, t,
    \label{eq:power_balance}
\end{equation}
directly linking individual technology outputs to the residual load introduced above. 
Analogous to the residual load definition, define the period-wise averages
\begin{equation}
\bar{P}_i^\ell(t) := \frac{1}{|I^\ell(t)|} \sum_{s \in I^\ell(t)} P_i(s),
\qquad
\bar{P}_i^h(t) := \frac{1}{|I^h(t)|} \sum_{s \in I^h(t)} P_i(s),
\end{equation}
where $\ell$ and $h$ denote the lower and higher temporal resolutions, respectively, with $h \ll \ell$.

The flexibility provision curve of technology $i$ is defined as
\begin{equation}
FlexProvCurve_i^{\ell|h}(t)
:= \bar{P}_i^h(t) - \bar{P}_i^\ell(t).
\end{equation}

The contribution of technology $i$ to flexibility needs at resolution pair $\ell|h$ is then
\begin{align}
FlexProv_i^{\ell|h}
&= \frac{1}{2} \sum_{t=0}^{T}
FlexProvCurve_i^{\ell|h}(t)\,
FlexSign^{\ell|h}(t).
\end{align}

A positive value indicates that the technology reduces flexibility needs at the respective timescale, whereas a negative value implies that it increases system variability.

By construction, the sum of all technology contributions equals total flexibility needs (\textit{Additivity}),
\begin{equation}
\sum_i FlexProv_i^{\ell|h} = FlexNeed^{\ell|h},
\end{equation}
such that flexibility requirements are exactly and consistently allocated across technologies (see Appendix~\cref{app:proof_additivity} for the full derivation). This formulation can be interpreted as a sign-weighted correlation between technology-specific deviations and the system-wide flexibility signal.

In contrast to the methodology for flexibility needs, which is well documented in several studies \cite{artelys_mainstreaming_2017, acer_flexibility_2025,trinomics_study_2022, european_commission_role_2019,rte_bilan_2015, artelys_optimal_2018, artelys_effect_2019, iea_managing_2023}, the methodology for flexibility provision remains only sparsely described in the literature. The approach presented here builds on \cite{iea_managing_2023} and on documentation from \textit{Artelys Crystal Super Grid} software, which is no longer publicly available \cite{artelys_artelys_2025}. 

To the best of the authors' knowledge, this formulation differs from the flexibility attribution approach used in the METIS project \cite{artelys_mainstreaming_2017}, which does not guarantee additivity, as interaction effects between technologies lead to path-dependent allocations (see Appendix~\ref{app:non_additivity_add_tech_method}). In contrast, the formulation proposed here directly decomposes the flexibility signal itself, ensuring an exact and consistent attribution across technologies.

\subsection{Modelling of the German energy system}
\label{sec:modelling}
This section first introduces the general modelling framework (Section~\ref{sec:general-model}), then presents the most relevant flexibility-related features of the model (Section~\ref{sec:methods:modelling:flexibility-options}) and concludes with the scenario creation in Section~\ref{sec:methods:modelling:scenarios}.

\subsubsection{Model of the German energy system}
\label{sec:general-model}
We use a sector-coupled model of the German energy system: PyPSA-DE \cite{lindner_pypsa-_2025}. It is based on the open-source, multi-vector energy system model PyPSA-Eur \cite{horsch_pypsa-eur_2018}. This, in turn, is built on the modelling framework Python for Power System Analysis (PyPSA) \cite{brown_pypsa_2018}.\newline
The model co-optimises the investment and operation of generation, storage, conversion and transmission infrastructures in a linear optimisation problem. It represents the energy system infrastructure for Germany and its neighbouring countries for the years 2025, 2035 and 2045. The model optimises supply, demand, storage, and transmission networks across a range of sectors including electricity, heating, transport, agriculture, and industry with perfect foresight over a weather year for each investment period, but myopic foresight between the investment periods. It is designed to minimise the total system costs by utilising a flexible number of regions and hourly resolution for full weather years. Germany is resolved in 30 regions, with the remaining 19 regions representing electricity-interconnected neighbouring countries as well as Italy and Spain. To capture the short-term storage operation and flexibility needs accurately, we use an hourly resolution for one full year (8760 hours) in all scenarios.

PyPSA-DE provides a comprehensive representation of all major CO$_2$-emitting sectors. The CO$_2$ pricing is done indirectly via a budget. Non-CO$_2$ greenhouse gas (GHG) emissions are taken into account by tightening the remaining CO$_2$ budget. This is necessary as PyPSA-DE only covers CO$_2$ emissions.\newline 
The heating sector is disaggregated into individual buildings (decentral) and district heating (central). The heating supply technologies include heat pumps, resistive heaters, gas boilers, and CHP plants, which are optimised separately for decentralised use and central district heating. District heating networks can also be supplemented with waste heat from various conversion processes, including electrolysis, methanol production, ammonia synthesis, and Fischer-Tropsch fuel synthesis.

The transport sector is divided into three main segments: land transport (by energy source), shipping, and aviation. The main assumptions of the future development are a high degree of electrification in the land transport and usage of hydrogen and carbonaceous fuels in the aviation and shipping segments. The Aladin model provides Germany-specific data on e.g. the share of electrification in the transport sector \cite{plotz_modelling_2014}. Within the industrial sector, major sub-sectors such as steel, chemicals, non-metallic minerals, and non-ferrous metals are modelled separately to account for various fuel and process-switching scenarios. Fuels and processes as well as their transition paths are exogenously set. The Forecast model delivers data on production volumes for industry processes \cite{fraunhofer_isi_forecasteload_2024}. PyPSA-DE also includes different energy carriers and materials, with a focus on managing the carbon cycle through methods like carbon capture, BECCS, direct air capture, and using captured carbon for synthetic fuels. Renewable potentials and time series for wind and solar electricity generation are calculated using \textit{atlite}, taking into account land eligibility constraints such as nature reserves and distance criteria from settlements \cite{hofmann_atlite_2021}.\\
\\

PyPSA-DE incorporates various Germany-specific policies. The CO$_2$ emission amounts are consistent with Germany's GHG targets. Nuclear power plants are not included, following the political phase-out in 2023, and after 2038, coal-fired power plants are banned. From 2030 onwards, there is the possibility to build hydrogen-fuelled OCGT, CCGT and CHP plants. 
Gas turbines built from 2035 are assumed to be hydrogen-ready and are exogenously switched to hydrogen operation in 2045.
The model includes minimum capacity constraints on the buildout of renewable energy which are oriented towards the German goals until 2030.

An important exogenous driver of flexibility needs is the representation of inelastic system loads (non-dispatchable demand). The load profiles for major demand categories in Germany in 2045 can be found in the Appendix (Fig.~\ref{fig:app:loads_by_carrier}). Household and industrial electricity consumption exhibit pronounced daily and weekly patterns. Household profiles are generated from synthetic ENTSO-E–based data, while industrial load levels follow data from the Hotmaps project \cite{simon_pezzutto_and_stefano_zambotti_and_silvia_croce_and_hotmaps_2018} and their temporal structure is derived from sector-specific load profiles \cite{kirstin_ganz_wie_2021}. Agricultural electricity demand is modeled as constant and remains negligible. BEV load profiles are shaped by hourly traffic counts, with availability defined as the inverse of traffic intensity to implicitly reflect post-arrival plug-in behaviour (see \ref{app:bev_modelling}).
Heat demand shows a strong temperature-driven seasonal pattern with winter peaks and additional daily morning and evening peaks. With increasing electrification, this heat-driven seasonality becomes a major contributor to long-term flexibility requirements. Other load categories are either assumed to be flat or too small to materially affect system flexibility.

\subsubsection{Flexibility options}
\label{sec:methods:modelling:flexibility-options}

According to Kondziella and Bruckner \cite{kondziella_flexibility_2016}, flexibility options in energy systems can be allocated across both the supply and demand sides. These include highly flexible power plants, energy storage, curtailment, DSM, grid extension, virtual power plants, and linkage of energy markets through sector coupling.
PyPSA-DE incorporates several key features that are particularly relevant for capturing flexibility in the energy system:\\

\begin{itemize}
    \item \textbf{Flexible generation:} 
    The model includes various dispatchable generation technologies, such as gas- and hydrogen-fired power plants (OCGT and CCGT), which can rapidly ramp up or down to balance supply and demand fluctuations.

    \item \textbf{Flexible demand:} 
    A key DSM option is the flexible charging of battery electric vehicles (BEVs). An exogenously defined share of BEV battery capacity can be used for flexible charging. In this case, the charging profile is not fixed but optimised endogenously by the model within each day. To prevent the use of BEV batteries as long-term storage and to ensure user satisfaction, the state of charge must reach at least 80\% every morning at 7:00. V2G is also included as an additional flexibility option for BEVs, allowing them to discharge electricity back to the grid. BEV flexible charging and V2G future behaviour is highly uncertain and there are different approaches in the literature to model the BEV load \cite{goke_how_2023,shariatzadeh_electric_2025}, charging and discharging behaviour \cite{schill_power_2015,hanemann_effects_2017, lauvergne_integration_2022} as well as the effect on the residual load \cite{muratori_impact_2018, fischer_electric_2019,taljegard_impacts_2019}. Refer to the Appendix \ref{app:bev_modelling} for a more detailed description of the assumptions and modelling of BEV flexibility.
    In addition, the model includes industrial DSM, where electricity demand can be shifted within the same day, constrained by both the shiftable power and maximum holding duration. The maximum load shifting potentials are 6~GW in 2035 and 10~GW in 2045 and the holding duration is up to 4~hours (following~\cite{ffe_regionale_2021}). The future potentials are highly uncertain and depend on factors such as the degree of electrification, the flexibility of industrial processes, and the willingness of industries to participate in DSM. Refer to the Appendix \ref{app:industrial_dsm} for a more detailed description of the assumptions and modelling of industrial DSM.
    \item \textbf{Energy storage:} 
    PyPSA-DE encompasses multiple storage technologies to manage different energy forms. Electricity can be stored in home and utility-scale batteries, pumped hydro storage (PHS), hydrogen (produced via electrolysis and stored in tanks or caverns), and synthetic energy carriers such as methane or methanol. Hydrogen can be re-electrified using gas turbines or fuel cells, while (synthetic) methane can be converted to electricity in gas turbines or combined heat and power (CHP) plants. Thermal energy storage is modeled as large-scale water pits or tanks for seasonal heat storage in district heating networks and short-term storage for individual buildings.

    \item \textbf{Sector coupling and Power-to-X (PtX):} 
    The model integrates multiple energy sectors, including electricity, heating, hydrogen, and synthetic fuels. The electrification of heating and the production of hydrogen introduce significant DSM potential. Heat storage enables power-to-heat (PtH) technologies to decouple heat generation from consumption, while hydrogen storage allows temporal shifting of electricity use for electrolysis. 
    The model also includes flexible synthetic fuel production processes (e.g., methanol synthesis, Fischer--Tropsch synthesis) that convert excess renewable electricity into storable fuels.
\end{itemize}

Another important provider of spatial flexibility is the electricity transmission network. Apart from network expansion, dynamic line rating (DLR) is modelled as a flexibility option for the transmission network. DLR allows for real-time adjustment of transmission line capacities based on environmental conditions, such as temperature and wind speed, which can enhance the grid's ability to accommodate variable renewable energy sources \cite{glaum_leveraging_2023}. Curtailment also represents a flexibility option, as it allows the system to reduce generation from renewable sources when supply exceeds demand or grid capacity. 
\subsubsection{Scenarios}
\label{sec:methods:modelling:scenarios}

\begin{table*}[ht]
\caption{Overview of flexibility scenarios and main configuration parameters}
\begin{center}
\begin{tabular}{lcccc}
\toprule
\textbf{Parameter} & \textbf{LowFlex (LF)} & \textbf{LowBattery (LB)} & \textbf{Base (BA)} & \textbf{HighFlex (HF)} \\
\midrule
Dynamic line rating & $\times$ & $\times$ & $\times$ & \checkmark \\
Iron-air battery technology & $\times$ & $\times$ & $\times$ & \checkmark \\
\midrule
BEV flexible charging (start/avail.) & 2045 / 20\% & 2045 / 20\% & 2035 / 50\% & 2035 / 80\% \\
V2G (start/avail.) & $\times$ & $\times$ & $\times$ & 2035 / 80\% \\
Industry DSM & $\times$ & $\times$ & $\times$ & \checkmark \\
\midrule
Electrolysis operation & 50\% inflexible & flexible & flexible & flexible \\
Other PtX & inflexible & flexible & flexible & flexible \\
\midrule
Battery charge/discharge capacity (GW) & 75\% of Base & 50\% of Base & unrestricted & unrestricted \\
Heat storage charge/discharge capacity (GW) & 75\% of Base & unrestricted & unrestricted & unrestricted \\
\midrule
H2 storage energy capacity (GWh) & 75\% of Base & unrestricted & unrestricted & unrestricted \\
\bottomrule
\end{tabular}
\begin{tablenotes}
\item[a] Scenario definitions based on the PyPSA-DE configuration. 
LowFlex assumes slow deployment of flexibility technologies across all sectors. LowBattery restricts battery capacity and BEV flexibility. Base represents the baseline scenario. HighFlex assumes an early and extensive rollout of flexibility options across all sectors. 
\end{tablenotes}
\label{tab:flex_scenarios}
\end{center}
\end{table*}

To capture the potential of different flexibility technologies and assess the system costs associated with limited flexibility expansion, four scenarios are defined: \textit{LowFlex (LF)}, \textit{LowBattery (LB)}, \textit{Base (BA)}, and \textit{HighFlex (HF)}. These scenarios differ in the availability and capacity limits of key flexibility technologies. The main configuration parameters are summarized in Table~\ref{tab:flex_scenarios}.\newline

BA represents a cost-efficient pathway to climate neutrality with a balanced mix of technologies and energy carriers. Renewable electricity expansion broadly meets political targets, with moderate import levels for energy carriers. This scenario is oriented towards the \textit{Technologiemix} scenario from the Ariadne project \cite{luderer_energiewende_2025}.

Transmission capacity expansion is identical across all scenarios, limited to 150\% of the 2025 infrastructure in terms of total capacity times transmission length. Dynamic line rating is only enabled in HF.\\ 
Battery electric vehicle (BEV) flexible charging varies significantly: in LF and LB, only 20\% of vehicles can charge flexibly starting in 2045, whereas in BA and HF, 50\% and 80\% respectively can charge flexibly from 2035 onwards. V2G is only available in HF from 2035 onwards for 80\% of the BEV fleet. Iron-air battery technology and industrial DSM are also exclusive to HF, with DSM limited to Germany due to data availability.\\
Electrolysis operates with reduced flexibility in LF. It has a minimum load of 50\% of its capacity, meaning it cannot operate below this threshold. Other PtX technologies like Fischer-Tropsch, methanolisation and methanation are inflexible in LF, meaning they must operate at full load throughout the year.\\ 
Storage technologies face varying capacity limits across scenarios. In LF, battery and heat storage discharge power capacities and H2 storage energy capacity are all limited to 75\% of the BA scenario. In LB, only battery discharge capacity is limited to 50\% of BA, while hydrogen and heat storage remain unrestricted. Note that these restrictions apply to total capacities per location in 2035 and 2045; existing capacities in 2025 are not reduced. Consequently, total installed capacity may slightly exceed the derived thresholds where a technology's 2025 capacity already surpasses the threshold-constrained 2045 capacity.\\

Although this study focuses on Germany, flexibility options are modeled for the entire interconnected European system, except for the industry DSM. This approach prevents neighbouring countries from implicitly subsidising German flexibility needs by providing cross-border flexibility without the associated costs being reflected in Germany. The cost-optimal solution does not distinguish between domestic and cross-border flexibility provision, which would otherwise lead to biased results.

\section{Results}
\label{sec:results}

\subsection{Structural changes in the energy system}
\label{sec:res:system-change}

\begin{figure*}[htb]
    \centering
    \includegraphics[width=0.9\textwidth]{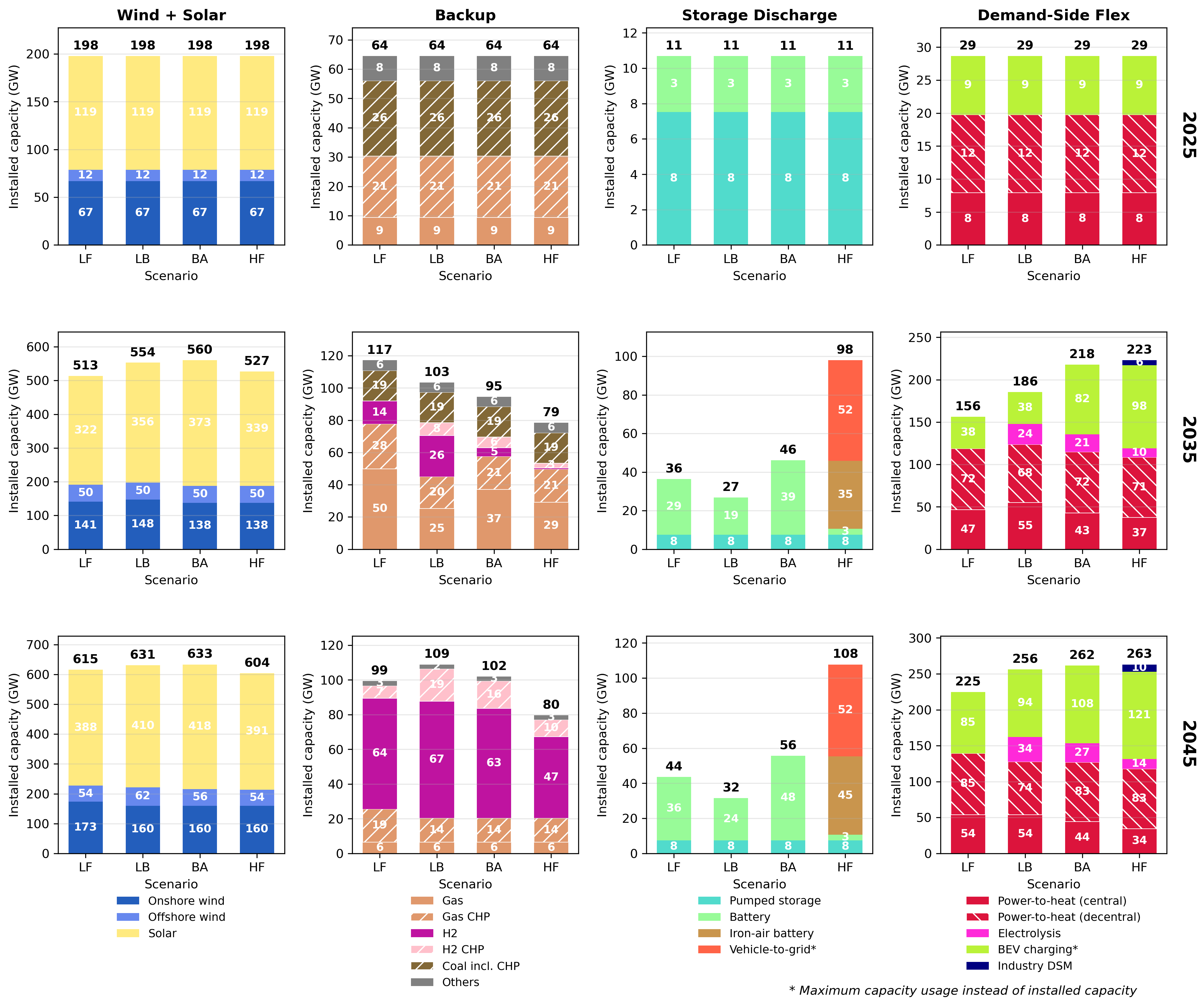}
    \caption{\textbf{Capacity comparison for different scenarios}
        The Figure shows the capacity of key technologies (Renewables, Backup, Storage, Demand-Side) for different scenarios on the development of the energy system from 2025 until 2045. LF = LowFlex, LB = LowBattery, BA = Base, HF = HighFlex.}
    \label{fig:capacity-comparison}
\end{figure*}

To contextualize the development of flexibility needs and provision, it is essential to examine the structural changes in the underlying energy systems. Between 2025 and 2045, the installed capacities of solar and wind (onshore+offshore) expand from 119~GW to 388-418~GW and from 79~GW to 214-227~GW across all scenarios. The land transport sector reaches 85\% electrification, resulting in an increased electricity consumption from 36~TWh in 2025 to 206~TWh in 2045. The consumption of directly electrified heat reaches 161-184~TWh in 2045. In the industrial sector, electrification leads to an increased electricity consumption of 165~TWh by 2045. While only 5\% of electricity demand is modeled as flexible in 2025, this share increases to approximately 35\% by 2045, and is likely to be even higher if potential flexibilities in industry and household are considered. 

In Figure~\ref{fig:capacity-comparison}, the installed power capacities of key technologies are compared across different scenarios. Refer to Table~\ref{tab:storage-capacities} for an overview on the storage energy capacities in 2045.  

The Base (BA) scenario serves as a reference point, with the other scenarios exhibiting similar capacity trends but with variations in specific technologies. Both the LowFlex (LF) and LowBattery (LB) scenarios have higher costs to provide flexibility and therefore struggle more to integrate especially solar generation. Wind power, which is less variable on the daily timescale, therefore reaches higher capacities in LF and LB. The HighFlex (HF) scenario can make better use of renewable resources and therefore manages with lower overall installed capacities compared to BA. 

The scenarios exhibit a decreasing amount of dispatchable backup capacity with increasing flexibility in 2035. LF in particular builds additional gas-fired capacity, as hydrogen is expensive due to inflexible domestic production and costly imports. LB, by contrast, exploits flexible electrolysis operation to increase system flexibility and therefore invests more in H2-fuelled backup capacity. Both lower-flexibility scenarios rely more heavily on central resistive heaters compared to BA and HF. This reflects two factors: the limited operational flexibility of the heat system combined with the inherently flexible operation of resistive heaters, and the high costs of gas and hydrogen generation and imports. In 2045, all gas capacity built from 2035 is exogenously retrofitted to H2 backup, after which LF, LB, and BA require a similar total amount of dispatchable backup capacity. The comparatively lower CHP backup capacity in LF is a result of higher utilisation of existing CHP plants and a greater contribution of central resistive heaters to heat supply. LF is also the only scenario in which CHP plants with carbon capture are built. The larger spread in backup capacity across scenarios in 2035 compared to 2045 reflects the fact that in 2035 the system is still in transition from a fossil-fuel based to a carbon-neutral, electrified system and therefore more sensitive to the availability of flexibility options. By 2045, the system has adapted to the available flexibility options, reducing cross-scenario differences. 
HF reduces the need for backup capacity by 16 and 22~GW compared to BA in 2035 and 2045 respectively, owing to additional storage discharge from V2G and iron-air batteries.

The installed storage capacity varies significantly across the scenarios. LF and LB are constrained to build only 75\%/50\% of the battery discharge capacity of the BA scenario which they fully utilise. BA builds 39~GW of battery dispatch capacity and 310~GWh of battery energy capacity in 2035, resulting in an energy-to-power ratio of 8 hours. PHS capacities are identical across the scenarios due to already installed assets and limits in the further expansion in Germany. In HF, battery dispatch capacity is not expanded beyond the 3~GW already installed in 2025. Instead, this is substituted by 45~GW of iron-air batteries by 2045, complemented by an additional maximum capacity contribution of 52~GW from V2G. The iron-air batteries have a storage energy capacity of 1765~GWh and an energy-to-power ratio of 39 hours, which allows them to provide flexibility on longer timescales (2045). 
For flexible electricity consumers, the 50\% minimum load constraint on electrolysis in LF leads to hydrogen production being relocated entirely outside Germany. In LB, 3/7~GW more electrolysis capacity is built compared to BA, as the flexible operation can partly compensate for the restricted battery expansion. In HF, 11/13~GW less electrolysis capacity is built, as the value of flexible electrolysis operation is reduced by the availability of other flexibility options such as industrial DSM.\\
The electrification of decentralised heat is comparable across all scenarios. Only LB exhibits slightly lower decentral heat capacities, as it strongly invests in decentral thermal storage to enable more flexible operation of decentralised heat technologies, thereby reducing the need for installed generation capacity, and shows slightly higher biomass utilisation in decentral heat. 
For central heat, LF and LB build more resistive heaters to exploit their flexible operation as a substitute for the limited flexibility from other technologies. HF, in contrast, builds fewer central resistive heaters, as the availability of other flexibility options allows more heat pumps to be operated, which are more efficient but less flexible.\\
BEV maximum charge and discharge capacity rather than installed capacity is reported in Figure~\ref{fig:capacity-comparison} and Table~\ref{tab:storage-capacities} in the Appendix, as the actual flexibility contribution is constrained by various charging and discharging constraints. LF and LB have a lower maximum charge power of 85/94~GW compared to 108/121~GW in BA/HF (2045), reflecting their lower availability of flexible charging.

\subsection{Flexibility needs and its causes}
\label{sec:res:flex-needs}

The importance of flexibility in the energy system is reflected in the increasing flexibility needs across all timescales. Figure~\ref{fig:flex} (A) illustrates the evolution of flexibility needs from 2025 to 2045 on daily, weekly, and annual timescales for the different scenarios. For BA, daily flexibility needs more than triple by 2035, rising from 51~TWh/a in 2025 to 165~TWh/a (factor 3.2), and nearly quadruple to 188~TWh/a by 2045 (factor 3.7). Note that these values, while expressed in TWh/a, represent a measure of flexibility potential rather than an energy demand or supply. Weekly flexibility needs also rise, albeit at a slower rate, more than doubling by 2035 (47$\to$101~TWh/a) and reaching 120~TWh/a by 2045, a 2.6-fold increase. Annual flexibility needs grow by a factor of 2.75 by 2035 and 3.25 by 2045 (28$\to$77$\to$91~TWh/a).
These upward trends are robust across different weather years and energy system configurations, as flexibility needs are directly determined by the residual load. However, the residual load is an endogenous outcome of the optimisation and therefore subject to several sensitivities. \cite{iea_managing_2023} find that daily flexibility needs vary little across weather years, while weekly and annual flexibility needs show variability of up to 13\% and 30\%, respectively, based on 30 weather years across Europe. A sensitivity analysis using weather years 2013, 2020, and 2023 confirms this ordering: daily flexibility needs vary by at most 15\%, weekly by 18\%, and annual by 20\% (see 
Figure~\ref{fig:app:flex-needs-weather-years}). Energy systems with fewer flexibility options struggle more to integrate renewable energy and therefore exhibit higher flexibility needs overall.

A major driver of rising flexibility needs across different timescales is the growing share of non-dispatchable supply from VRES. See Figure~\ref{fig:app:heatmap_non-disp-supply} in the Appendix for the generation profiles of VRES for BA in 2045. The strong diurnal and seasonal patterns of solar generation are clearly visible, with high generation during midday and in the summer months and low generation in winter. Wind generation exhibits variability on longer (daily to weekly) timescales, as large weather systems move across the continent, and tends to produce more in winter, while also being exposed to short-term hourly fluctuations, albeit to a much lesser extent than solar. Ror generation plays only a minor role, with somewhat higher generation in spring. 

The increasing penetration of solar generation primarily drives the sharp rise in daily flexibility needs, whereas the growing share of wind generation contributes to higher variability on weekly and annual timescales. Figure~\ref{fig:flex} (A) illustrates the impact of different technologies across scenarios. For BA, more than 70\% of the flexibility needs caused by solar are daily. These daily flexibility needs are not evenly distributed over the year. In the summer season (April--September), the average daily flexibility needs in 2045 are 619~GWh/d compared to 408~GWh/d in the winter season (October--March), with a maximum of approximately 1075~GWh/d in 2045 compared to 300~GWh/d in 2025. Wind contributes approximately 70\% of its total flexibility to weekly and annual timescales, which are more evenly distributed over the year but still on average 30\% higher in winter.\newline

The other side of the coin is the non-dispatchable demand which was described in Sec.~\ref{sec:general-model}. The diurnal demand pattern of households, industry and especially BEVs (see Fig.~\ref{fig:app:heatmap_non-disp-demand} in the Appendix) has a positive effect on the flexibility needs, meaning it reduces the need for flexibility on daily timescales (see Figure~\ref{fig:flex} (A)). This is due to the fact that the demand is on average higher during the day than at night which aligns with the solar generation pattern. However, the loads increase the weekly flexibility needs. This finding indicates that the lower demand on the weekends compared to weekdays shifts more generation into the week, which increases the need for flexibility on these timescales. The electricity demand from decentral heat pumps and resistive heaters is also considered non-dispatchable due to limited thermal storage capacity. Both technologies increase weekly and annual flexibility needs, but their effect on daily flexibility needs is ambiguous and generally smaller.

The combined effect of non-dispatchable supply and demand can be seen in the development of the residual load. Fig.~\ref{fig:app:residual-load} in the Appendix shows the evolution of the residual load for two representative months (February and July) as well as the residual load duration curve. The residual load fluctuations between consecutive hours can reach up to ±149~GW in 2045 (±39~GW in 2025). The maximum increases from 80 to 193~GW from 2025 until 2045 and the minimum from -35~GW to -228~GW in 2045. The maximum is particularly important to determine the need for battery discharge capacity and backup power plants whereas the minimum is relevant for flexible demand side technologies like PtH, PtG and battery / BEV charging. 

The spatial distribution of drivers of flexibility needs across the 30 regions in Germany for BA in 2045 can be found in the Appendix in Figure~\ref{fig:app:flex-spatial} (A). Solar generation is relatively evenly distributed across Germany, whereas wind generation is more concentrated in the north and coastal areas. Consequently, regions with high solar generation exhibit more pronounced daily flexibility needs, while areas with significant wind resources show greater weekly and seasonal flexibility requirements.

\subsection{Flexibility provision}
\label{sec:res:flex-provision}

\begin{figure*}[htb!]
    \centering
    \footnotesize

    \begin{subfigure}{\textwidth}
        \centering
        \includegraphics[width=\textwidth]{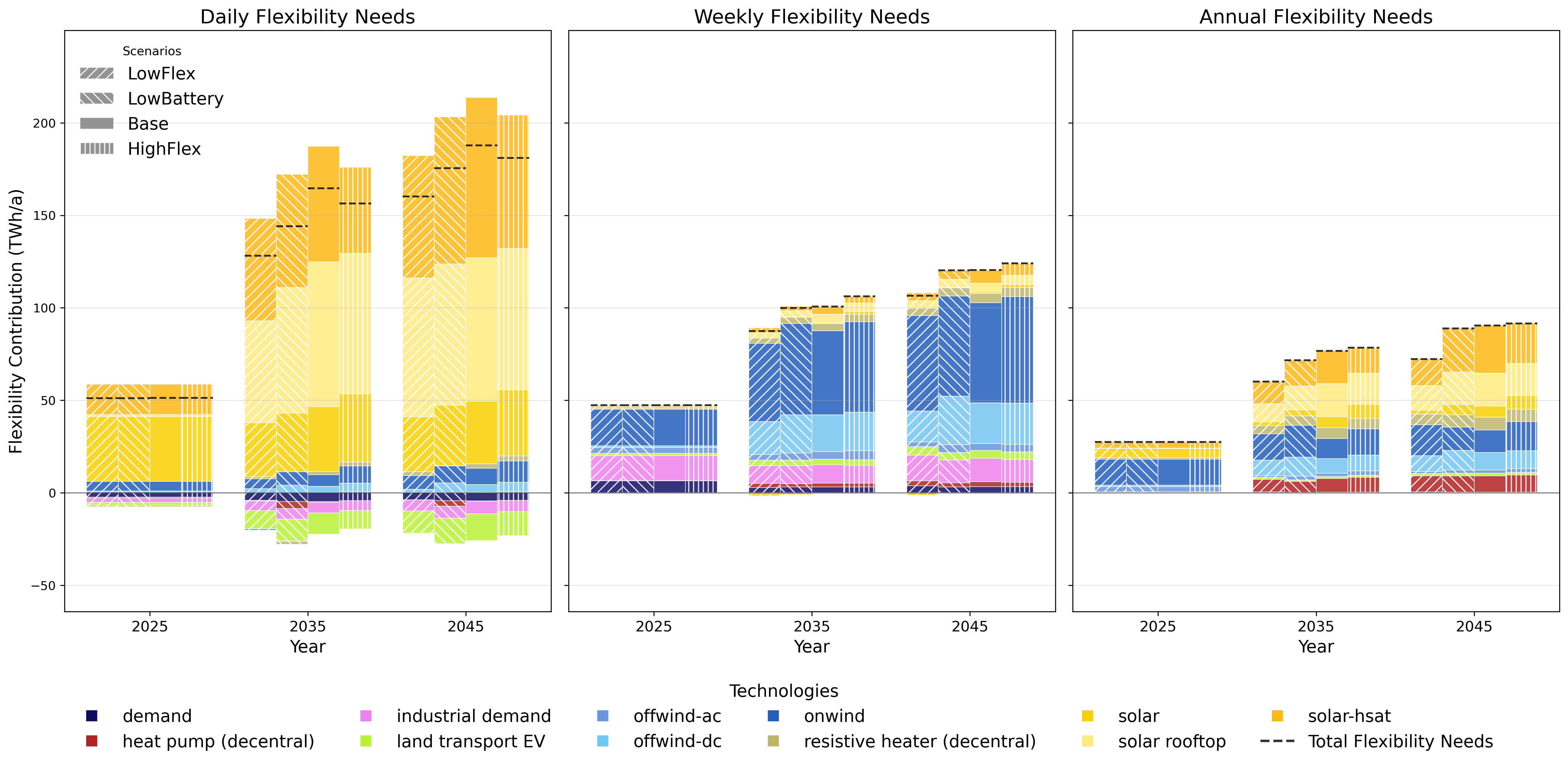}
        \caption{Flexibility needs across different scenarios}
    \end{subfigure}

    \vspace{0.8em}

    \begin{subfigure}{\textwidth}
        \centering
        \includegraphics[width=\textwidth]{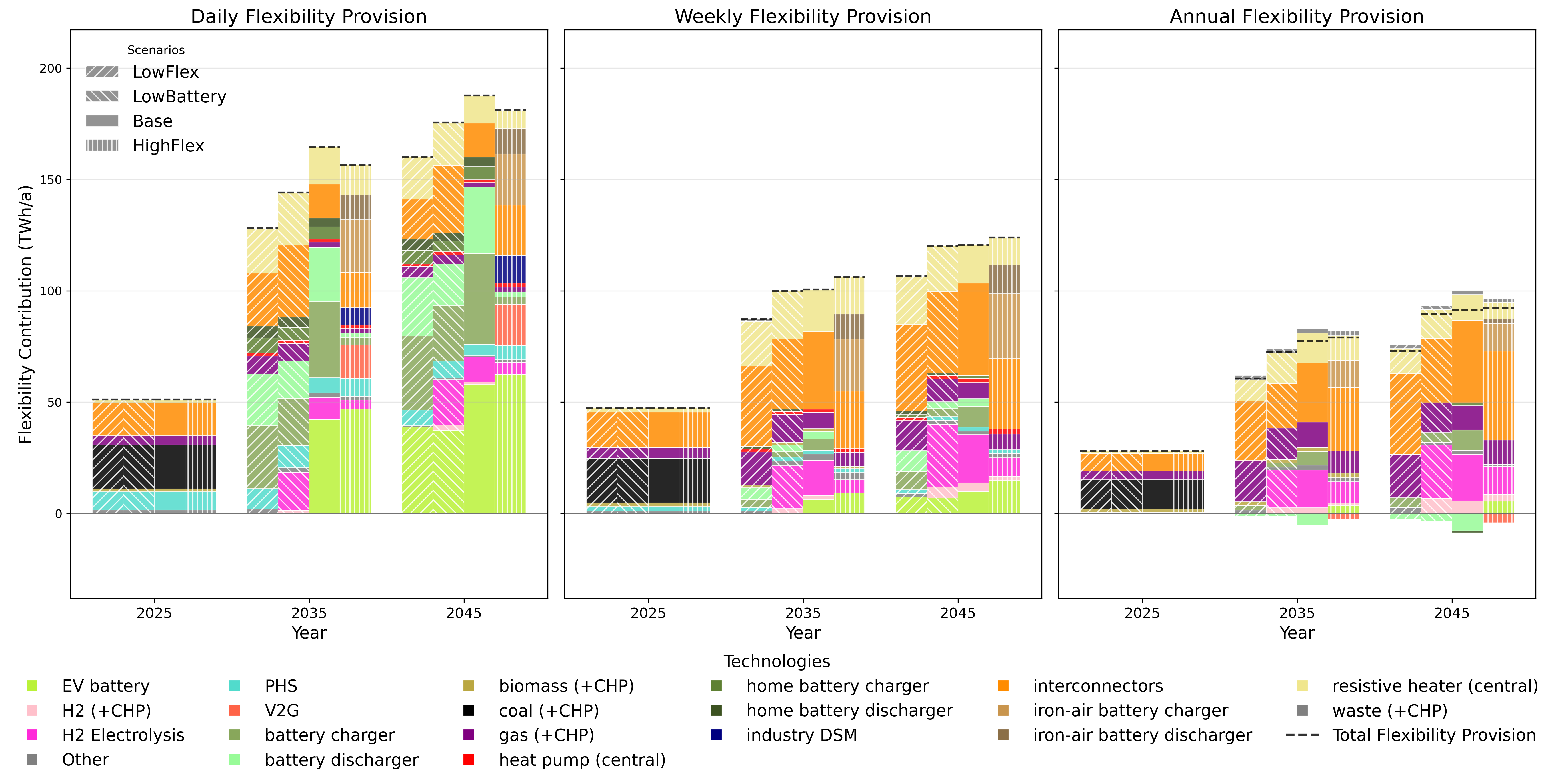}
        \caption{Flexibility provision across different scenarios}
    \end{subfigure}

    \caption{\textbf{Flexibility causes and provision across different scenarios.} 
    The figure shows flexibility needs (A) and provision (B) for different scenarios across various timescales.}
    \label{fig:flex}
\end{figure*}

A major advantage of the presented method for calculating flexibility needs and provision is that it enables a clear illustration of how different technologies contribute to flexibility across various timescales. Figure~\ref{fig:flex} (B) shows the flexibility provision across scenarios and timescales for the years 2025 to 2045.

\subsubsection{Daily}
Daily flexibility refers to the shifting of energy within a single day. In BA, a substantial share of 38\% is provided by  stationary batteries, with 22\% from charging and 16\% from discharging. The second-largest contribution of 30\% stems from flexible BEV charging. Resistive heaters represent another important source of flexible demand (7\%), followed by electrolysis (6\%), cross-border electricity exchange (5\% imports, 3\% exports), and pumped hydro storage (PHS). Gas-fired plants and heat pumps contribute only minor shares to daily flexibility.\\
While the flexibility mix in 2035 and 2045 exhibits a similar composition, the overall magnitude is lower in 2035. Both years differ substantially from 2025, when flexibility provision is dominated by coal- and gas-fired generation as well as cross-border exchanges, and total flexibility requirements are considerably smaller. It should be noted that the relatively high contribution of coal partly results from the absence of unit commitment constraints in the model.

Comparing the scenarios reveals that the availability of flexibility options leads to substantial changes in the overall flexibility provision mix. In LF and LB, where flexibility options are limited, the system relies more heavily on cross-border exchanges to provide daily flexibility. In LB, no flexibility contribution from electrolysis is observed, as hydrogen production largely relocates to Spain when flexible operation is no longer possible. This indicates that economically viable electrolysis operation in Germany requires sufficient flexibility to capitalise on low electricity prices during periods of excess renewable generation. Despite the 75/50\% restriction on the discharging capacity of batteries, they still contribute a considerable share to daily flexibility in LF and LB. PHS and central resistive heaters are utilized more intensively to compensate for the absence of other storage options.\\ 
LB exhibits a composition similar to BA but compensates for the reduced flexibility from stationary batteries with higher contributions from electrolysis, PHS, gas turbines, central resistive heaters, and trade.\\
In contrast, HF illustrates the potential of an energy system endowed with extensive flexibility options. Its lower overall daily flexibility need (compared to BA) does not stem from limited renewable integration, as in LF or LB, but rather from the system's ability to utilize renewable generation more efficiently, requiring less installed VRES capacity and, consequently, reducing the flexibility needs driven by variable renewable output. Flexible BEV charging and V2G operation together provide more than 40\% of daily flexibility in 2045. The deployment of iron--air batteries further enhances the system’s capability to balance short-term fluctuations, contributing 19\%. In addition, industrial DSM contributes around 7\% of daily flexibility, representing a significant share of the overall provision.

\subsubsection{Weekly}
On weekly timescales, the flexibility provision mix changes significantly. Batteries and BEVs play only a minor role, as their storage capacities are insufficient to cover fluctuations over several days. A major contributor in BA is electricity trade, with 22\% from imports and 12\% from exports (2045). Trade between European countries helps to balance the uneven wind production across the continent as large weather systems sweep across it \cite{frysztacki_strong_2021}. Electrolysis provides 18\% of weekly flexibility, followed by central resistive heaters (14\%) and gas+H2 turbines (9\%). LF and LB require higher contributions from gas generation and electricity trade to compensate for their limited flexibility options. In HF, the iron--air battery has the largest impact, contributing 34\%. This reduces the need for electrolysis and gas-based backup generation.

\subsubsection{Annual}
For annual flexibility provision, electricity exchange via interconnectors is the dominant contributor across all scenarios and years, accounting for more than 40\% in all cases. In BA, the remaining flexibility is provided by electrolysis (23\%), gas- and H2-based backup capacity (17\%), and central resistive heaters (13\%). The overall mix is broadly similar across scenarios, with the notable exception of LF, where electrolysis is absent and its contribution is instead substituted by a larger share of gas-based backup capacity. Notably, battery storage exhibits a negative annual flexibility contribution of around 8\% in 2045, as its short-term intra-day cycling pattern amplifies rather than smooths seasonal residual load variations.

\subsubsection{Spatial flexibility provision}
The spatial distribution of flexibility provision shows a clear geographic pattern that mirrors the causes (see Figure~\ref{fig:app:flex-spatial} (B) in the Appendix). In the northern and coastal regions, where wind generation is concentrated, flexibility is predominantly provided by interconnectors and H2 electrolysis across all timescales, reflecting the co-location of large-scale electrolysis with cheap offshore wind electricity. The northern offshore node exhibits by far the largest total flexibility provision, underlining its role as the primary balancing hub of the system. For daily flexibility, battery storage and BEV charging contribute more visibly in central and southern regions, where solar-driven intra-day imbalances are more pronounced. Interconnectors dominate the flexibility provision mix at all timescales and across virtually all nodes, but their relative share increases further at the annual timescale, where domestic storage options are largely insufficient to bridge seasonal imbalances.

\subsubsection{Aggregated vs. Per-Node}
The methodology has been applied so far to the aggregated statistics of the energy system for Germany. However, it can also be applied to the individual 30 nodes of the German system and be aggregated afterwards. The results can be found in the Appendix in Figure~\ref{fig:app:flex-per-node}. The general trends remain; however, the flexibility needs increase across all timescales and all years. The role of electricity transmission is much more pronounced, especially for weekly and annual timescales.

\subsection{Flexible technology operation}
\label{sec:res:flexible-technology-operation}

\begin{figure*}[htbp]
    \centering
    \includegraphics[width=\textwidth]{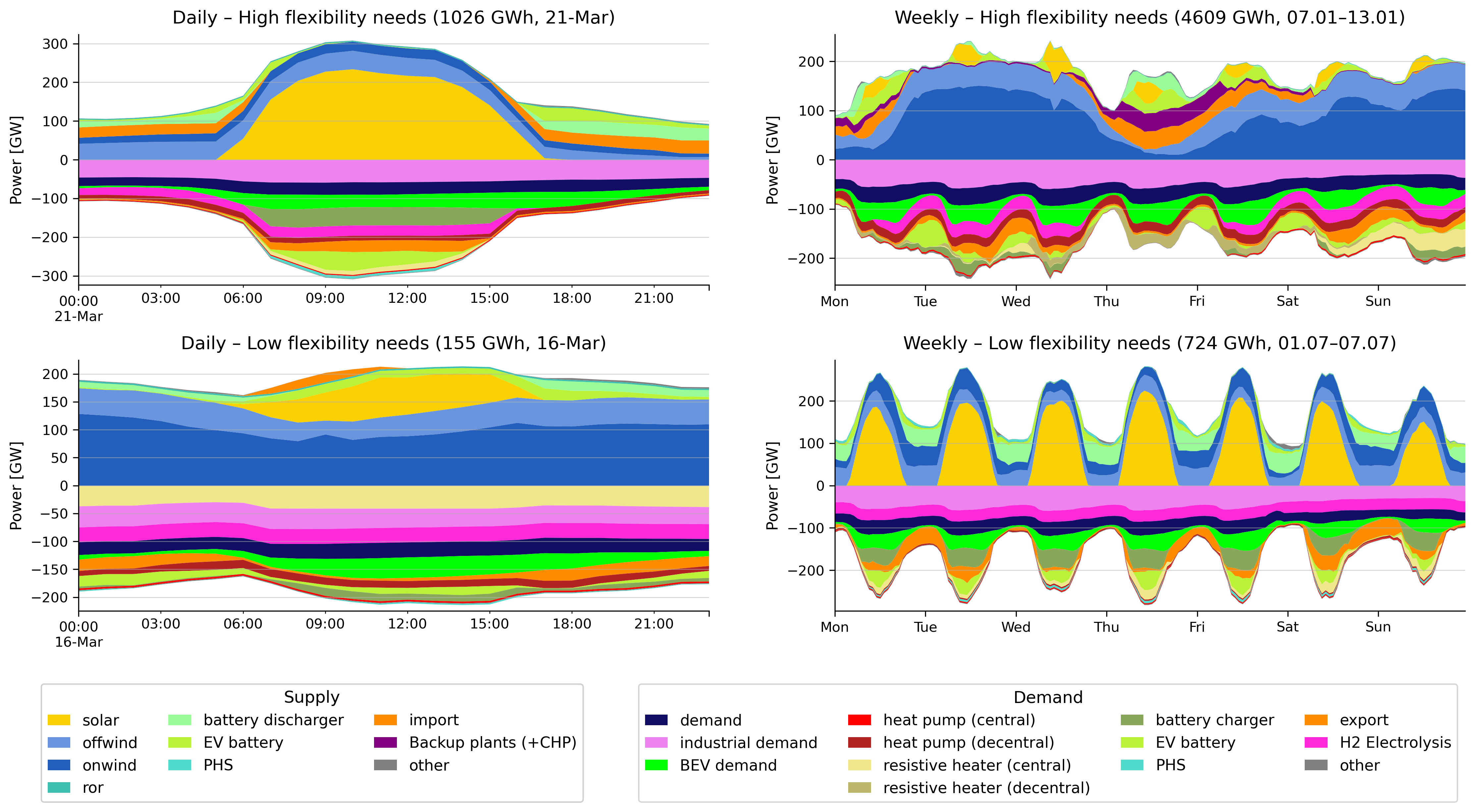}
    \caption{Energy balance for the Base scenario in 2045 showing representative periods with high and low flexibility needs at daily (left) and weekly (right) granularity.}
    \label{fig:app:days-weeks-balances}
\end{figure*}

To provide a clearer understanding of how the system operates under varying flexibility requirements, Figure~\ref{fig:app:days-weeks-balances} illustrates the operation of the energy system in BA for representative days (left) and weeks (right) characterised by high and low flexibility needs. Additionally, Figure~\ref{fig:app:tech-profiles} in the Appendix illustrates the operation profiles of key flexibility technologies for these representative periods.

\subsubsection{Daily flexibility dynamics}
A typical high-flexibility day (Figure~\ref{fig:app:days-weeks-balances} a) is characterised by substantial solar generation around midday, combined with an alternating wind profile throughout the day that correlates with the solar output. The system utilizes the excess solar generation to charge BEVs and stationary batteries, operate electrolysis, export electricity, drive power-to-heat (PtH) units, and pump water into PHS. In this situation, the entire portfolio of short-term flexibility options is activated. After the midday solar peak, the system discharges batteries and imports electricity to meet the residual electricity and heat demand of households and industry. Solar generation orchestrates the operation of all other components of the energy system. 

On an average flexibility day, solar generation is lower and/or the wind profile complements the reduced solar generation during the night. The overall pattern resembles that of the high-flexibility case but with lower intensity of flexibility operation. Typically, BEV charging, electrolysis, and batteries absorb the excess generation in summer, while in winter PtH units also play a relevant role. 

A low-flexibility day, by contrast, is characterised by a relatively stable combined output of solar and wind generation (Figure~\ref{fig:app:days-weeks-balances} c). This can occur either under very low solar generation with steady wind conditions or, as illustrated in the figure, through a complementary interplay of solar and wind production. In this case, the system still employs several flexibility technologies — particularly electrolysis and batteries — but these operate in a near-baseload mode rather than being used dynamically to balance fluctuations. 

These technologies exhibit a characteristic horizontal ellipse in their operation profiles, reflecting the daily rhythm of solar generation: charging and consuming during peak hours, discharging and reducing output overnight. This elliptical pattern is the hallmark of technologies operating on a daily timescale. 

As daily flexibility needs fluctuate throughout the year, the contributions of individual technologies also vary considerably. Figure~\ref{fig:app:temporal-flex-daily} in the Appendix shows the daily flexibility contributions for each day of 2045 in BA for selected technologies. BEVs, batteries, and PHS provide most of their daily flexibility during the summer months, when needs are elevated due to high solar penetration. Electrolysis also contributes significantly in winter, when wind generation exhibits stronger daily variability. Import and export show no clear seasonal pattern, alternating between providing and causing daily flexibility needs, but with a higher total provision. Resistive heaters in district heating networks provide positive flexibility throughout the year, with a slight reduction in winter. In contrast, resistive heaters in rural and decentralised areas create flexibility needs in winter, owing to limited storage capacity — a pattern similarly observed for heat pumps. Backup capacity contributes primarily in winter, coinciding with the highest residual load peaks.

\subsubsection{Weekly flexibility dynamics}
A week with high weekly flexibility needs typically occurs during winter, when wind generation exhibits strong variability. In such periods, the system relies heavily on imports, gas- and hydrogen-based backup generation, as well as batteries, to meet electricity and electrified heat demand. In Figure \ref{fig:app:days-weeks-balances} b), the second week of January is shown as an example. Here, wind generation is high at the beginning of the week but drops to very low levels on Thursday. This is precisely when flexible backup capacities, grid imports, and battery storage play a crucial role in maintaining system balance. In addition, flexible demand from technologies such as BEV charging, battery charging, and electrolysis is reduced.

In a medium-flexibility week, the wind profile is less variable, allowing the system to meet demand without activating backup generation. In this case, batteries and imports, together with reduced demand from BEV charging and electrolysis, are sufficient to balance the system.

In a low-flexibility week, both solar and wind generation are relatively stable. The system can meet demand using primarily batteries, PHS, and modest imports. Cheap solar electricity is absorbed by BEV charging, battery storage, and PHS, with surplus energy exported. During the night, batteries provide low-cost electricity for electrolysis, enabling electrolyzers to operate at near-baseload levels during these hours.

A weekly pattern is found especially in PtX and backup technologies, but also in some storage technologies, particularly PHS and BEV charging, which exhibit phases of weekly operation. In contrast to the diurnal profile, a weekly pattern can be identified as pronounced vertical lines in Figure~\ref{fig:app:tech-profiles} (Appendix), indicating that a technology supplies or consumes at high volumes over several consecutive days. Gas- and hydrogen-based backup plants exhibit this behaviour especially in winter. PtH also operates on a weekly basis, particularly in winter, while PtG (primarily electrolysis) exploits low-price periods both in summer and winter. Among storage technologies, BEV charging and PHS show, on top of their pronounced daily operation, additional phases of weekly-timescale operation.

Figure~\ref{fig:app:temporal-flex-weekly} in the Appendix shows the weekly flexibility contributions for each week of 2045 in BA. Imports and exports provide weekly flexibility throughout the year with no clear seasonal pattern. Backup capacity contributes most strongly in winter, when wind generation fluctuates on weekly timescales. Electrolysis provides weekly flexibility throughout the year, while PtH technologies cause flexibility needs in winter and contribute positively in summer.

\subsection{Impact of energy system flexibility}
\label{sec:res:degrees-of-flexibility}

\begin{figure*}[ht!]
    \centering
    \small  %
    \setlength{\tabcolsep}{8pt}
    {\large \textbf{2035}} \hspace{0.35\textwidth} {\large \textbf{2045}} \\[4pt]
    \begin{tabular}{c c}
    \multicolumn{2}{c}{\normalsize \textbf{(A) System Cost}} \\[3pt]
    \begin{subfigure}{0.46\textwidth}
        \centering
        \includegraphics[width=\textwidth]{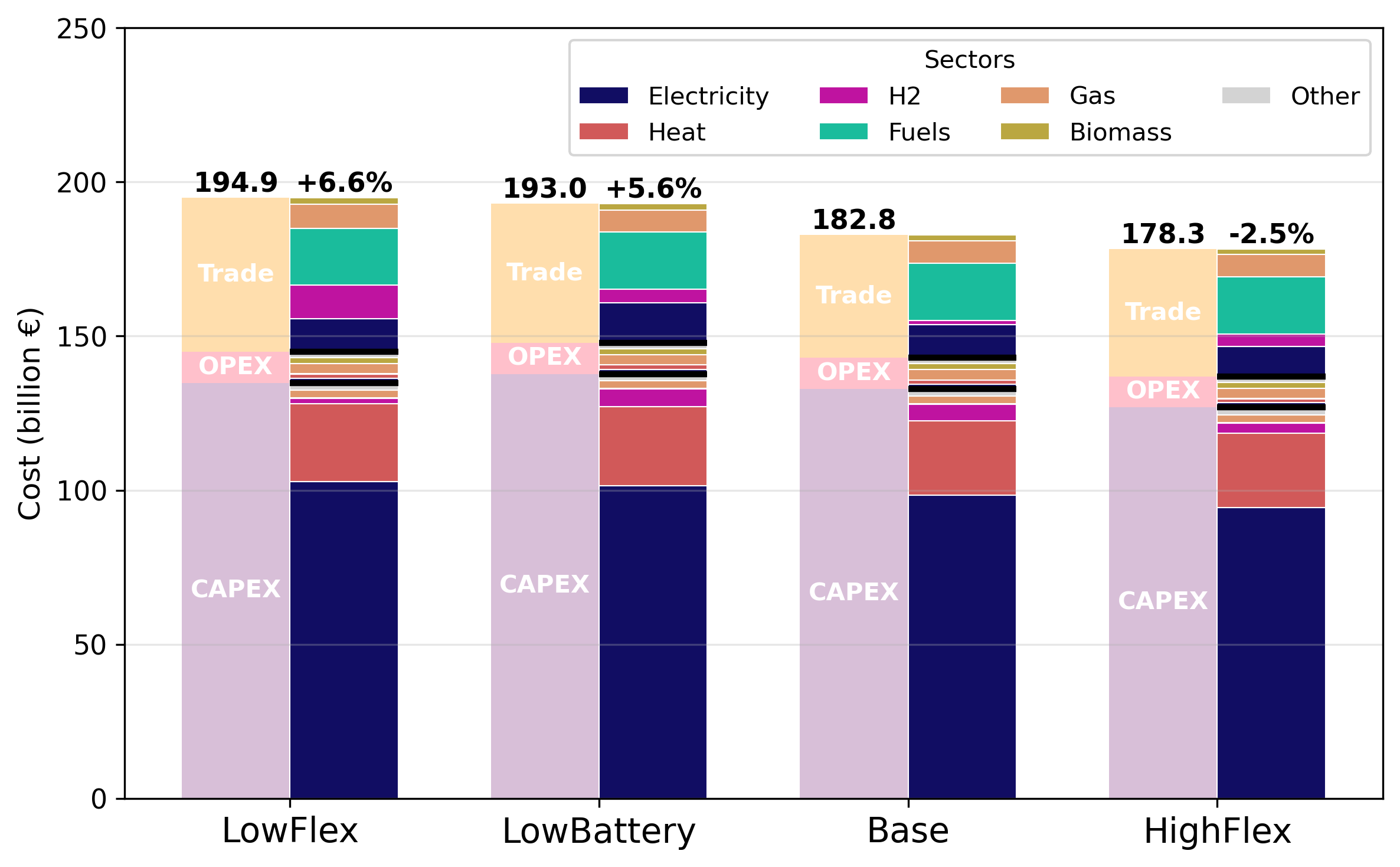}
    \end{subfigure} &
    \begin{subfigure}{0.46\textwidth}
        \centering
        \includegraphics[width=\textwidth]{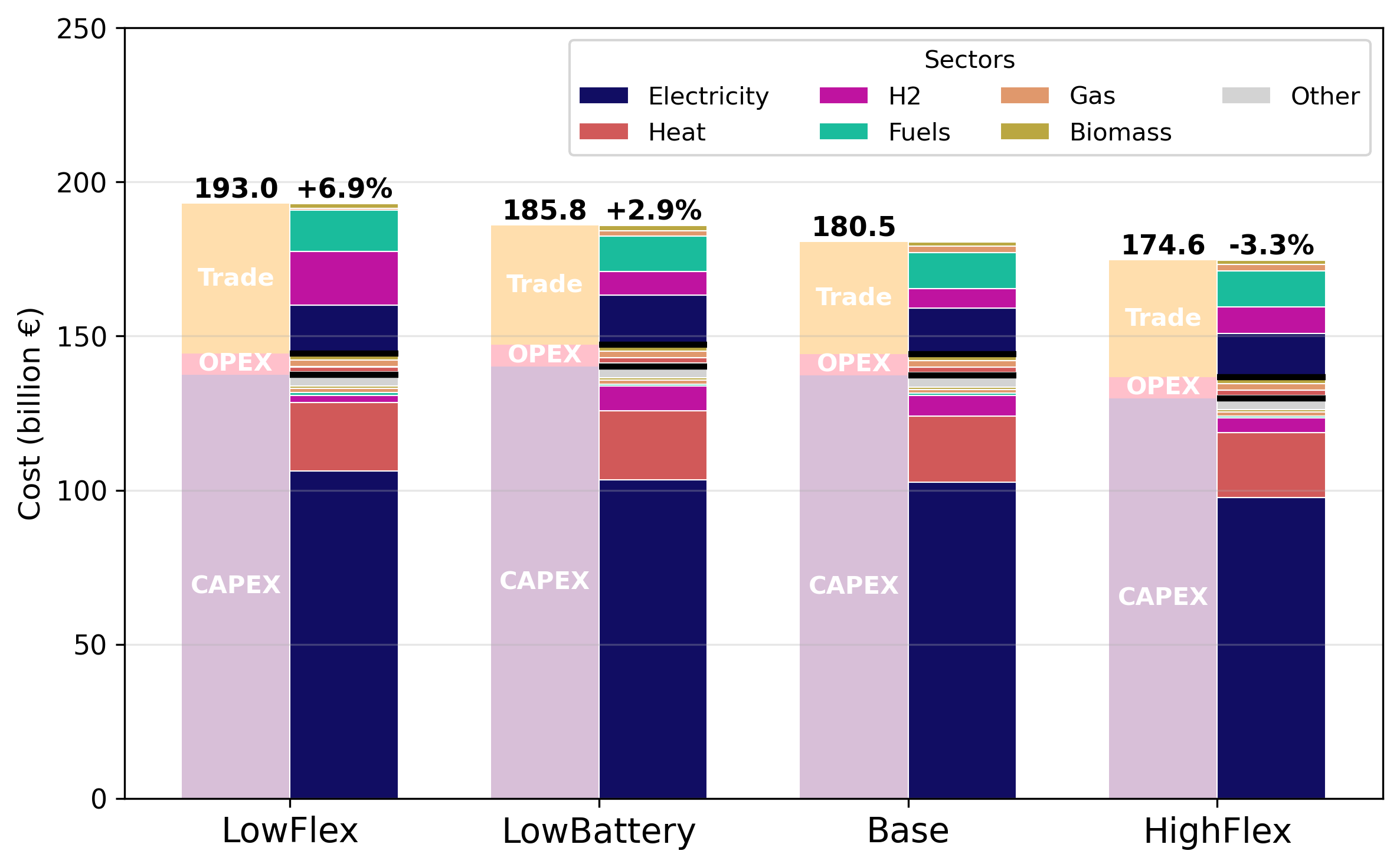}
    \end{subfigure}
    \\[10pt]
    \multicolumn{2}{c}{\normalsize \textbf{(B) Electricity Price}} \\[3pt]
    \begin{subfigure}{0.46\textwidth}
        \centering
        \includegraphics[width=\textwidth]{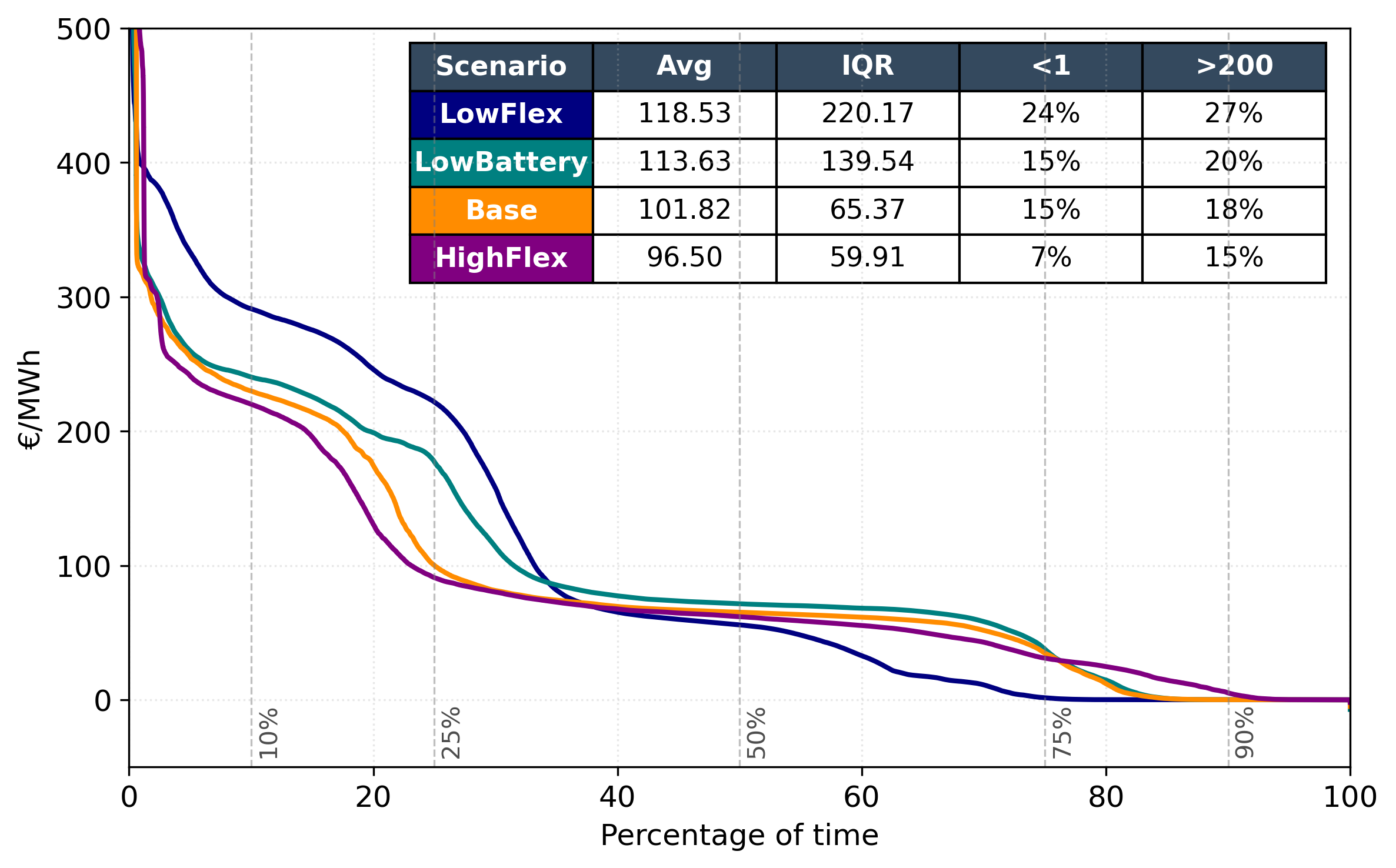}
    \end{subfigure} &
    \begin{subfigure}{0.46\textwidth}
        \centering
        \includegraphics[width=\textwidth]{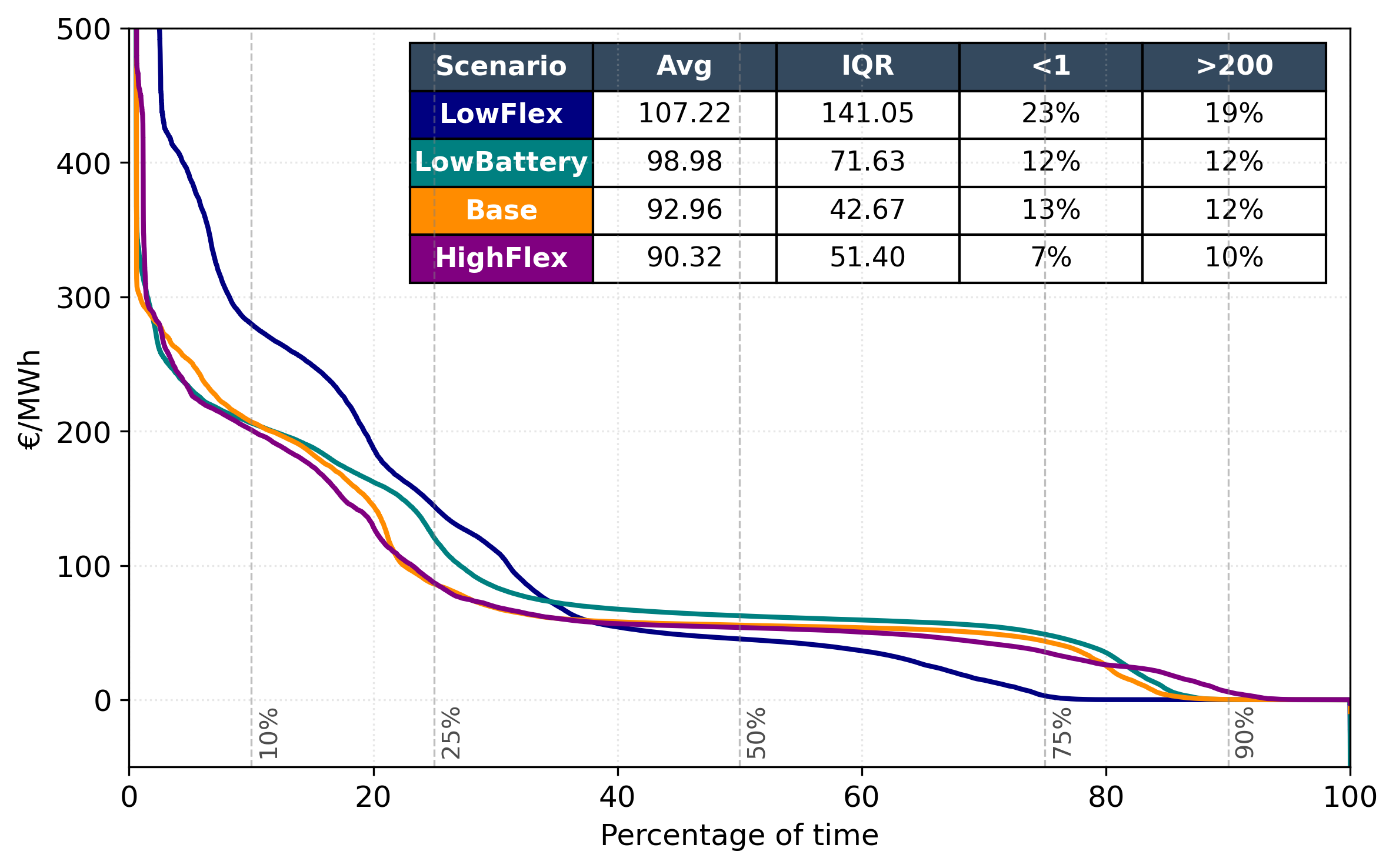}
    \end{subfigure}
    \\[10pt]
    \multicolumn{2}{c}{\normalsize \textbf{(C) Import}} \\[3pt]
    \begin{subfigure}{0.46\textwidth}
        \centering
        \includegraphics[width=\textwidth]{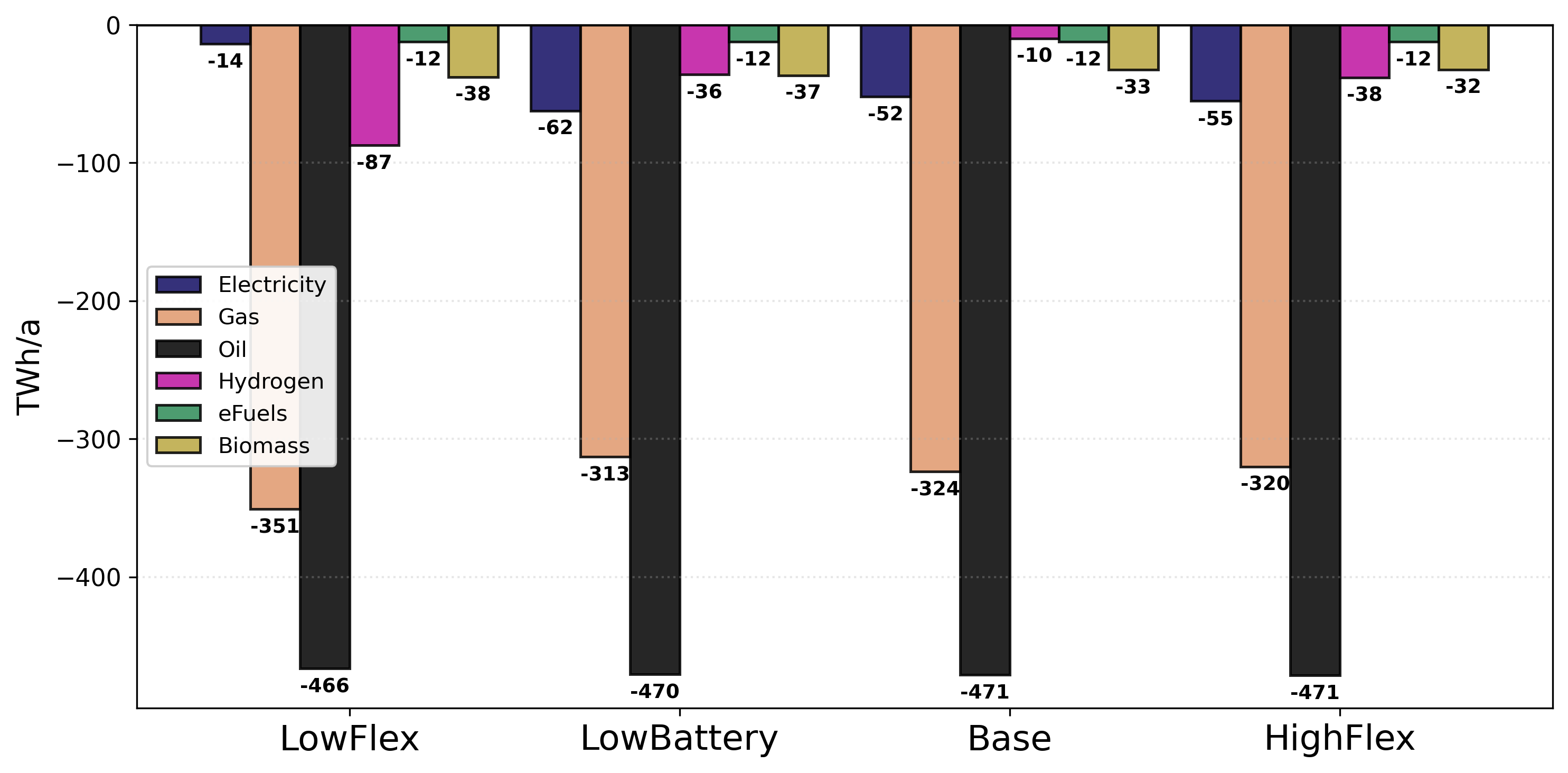}
    \end{subfigure} &
    \begin{subfigure}{0.46\textwidth}
        \centering
        \includegraphics[width=\textwidth]{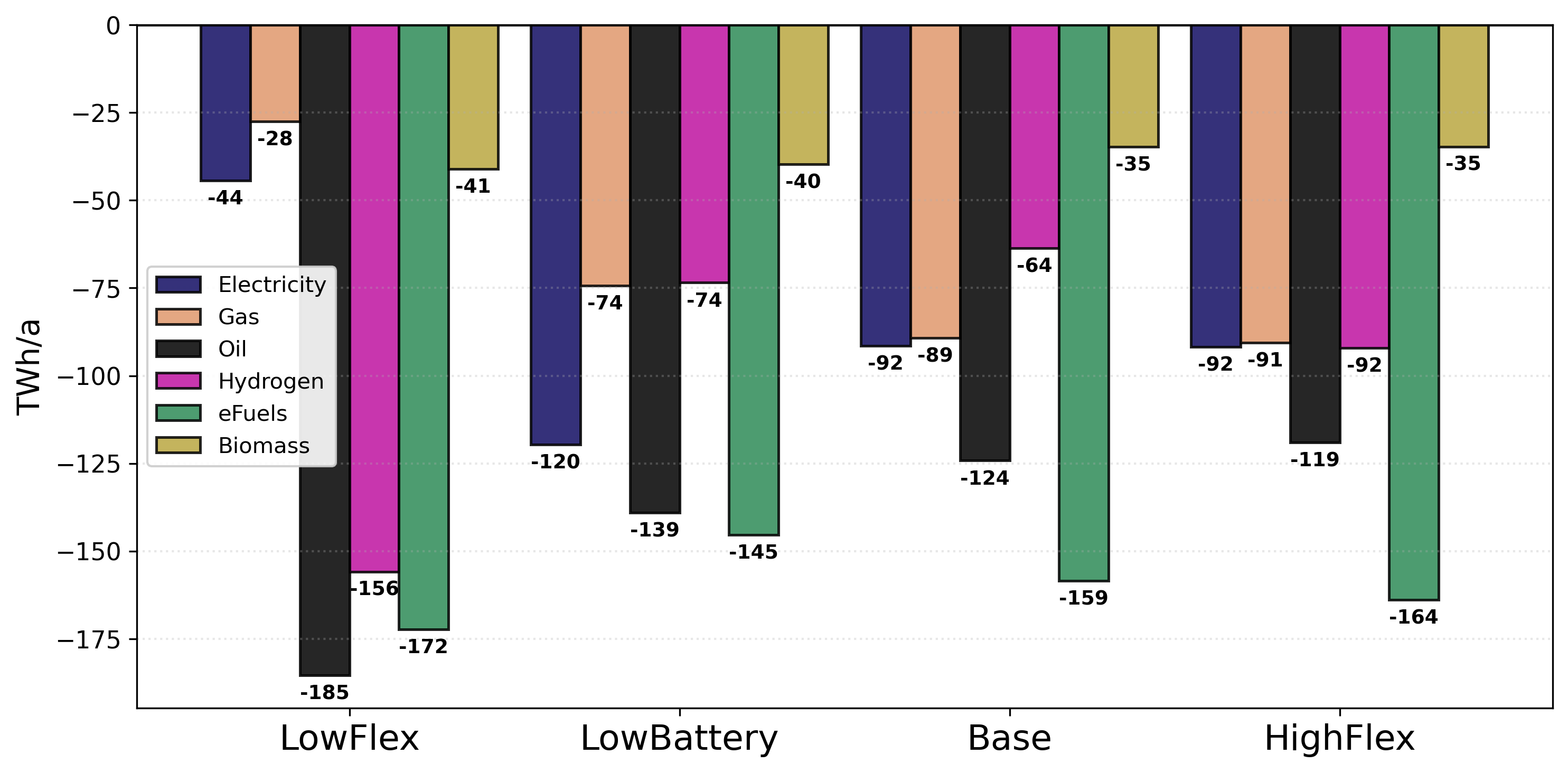}
    \end{subfigure}
    \\
    \end{tabular}
    \caption{Overview of system results by category for 2035 and 2045. (A) Total annualized system costs broken down by sector and cost component (CAPEX, OPEX, Trade). (B) Electricity price duration curves showing price ranges and distribution across different flexibility scenarios. (C) Net import volumes for electricity, gas, oil, hydrogen, e-fuels and biomass.%
    }
    \label{fig:system-results-grid}
\end{figure*}

The degree of flexibility available in the energy system has significant implications for various system-level outcomes. Figure~\ref{fig:system-results-grid} provides an overview of key system results across different flexibility scenarios for the years 2035 and 2045, including system costs, electricity prices and import levels.

\subsubsection{System costs}
Total energy system costs reveal distinct impacts of different flexibility degrees across scenarios. In 2035, LF incurs 6.6\% higher costs (195 billion €) compared to BA (183 billion €), driven primarily by substantially increased hydrogen imports (10.93 billion € vs. 1.3 billion €), biomass imports and elevated electricity CAPEX. The increased cost of electricity CAPEX results from higher investment in backup generation as well as transmission and distribution grid infrastructure to mitigate the effect of missing short-term flexibility. Restricting the battery expansion to 50\% of BA leads to a substantial increase of 5.6\%, due to higher electricity and H2 imports and increased investment in backup generation. In HF, the system costs can be reduced by 2.5\%. The reduction is driven by lower CAPEX for battery, renewables and backup generation, as well as reduced electrolysis infrastructure.\\ 

By 2045, cost differentials persist but narrow for LB (LF +6.9\%; LB +2.9\%). Hydrogen import costs remain very high for LF (17.4 billion € vs. 6.3 billion € in BA). Conversely, HF achieves system cost reductions of 3.3\% through early deployment of V2G capabilities, industrial demand-side management, and dynamic line rating, which enhance infrastructure utilisation and reduce both generation overcapacity and cross-border energy dependencies. 
The total system costs, including all modelled neighbouring countries, show similar trends. LF and LB have higher costs of 6.5\% and 3.1\% in 2035, and 7.3\% and 2.3\% in 2045, respectively, while HF achieves cost reductions of 1.6\% in 2035 and 2.1\% in 2045 compared to BA.
These findings underscore that delayed flexibility deployment imposes persistent cost penalties primarily through increased trade exposure and suboptimal asset utilisation, while diversified flexibility portfolios prove more cost-effective. This emphasises that investing in flexibility options shifts the cost from operational expenditures paid to foreign countries towards capital expenditures inside Germany, leading to overall cost savings.

\subsubsection{Prices}

Figure~\ref{fig:system-results-grid} (B) illustrates electricity price duration curves for the different scenarios in 2035 and 2045. In both years, LF exhibits the highest average electricity price and the highest interquartile range. The absence of sufficient heat storage and batteries results in a very high number of zero-price hours. Meanwhile, the increased reliance on hydrogen imports and backup capacities to cover peak demand drives up the frequency of high-price hours. Overall the electricity price is 14--17~\euro/MWh higher than in BA. 
LB, lacking only batteries, shows a particularly increased need for backup generation, which elevates the high-price side of the duration curve. However, the rest of the price distribution remains similar to BA. 
In contrast, the high flexibility in HF results in very few zero-price hours (7\% in 2035 and 2045). The need for peak-price backup generation is also reduced, leading to the overall lowest average prices (3--5~\euro/MWh lower than BA). 
This demonstrates that the amount of flexibility in the system largely influences electricity prices. This is consistent with \cite{helisto_impact_2023}, who find for pan-European scenarios that additional flexible demand stabilises prices, particularly reducing the number of hours with very low prices, while limited flexibility leads to more extreme price duration curves.

A similar picture arises for hydrogen prices. LF has significantly higher average hydrogen prices (+27/25~\euro/MWh in 2035/2045 compared to BA) due to inflexible domestic production, limited storage options, high imports and need for backup generation. The average heat prices are rather similar across the scenarios but LF has the highest volatility and highest average prices in 2045.

\subsubsection{Autarky}

Comparing net imports (Figure~\ref{fig:system-results-grid} (C)) highlights the strong dependency in LF. In 2045, biomass imports increase by 17\% in LF (15\% in 2035) and by 14\% in LB (12\% in 2035) relative to BA, while remaining unchanged in HF. Combined fossil fuel imports (oil and gas) are largely unaffected overall, although individual components vary across scenarios. In contrast, hydrogen and e-fuel imports respond more strongly to system flexibility: imports increase by 47\% in LF and by 15\% in HF, while remaining close to BA levels in LB. Considering all traded fuels and gases (oil, gas, hydrogen, and e-fuels), total imports increase by 24\% in LF and 6.9\% in HF, whereas LB remains nearly unchanged. Overall, these results indicate that reduced flexibility primarily drives higher import requirements for synthetic fuels, while fossil fuel imports remain comparatively robust across scenarios. The increase in hydrogen and e-fuel imports in HF can be explained by reduced incentives for domestic production, as cheaper flexibility options displace electrolysis and e-fuel production, shifting supply abroad and thereby increasing net imports.

The LF scenario exhibits the lowest net electricity imports despite the highest gross exports (186~TWh in 2045), as excess renewable generation cannot be stored domestically and must be exported, while costly imports remain necessary during system stress. Consequently, import costs in LF are 5\% higher than in BA. LB shows the largest deviation, with net imports rising by 30\% in 2045 due to the absence of battery storage, and gross imports also increase (280~TWh). BA itself (280~TWh gross, 89~TWh net imports) aligns broadly with \cite{goke_how_2023}, who report 239~TWh gross and 124~TWh net imports for a fully renewable European system. Across all scenarios, high gross exchanges relative to modest net positions indicate that cross-border transmission primarily balances local VRES fluctuations rather than establishing persistent directional flows. HF achieves the lowest imports and import costs (-6\% vs.\ BA in 2045) thanks to its high domestic flexibility.

The usable net transfer capacity (NTC) of AC and DC transmission lines increases by a factor of 2.4 from 52~GW in 2025 to 123~GW in 2045 in BA, with AC NTC increasing from 39 to 87~GW and DC from 14 to 35~GW. This is consistent with \cite{goke_how_2023}, who find similar relative increases for a fully renewable European system.\footnote{Usable AC cross-border capacity is estimated at 50\% (60\% in 2045) of installed NTC, reflecting a 70\% security margin on AC capacity of which 70\% is available for international trade ($0.7 \times 0.7 \approx 0.5$), while DC capacity is assumed to be fully usable.}  Across scenarios, NTC is highest in LB (+7\% vs. BA in 2045), where it substitutes for missing short-term flexibility, and lowest in HF (-5\% in 2045), which relies less on short-term cross-border exchange due to high domestic flexibility.

\subsubsection{Additional results}

The curtailment of wind and solar generation is highest in LF, reaching  43/45~TWh for wind and 26/25~TWh for solar in 2035 and 2045. Especially the high wind curtailment is due to the restricted flexibility operation of electrolysis and the limited storage capacity for H2 and batteries. LB has similar curtailment rates as BA for wind but 4-9~TWh higher for solar, due to the limited possibility of short-term battery storage. The HF scenario is able to reduce total curtailment by 65/60\% in 2035/2045 compared to BA. Curtailment, even though it can be considered a flexibility option, is not part of the flexibility provision in this analysis, as it is not part of the residual load.

\section{Discussion}
\label{sec:discussion}

\subsection{Comparison with existing literature}
\label{discussion:comparison_literature}

The increasing importance of flexibility is reflected in rising flexibility needs across all timescales (Table~\ref{tab:flex-comparison}). A direct comparison of growth rates is limited by the absence of 2035 values in most studies; therefore, literature values for 2020--2030 are compared to our 2025--2035 results. Over this period, our Base scenario shows substantially stronger growth (daily $\sim$3.2$\times$, weekly $\sim$2.2$\times$, annual $\sim$2.8$\times$) than the literature, which typically reports increases of only $\sim$1.1--1.7 across timescales \cite{artelys_mainstreaming_2017,artelys_effect_2019,trinomics_power_2023,
european_commission_joint_research_centre_flexibility_2023}.  
For later periods, comparing 2030--2045/2050 in the literature with 2035--2045 in this study reveals the opposite pattern: literature values continue to grow ($\approx$1.2--2.5), whereas our results increase only marginally ($\approx$1.1--1.2), indicating a saturation of flexibility needs after 2035. Over the full horizon, our growth (2025--2045) lies in the upper range of reported values.  
In absolute terms, our 2025 flexibility needs already exceed most literature estimates for 2020--2030 across all timescales, and by 2045 particularly daily flexibility is at or above the upper end of reported ranges, while annual values remain comparable. Across all comparisons, daily flexibility shows the strongest growth, followed by annual and weekly flexibility, highlighting the increasing importance of short-term balancing.\newline

The consistently lower values in other studies can be attributed to their single-node representation and lower renewable shares in earlier scenarios. Our own single-node, 1-hour model yields slightly lower flexibility needs than the spatially resolved Base model, confirming that network constraints and spatial heterogeneity amplify flexibility requirements. A further source of discrepancy in absolute flexibility needs lies in the definition of which technologies are counted as dispatchable demand; most studies do not clearly state whether, for instance, electrified heat demand for individual buildings is treated as dispatchable or not. Methodological differences in the calculation of annual flexibility may also play a role, as some studies define it as the difference between monthly averages rather than weekly averages as used here. Beyond methodology, the choice of weather year introduces uncertainty up to 15\% (daily), 18\% (weekly) and 30\% for annual needs (see Figure \ref{fig:app:flex-needs-weather-years} in the Appendix; \cite{iea_managing_2023}). Finally, as demonstrated in this study, differing scenario frameworks, in terms of technology assumptions, demand levels, and sector coupling, inherently lead to different flexibility requirements. \newline

\begin{table}[ht]
\centering
\footnotesize
\setlength{\tabcolsep}{4pt}
\caption{Flexibility needs (TWh/a) for Germany across timescales and studies. $^\dagger$Spatial resolution: one node per country or lower.$^\ddagger$Flexibility needs method per node (aggregated over all nodes within the country).$^*$Values for 2050.}\label{tab:flex-comparison}
\begin{tabular}{llrrrrrr}
\toprule
\textbf{Source} & \textbf{Scale} & \textbf{2020} & \textbf{2025} & \textbf{2030} & \textbf{2035} & \textbf{2040} & \textbf{2045} \\
\midrule
\multirow{3}{*}{\cite{artelys_mainstreaming_2017}$^\dagger$}
& Daily  & 28 & 33 & 42 & — & — & — \\
& Weekly & 31 & 33 & 37 & — & — & — \\
& Annual & 17 & 22 & 27 & — & — & — \\
\midrule
\multirow{3}{*}{\cite{artelys_effect_2019}$^\dagger$}
& Daily  & 30 & — & 35-45 & — & — & — \\
& Weekly & 29 & — & 35-38 & — & — & — \\
& Annual & 18 & — & 23-29 & — & — & — \\
\midrule
\multirow{3}{*}{\cite{european_commission_role_2019}$^\dagger$}
& Daily  & — & — & 31-37 & — & — & — \\
& Weekly & — & — & 36-47 & — & — & — \\
& Annual & — & — &  —    & — & — & — \\
\midrule
\multirow{3}{*}{\cite{trinomics_power_2023}$^\dagger$}
& Daily  & 25 & — & 90-100 & — & 105-135 &  115-120$^*$ \\
& Weekly & 30 & — & 62-63  & — & 85-87   &  83-110$^*$ \\
& Annual & 18 & — & 45-55  & — & 58-70   &  59-62$^*$ \\
\midrule
\multirow{3}{*}{\cite{european_commission_joint_research_centre_flexibility_2023}$^\dagger$}
& Daily  & 25 & — & 50 & — & — & 125 \\
& Weekly & 30 & — & 55 & — & — & 135 \\
& Annual & 17 & — & 47 & — & — &  90 \\
\midrule
\multirow{3}{*}{\shortstack[l]{This study\\(Base)}}
& Daily  & — & 51 & — & 165 & — & 188 \\
& Weekly & — & 47 & — &  101 & — & 120 \\
& Annual & — & 28 & — &  77 & — &  91 \\
\midrule
\multirow{3}{*}{\shortstack[l]{This study\\(Base)$^\dagger$}}
& Daily  & — & 50 & — & 152 & — & 161 \\
& Weekly & — & 48 & — &  104 & — & 138 \\
& Annual & — & 28 & — &  73 & — &  95 \\
\midrule
\multirow{3}{*}{\shortstack[l]{This study\\(Base)$^\ddagger$}}
& Daily  & — & 70 & — & 224 & — & 257 \\
& Weekly & — & 61 & — & 145 & — & 169 \\
& Annual & — & 43 & — & 136 & — & 157 \\
\bottomrule
\end{tabular}
\end{table}

The dominance of BEVs and batteries in daily flexibility provision is broadly consistent with \cite{trinomics_power_2023}, who also find growing BEV and electrolysis contributions toward 2050, while interconnectors remain important across all timescales in both studies. The central role of electrolysis on weekly and annual timescales aligns with \cite{trinomics_power_2023}, \cite{gils_interaction_2021}, and \cite{goke_how_2023}, who highlight flexible electrolysis operation as the single most impactful balancing option. The importance of resistive heaters and heat-based flexibility in our results is consistent with \cite{gaafar_system_2024}, who identify centralised heat as a major contributor alongside electrolysis. The high share of electricity trade in weekly and annual flexibility is in line with \cite{trinomics_power_2023} and \cite{artelys_mainstreaming_2017}, though our interconnector shares are lower, possibly reflecting differences in assumed cross-border capacities or domestic flexibility availability.

\subsection{Limitations}
\label{discussion:methods_limitations}
The modelling of the energy system and the flexibility assessment method involve several limitations that should be acknowledged. The model does not represent unit commitment constraints such as minimum up and down times or ramp rate limitations of dispatchable generators, which may lead to an overestimation of system flexibility provision. Cost assumptions for emerging technologies, particularly iron-air batteries, V2G systems, and industrial DSM, remain highly uncertain. In addition, the available capacity of flexible BEV charging, the degree of electrification and thus DSM potential, and the deployable volumes of V2G and industrial DSM are uncertain. The assumption of perfect foresight within each optimisation year implies fully optimised storage operation and investment decisions, neglecting behavioural barriers, policy uncertainty, and infrastructure path dependencies.\\ 

The correlation-based flexibility metric used here (see Section~\ref{sec:methods:flexibility-correlation-method}), while intuitive, has several limitations. Although the model accounts for the time span within which energy must be shifted, it does not capture the duration of the shift itself. For example, in the case of daily flexibility needs, the metric indicates that energy must be shifted within 24 hours but does not specify how much energy must be shifted or for how long within that period. Moreover, it provides no information on the maximum volume of energy that must be shifted and therefore does not indicate the required shifting capacity. This would be important for dimensioning storage requirements or quantifying how long disruptions to consumers would persist if insufficient flexibility capacity were available. If a technology’s flexibility provision exceeds system flexibility needs, the method also counts the excess provision as a flexibility contribution, potentially leading to an overestimation of the actual contribution. While this represents a theoretical edge case that is unlikely to occur frequently in practice, it remains a methodological limitation.
The application and comparability of systems with different spatial resolutions are also unclear. The method can either be applied at the regional level and subsequently aggregated, or system characteristics can first be aggregated across regions and the method applied once at the system level. These approaches may yield different results. In its original form, the metric measures only the total amount of energy provided as flexibility. Consequently, technologies with higher overall generation or consumption tend to show larger flexibility contributions. To assess the relative efficiency of a technology’s flexibility provision, the contribution could be normalised by installed capacity, total generation or consumption, or alternatively by total capital and operational expenditures to reflect the cost of flexibility provision.

\section{Summary and conclusions}

This study quantifies flexibility needs and technology-specific contributions in climate-neutral energy systems using a correlation-based flexibility assessment method applied to a comprehensive sector-coupled energy system model. The approach is illustrated through Germany's pathway to climate neutrality by 2045. We provide quantitative answers to three key research questions:

\begin{itemize}
\item Daily flexibility needs nearly quadruple from 51~TWh/a in 2025 to 188~TWh/a by 2045, with weekly and annual needs growing by factors of 2.6 and 3.25, respectively, driven primarily by the expanding shares of solar and wind generation.
\item On daily timescales, BEV charging and stationary batteries dominate, while electrolysis and electricity trade are the main providers on weekly and annual timescales, with the technology mix shifting substantially across scenarios depending on available flexibility options.
\item Compared to BA, limited flexibility raises system costs by up to 6.9\%, increases price ranges, and creates strong import dependence. Conversely, higher flexibility reduces system costs by up to 3.3\% and electricity prices by 3--5~\euro/MWh, while decreasing the required backup capacity by 22\% (22~GW) and strengthening energy self-sufficiency.
\end{itemize}

These findings establish flexibility as a quantifiable, cost-critical, and strategically decisive dimension of the energy transition. Without adequate mechanisms to bridge temporal mismatches between variable supply and inelastic demand, the economic benefits of low-cost renewable generation cannot be fully captured. Cost-effective decarbonisation therefore requires the integrated expansion of storage, demand response, sector coupling, and transmission infrastructure alongside renewable capacity. The energy transition is inherently a flexibility transition. Crucially, however, such conclusions presuppose a rigorous methodological foundation: as highlighted in the introduction, flexibility remains poorly defined and inconsistently quantified across the literature, and a central contribution of this work is to provide a transparent, additive framework that makes these findings reproducible and comparable.\\
Flexibility deployment demands the same urgency as renewable expansion: every percentage point of missing flexibility translates directly into higher system costs, greater import exposure, and underutilised renewable assets. Demand-side flexibility through BEV charging, V2G, and industrial DSM emerges as particularly high-value and should be unlocked through targeted market design and regulatory frameworks. Flexibility policies must explicitly address all timescales, since daily, weekly, and seasonal needs require fundamentally distinct technological solutions. Above all, flexibility infrastructure is an energy security asset: the massive increase in hydrogen imports under constrained flexibility in 2035 demonstrates that domestic flexibility investments are not merely cost optimisation but a prerequisite for strategic autonomy.
 
\section*{Acknowledgements}
J.G. and M.L. gratefully acknowledge funding from the Kopernikus-Ariadne project by the German Federal Ministry of Research, Technology and Space (Bundesministerium für Forschung, Technologie und Raumfahrt, BMFTR), grant number 03SFK5R0-2.

\section*{Author contributions}

\textbf{J.G.}:
Conceptualisation --
Data curation --
Formal Analysis --
Investigation --
Methodology --
Software --
Validation --
Visualisation --
Writing - original draft --
\textbf{M.L.}:
Investigation --
Methodology --
Software --
Supervision --
Validation --
Writing - review \& editing --
\textbf{T.B.}:
Conceptualization --
Formal Analysis --
Funding acquisition --
Methodology --
Project administration --
Supervision --
Validation --
Writing - review \& editing

\section*{Declaration of interests}

The authors declare no competing interests.

\section*{Declaration of generative AI and AI-assisted technologies in the writing
process}

During the preparation of this work the authors used Claude and ChatGPT in order
to improve wording. After using this tool/service, the authors reviewed and edited
the content as needed and take full responsibility for the content of the published
article.

\section*{Data and code availability}
A dataset of the model results will be made available on \url{zenodo} after peer-review.
The code to reproduce the experiments is available at \url{https://github.com/JulianGeis/pypsa-de-flex/tree/feature-flex}.

\renewcommand{\ttdefault}{\sfdefault}
\bibliography{flex}

\begin{thebibliography}{10}
\expandafter\ifx\csname url\endcsname\relax
  \def\url#1{\texttt{#1}}\fi
\expandafter\ifx\csname urlprefix\endcsname\relax\def\urlprefix{URL }\fi
\expandafter\ifx\csname href\endcsname\relax
  \def\href#1#2{#2} \def\path#1{#1}\fi

\bibitem{papaefthymiou_power_2018}
G.~Papaefthymiou, E.~Haesen, T.~Sach,
  \href{https://linkinghub.elsevier.com/retrieve/pii/S0960148118305196}{Power
  {System} {Flexibility} {Tracker}: {Indicators} to track flexibility progress
  towards high-{RES} systems}, Renewable Energy 127 (2018) 1026--1035.
\newblock \href {https://doi.org/10.1016/j.renene.2018.04.094}
  {\path{doi:10.1016/j.renene.2018.04.094}}.
\newline\urlprefix\url{https://linkinghub.elsevier.com/retrieve/pii/S0960148118305196}

\bibitem{IRENA_2018}
E.~Taibi, T.~Nikolakakis, Power system flexibility for the energy transition,
  {Part} 1: {Overview} for policy makers, Tech. rep., IRENA (2018).

\bibitem{denholm_grid_2011}
P.~Denholm, M.~Hand,
  \href{https://linkinghub.elsevier.com/retrieve/pii/S0301421511000292}{Grid
  flexibility and storage required to achieve very high penetration of variable
  renewable electricity}, Energy Policy 39~(3) (2011) 1817--1830.
\newblock \href {https://doi.org/10.1016/j.enpol.2011.01.019}
  {\path{doi:10.1016/j.enpol.2011.01.019}}.
\newline\urlprefix\url{https://linkinghub.elsevier.com/retrieve/pii/S0301421511000292}

\bibitem{heggarty_quantifying_2020}
T.~Heggarty, J.-Y. Bourmaud, R.~Girard, G.~Kariniotakis,
  \href{https://linkinghub.elsevier.com/retrieve/pii/S030626192031326X}{Quantifying
  power system flexibility provision}, Applied Energy 279 (2020) 115852.
\newblock \href {https://doi.org/10.1016/j.apenergy.2020.115852}
  {\path{doi:10.1016/j.apenergy.2020.115852}}.
\newline\urlprefix\url{https://linkinghub.elsevier.com/retrieve/pii/S030626192031326X}

\bibitem{lund_review_2015}
P.~D. Lund, J.~Lindgren, J.~Mikkola, J.~Salpakari,
  \href{https://www.sciencedirect.com/science/article/pii/S1364032115000672}{Review
  of energy system flexibility measures to enable high levels of variable
  renewable electricity}, Renewable and Sustainable Energy Reviews 45 (2015)
  785--807.
\newblock \href {https://doi.org/10.1016/j.rser.2015.01.057}
  {\path{doi:10.1016/j.rser.2015.01.057}}.
\newline\urlprefix\url{https://www.sciencedirect.com/science/article/pii/S1364032115000672}

\bibitem{emmanuel_review_2020}
M.~Emmanuel, K.~Doubleday, B.~Cakir, M.~Marković, B.-M. Hodge,
  \href{https://linkinghub.elsevier.com/retrieve/pii/S0038092X20307489}{A
  review of power system planning and operational models for flexibility
  assessment in high solar energy penetration scenarios}, Solar Energy 210
  (2020) 169--180.
\newblock \href {https://doi.org/10.1016/j.solener.2020.07.017}
  {\path{doi:10.1016/j.solener.2020.07.017}}.
\newline\urlprefix\url{https://linkinghub.elsevier.com/retrieve/pii/S0038092X20307489}

\bibitem{brunner_future_2020}
C.~Brunner, G.~Deac, S.~Braun, C.~Zöphel,
  \href{https://linkinghub.elsevier.com/retrieve/pii/S096014811931626X}{The
  future need for flexibility and the impact of fluctuating renewable power
  generation}, Renewable Energy 149 (2020) 1314--1324.
\newblock \href {https://doi.org/10.1016/j.renene.2019.10.128}
  {\path{doi:10.1016/j.renene.2019.10.128}}.
\newline\urlprefix\url{https://linkinghub.elsevier.com/retrieve/pii/S096014811931626X}

\bibitem{huber_integration_2014}
M.~Huber, D.~Dimkova, T.~Hamacher,
  \href{https://linkinghub.elsevier.com/retrieve/pii/S0360544214002680}{Integration
  of wind and solar power in {Europe}: {Assessment} of flexibility
  requirements}, Energy 69 (2014) 236--246.
\newblock \href {https://doi.org/10.1016/j.energy.2014.02.109}
  {\path{doi:10.1016/j.energy.2014.02.109}}.
\newline\urlprefix\url{https://linkinghub.elsevier.com/retrieve/pii/S0360544214002680}

\bibitem{deetjen_impacts_2017}
T.~A. Deetjen, J.~D. Rhodes, M.~E. Webber,
  \href{https://linkinghub.elsevier.com/retrieve/pii/S0360544217301950}{The
  impacts of wind and solar on grid flexibility requirements in the {Electric}
  {Reliability} {Council} of {Texas}}, Energy 123 (2017) 637--654.
\newblock \href {https://doi.org/10.1016/j.energy.2017.02.021}
  {\path{doi:10.1016/j.energy.2017.02.021}}.
\newline\urlprefix\url{https://linkinghub.elsevier.com/retrieve/pii/S0360544217301950}

\bibitem{lannoye_power_2012}
E.~Lannoye, D.~Flynn, M.~O'Malley,
  \href{http://ieeexplore.ieee.org/document/6345375/}{Power system flexibility
  assessment \&\#x2014; {State} of the art}, in: 2012 {IEEE} {Power} and
  {Energy} {Society} {General} {Meeting}, IEEE, San Diego, CA, 2012, pp. 1--6.
\newblock \href {https://doi.org/10.1109/PESGM.2012.6345375}
  {\path{doi:10.1109/PESGM.2012.6345375}}.
\newline\urlprefix\url{http://ieeexplore.ieee.org/document/6345375/}

\bibitem{epri_metrics_2014}
{EPRI}, \href{https://www.epri.com/research/products/3002004243}{Metrics for
  {Quantifying} {Flexibility} in {Power} {System} {Planning}}, Tech. Rep.
  3002004243, Electric Power Research Institute (EPRI), Palo Alto, CA,
  publisher: Electric Power Research Institute (2014).
\newline\urlprefix\url{https://www.epri.com/research/products/3002004243}

\bibitem{dvorkin_assessing_2014}
Y.~Dvorkin, D.~S. Kirschen, M.~A. Ortega‐Vazquez,
  \href{https://ietresearch.onlinelibrary.wiley.com/doi/10.1049/iet-gtd.2013.0720}{Assessing
  flexibility requirements in power systems}, IET Generation, Transmission \&
  Distribution 8~(11) (2014) 1820--1830.
\newblock \href {https://doi.org/10.1049/iet-gtd.2013.0720}
  {\path{doi:10.1049/iet-gtd.2013.0720}}.
\newline\urlprefix\url{https://ietresearch.onlinelibrary.wiley.com/doi/10.1049/iet-gtd.2013.0720}

\bibitem{zerrahn_long-run_2017}
A.~Zerrahn, W.-P. Schill,
  \href{https://linkinghub.elsevier.com/retrieve/pii/S1364032116308619}{Long-run
  power storage requirements for high shares of renewables: review and a new
  model}, Renewable and Sustainable Energy Reviews 79 (2017) 1518--1534.
\newblock \href {https://doi.org/10.1016/j.rser.2016.11.098}
  {\path{doi:10.1016/j.rser.2016.11.098}}.
\newline\urlprefix\url{https://linkinghub.elsevier.com/retrieve/pii/S1364032116308619}

\bibitem{oh_energy-storage_2018}
E.~Oh, S.-Y. Son,
  \href{https://linkinghub.elsevier.com/retrieve/pii/S0960148117309904}{Energy-storage
  system sizing and operation strategies based on discrete {Fourier} transform
  for reliable wind-power generation}, Renewable Energy 116 (2018) 786--794.
\newblock \href {https://doi.org/10.1016/j.renene.2017.10.028}
  {\path{doi:10.1016/j.renene.2017.10.028}}.
\newline\urlprefix\url{https://linkinghub.elsevier.com/retrieve/pii/S0960148117309904}

\bibitem{heggarty_multi-temporal_2019}
T.~Heggarty, J.-Y. Bourmaud, R.~Girard, G.~Kariniotakis,
  \href{https://linkinghub.elsevier.com/retrieve/pii/S0306261919302107}{Multi-temporal
  assessment of power system flexibility requirement}, Applied Energy 238
  (2019) 1327--1336.
\newblock \href {https://doi.org/10.1016/j.apenergy.2019.01.198}
  {\path{doi:10.1016/j.apenergy.2019.01.198}}.
\newline\urlprefix\url{https://linkinghub.elsevier.com/retrieve/pii/S0306261919302107}

\bibitem{huclin_methodological_2023}
S.~Huclin, A.~Ramos, J.~P. Chaves, J.~Matanza, M.~González-Eguino,
  \href{https://linkinghub.elsevier.com/retrieve/pii/S0360544223028852}{A
  methodological approach for assessing flexibility and capacity value in
  renewable-dominated power systems: {A} {Spanish} case study in 2030}, Energy
  285 (2023) 129491.
\newblock \href {https://doi.org/10.1016/j.energy.2023.129491}
  {\path{doi:10.1016/j.energy.2023.129491}}.
\newline\urlprefix\url{https://linkinghub.elsevier.com/retrieve/pii/S0360544223028852}

\bibitem{artelys_mainstreaming_2017}
{Artelys}, {Directorate-General for Energy (European Commission)}, R.~Bardet,
  P.~Khallouf, L.~Fournié, C.~Andrey, P.~Attard,
  \href{https://data.europa.eu/doi/10.2833/97595}{Mainstreaming {RES}:
  flexibility portfolios : design of flexibility portfolios at {Member} {State}
  level to facilitate a cost efficient integration of high shares of
  renewables}, Publications Office of the European Union, 2017.
\newline\urlprefix\url{https://data.europa.eu/doi/10.2833/97595}

\bibitem{iea_managing_2023}
IEA, Managing the {Seasonal} {Variability} of {Electricity} {Demand} and
  {Supply}, Tech. rep., IEA (2023).

\bibitem{trinomics_power_2023}
{Trinomics}, {Artelys},
  \href{https://www.artelys.com/app/uploads/2023/03/Power-System-Flexibility-Penta-region-%E2%80%93-Current-State-and-Challenges-for-a-Future-Decarbonised-Energy-System.pdf}{Power
  {System} {Flexibility} in the {Penta} region – {Current} {State} and
  {Challenges} for a {Future} {Decarbonised} {Energy} {System}}, Final report
  TEC1305EU, Benelux General Secretariat (Mar. 2023).
\newline\urlprefix\url{https://www.artelys.com/app/uploads/2023/03/Power-System-Flexibility-Penta-region-%E2%80%93-Current-State-and-Challenges-for-a-Future-Decarbonised-Energy-System.pdf}

\bibitem{solarpower_europe_mission_2024}
{SolarPower Europe},
  \href{https://www.artelys.com/app/uploads/2024/07/SPE_Mission_Solar_Report_Final_062024.pdf}{Mission
  {Solar} 2040: {Europe}'s flexibility revolution}, Tech. rep., SolarPower
  Europe, Brussels, Belgium (Jun. 2024).
\newline\urlprefix\url{https://www.artelys.com/app/uploads/2024/07/SPE_Mission_Solar_Report_Final_062024.pdf}

\bibitem{rte_bilan_2015}
{RTE},
  \href{https://assets.rte-france.com/prod/public/2020-06/bp2015_synthese.pdf}{Bilan
  prévisionnel de l'équilibre offre-demande d'électricité en {France}},
  Tech. Rep. Édition 2015, RTE, publisher: RTE (2015).
\newline\urlprefix\url{https://assets.rte-france.com/prod/public/2020-06/bp2015_synthese.pdf}

\bibitem{european_commission_metis_2015}
{European Commission},
  \href{https://energy.ec.europa.eu/data-and-analysis/energy-modelling/metis_en}{{METIS}
  - {Modelling} the {Energy} {Transition} in the {European} {Energy} {System}}
  (2015).
\newline\urlprefix\url{https://energy.ec.europa.eu/data-and-analysis/energy-modelling/metis_en}

\bibitem{artelys_optimal_2018}
{Artelys}, \href{https://data.europa.eu/doi/10.2833/642476}{Optimal flexibility
  portfolios for a high-{RES} 2050 scenario: {METIS} {Studies} : study {S1}.},
  Tech. rep., European Comission, LU (Dec. 2018).
\newline\urlprefix\url{https://data.europa.eu/doi/10.2833/642476}

\bibitem{entso-e_system_2024}
{ENTSO-E}, System {Flexibility} {Needs} for the {Energy} {Transition}, Tech.
  rep., ENTSO-E (2024).

\bibitem{schill_residual_2014}
W.-P. Schill,
  \href{https://www.sciencedirect.com/science/article/pii/S0301421514003310}{Residual
  load, renewable surplus generation and storage requirements in {Germany}},
  Energy Policy 73 (2014) 65--79.
\newblock \href {https://doi.org/10.1016/j.enpol.2014.05.032}
  {\path{doi:10.1016/j.enpol.2014.05.032}}.
\newline\urlprefix\url{https://www.sciencedirect.com/science/article/pii/S0301421514003310}

\bibitem{schlachtberger_backup_2016}
D.~Schlachtberger, S.~Becker, S.~Schramm, M.~Greiner,
  \href{https://linkinghub.elsevier.com/retrieve/pii/S0196890416302606}{Backup
  flexibility classes in emerging large-scale renewable electricity systems},
  Energy Conversion and Management 125 (2016) 336--346.
\newblock \href {https://doi.org/10.1016/j.enconman.2016.04.020}
  {\path{doi:10.1016/j.enconman.2016.04.020}}.
\newline\urlprefix\url{https://linkinghub.elsevier.com/retrieve/pii/S0196890416302606}

\bibitem{ruhnau_storage_2022}
O.~Ruhnau, S.~Qvist, \href{https://doi.org/10.1088/1748-9326/ac4dc8}{Storage
  requirements in a 100\% renewable electricity system: extreme events and
  inter-annual variability}, Environmental Research Letters 17~(4) (2022)
  044018, publisher: IOP Publishing.
\newblock \href {https://doi.org/10.1088/1748-9326/ac4dc8}
  {\path{doi:10.1088/1748-9326/ac4dc8}}.
\newline\urlprefix\url{https://doi.org/10.1088/1748-9326/ac4dc8}

\bibitem{thimet_what-where-when_2023}
P.~Thimet, G.~Mavromatidis,
  \href{https://linkinghub.elsevier.com/retrieve/pii/S0306261923011285}{What-where-when:
  {Investigating} the role of storage for the {German} electricity system
  transition}, Applied Energy 351 (2023) 121764.
\newblock \href {https://doi.org/10.1016/j.apenergy.2023.121764}
  {\path{doi:10.1016/j.apenergy.2023.121764}}.
\newline\urlprefix\url{https://linkinghub.elsevier.com/retrieve/pii/S0306261923011285}

\bibitem{open_energy_transition_role_2025}
{Open Energy Transition}, The {Role} of {Energy} {Storage} in {Germany} (May
  2025).

\bibitem{gillich_schlusselrolle_2024}
A.~Gillich, H.~Brand, T.~Schmid, K.~Hufendiek,
  \href{https://publications.pik-potsdam.de/pubman/item/item_29603}{Die
  {Schlüsselrolle} von {Flexibilität} im {Stromsystem} 2030: {Nutzenanalyse}
  und kritische {Flex}-{Technologien}}, Tech. rep., Potsdam Institute for
  Climate Impact Research, artwork Size: 36 pages, 1 MB (Feb. 2024).
\newblock \href {https://doi.org/10.48485/PIK.2024.004}
  {\path{doi:10.48485/PIK.2024.004}}.
\newline\urlprefix\url{https://publications.pik-potsdam.de/pubman/item/item_29603}

\bibitem{buttner_influence_2024}
C.~Büttner, K.~Esterl, I.~Cußmann, C.~A. Epia~Realpe, J.~Amme, A.~Nadal,
  \href{https://linkinghub.elsevier.com/retrieve/pii/S2667095X24000060}{Influence
  of flexibility options on the {German} transmission grid — {A}
  sector-coupled mid-term scenario}, Renewable and Sustainable Energy
  Transition 5 (2024) 100082, publisher: Elsevier BV.
\newblock \href {https://doi.org/10.1016/j.rset.2024.100082}
  {\path{doi:10.1016/j.rset.2024.100082}}.
\newline\urlprefix\url{https://linkinghub.elsevier.com/retrieve/pii/S2667095X24000060}

\bibitem{ackermann_cost-benefit_2024}
L.~Ackermann, N.~Gabrek, B.~Zachmann, A.~Neitz-Regett, S.~Seifermann,
  \href{https://www.sciencedirect.com/science/article/pii/S1755008424000346}{Cost-benefit
  analysis and comparison of grid-stabilizing energy flexibility options and
  their applications in relation to the {German} energy system}, Renewable
  Energy Focus 49 (2024) 100570.
\newblock \href {https://doi.org/10.1016/j.ref.2024.100570}
  {\path{doi:10.1016/j.ref.2024.100570}}.
\newline\urlprefix\url{https://www.sciencedirect.com/science/article/pii/S1755008424000346}

\bibitem{frank_potential_2025}
F.~Frank, T.~Gnann, D.~Speth, B.~Weißenburger, B.~Lux,
  \href{https://linkinghub.elsevier.com/retrieve/pii/S2666792425000216}{Potential
  impact of controlled electric car charging and vehicle-to-grid on
  {Germany}’s future power system}, Advances in Applied Energy 19 (2025)
  100227.
\newblock \href {https://doi.org/10.1016/j.adapen.2025.100227}
  {\path{doi:10.1016/j.adapen.2025.100227}}.
\newline\urlprefix\url{https://linkinghub.elsevier.com/retrieve/pii/S2666792425000216}

\bibitem{ghaiurane_integrating_2025}
V.~Ghaiurane, {Patrick Jochem}, {INTEGRATING} {E}-{TRUCKS} {INTO} {ELECTRICITY}
  {MARKETS} {BY} {OPTIMIZED} {CHARGING}, Paris, 2025.

\bibitem{gombodshaw-johann_bos_bidirektionales_2025}
{Gombodshaw-Johann Boß}, {Julian Brendel}, {Jakob Gemassmer}, Bidirektionales
  {Laden} {Wie} es sich finanziell auszahlen kann, die {Antriebsbatterien} von
  {Elektrofahrzeugen} als {Speicher} für das {Stromnetz} einzusetzen, Tech.
  rep., Agora Verkehrswende (Nov. 2025).

\bibitem{brunner_what_2025}
C.~Brunner, S.~Misconel, P.~Hauser, D.~Möst,
  \href{https://www.sciencedirect.com/science/article/pii/S0301421525000588}{To
  what extent can flexibility options reduce the need for hydrogen backup power
  plants?}, Energy Policy 201 (2025) 114551.
\newblock \href {https://doi.org/10.1016/j.enpol.2025.114551}
  {\path{doi:10.1016/j.enpol.2025.114551}}.
\newline\urlprefix\url{https://www.sciencedirect.com/science/article/pii/S0301421525000588}

\bibitem{scharf_gas_2024}
H.~Scharf, D.~Möst,
  \href{https://www.sciencedirect.com/science/article/pii/S0301421524000466}{Gas
  power — {How} much is needed on the road to carbon neutrality?}, Energy
  Policy 187 (2024) 114026.
\newblock \href {https://doi.org/10.1016/j.enpol.2024.114026}
  {\path{doi:10.1016/j.enpol.2024.114026}}.
\newline\urlprefix\url{https://www.sciencedirect.com/science/article/pii/S0301421524000466}

\bibitem{gils_interaction_2021}
H.~C. Gils, H.~Gardian, J.~Schmugge,
  \href{https://www.sciencedirect.com/science/article/pii/S0960148121011769}{Interaction
  of hydrogen infrastructures with other sector coupling options towards a
  zero-emission energy system in {Germany}}, Renewable Energy 180 (2021)
  140--156.
\newblock \href {https://doi.org/10.1016/j.renene.2021.08.016}
  {\path{doi:10.1016/j.renene.2021.08.016}}.
\newline\urlprefix\url{https://www.sciencedirect.com/science/article/pii/S0960148121011769}

\bibitem{maruf_open_2021}
M.~N.~I. Maruf,
  \href{https://www.sciencedirect.com/science/article/pii/S0306261921001549}{Open
  model-based analysis of a 100\% renewable and sector-coupled energy
  system–{The} case of {Germany} in 2050}, Applied Energy 288 (2021) 116618.
\newblock \href {https://doi.org/10.1016/j.apenergy.2021.116618}
  {\path{doi:10.1016/j.apenergy.2021.116618}}.
\newline\urlprefix\url{https://www.sciencedirect.com/science/article/pii/S0306261921001549}

\bibitem{nebel_role_2022}
A.~Nebel, J.~Cantor, S.~Salim, A.~Salih, D.~Patel,
  \href{https://www.mdpi.com/2071-1050/14/16/10379}{The {Role} of {Renewable}
  {Energies}, {Storage} and {Sector}-{Coupling} {Technologies} in the {German}
  {Energy} {Sector} under {Different} {CO2} {Emission} {Restrictions}},
  Sustainability 14~(16) (2022) 10379, number: 16 Publisher: Multidisciplinary
  Digital Publishing Institute.
\newblock \href {https://doi.org/10.3390/su141610379}
  {\path{doi:10.3390/su141610379}}.
\newline\urlprefix\url{https://www.mdpi.com/2071-1050/14/16/10379}

\bibitem{goke_how_2023}
L.~Göke, J.~Weibezahn, M.~Kendziorski,
  \href{https://linkinghub.elsevier.com/retrieve/pii/S0360544223012264}{How
  flexible electrification can integrate fluctuating renewables}, Energy 278
  (2023) 127832, publisher: Elsevier BV.
\newblock \href {https://doi.org/10.1016/j.energy.2023.127832}
  {\path{doi:10.1016/j.energy.2023.127832}}.
\newline\urlprefix\url{https://linkinghub.elsevier.com/retrieve/pii/S0360544223012264}

\bibitem{gaafar_system_2024}
N.~Gaafar, P.~Jürgens, J.~Sepúlveda~Schweiger, C.~Kost,
  \href{https://dx.doi.org/10.1088/2753-3751/ad5726}{System flexibility in the
  context of transition towards a net-zero sector-coupled renewable energy
  system—case study of {Germany}}, Environmental Research: Energy 1~(2)
  (2024) 025007, publisher: IOP Publishing.
\newblock \href {https://doi.org/10.1088/2753-3751/ad5726}
  {\path{doi:10.1088/2753-3751/ad5726}}.
\newline\urlprefix\url{https://dx.doi.org/10.1088/2753-3751/ad5726}

\bibitem{acer_flexibility_2025}
ACER, Flexibility needs assessment methodology 2025, Tech. rep., ACER (2025).

\bibitem{artelys_effect_2019}
{Artelys}, \href{https://data.europa.eu/doi/10.2833/021194}{Effect of high
  shares of renewables on power systems: study {S11}.}, Tech. rep., European
  Comission, LU (2019).
\newline\urlprefix\url{https://data.europa.eu/doi/10.2833/021194}

\bibitem{artelys_metis_2018}
{Artelys},
  \href{https://www.artelys.com/app/uploads/2019/04/METIS_S8_The-role-and-potential-of-power-to-X-in-2050.pdf}{{METIS}
  {Studies} {Study} {S8} {The} role and potential of {Power}-to-{X} in 2050},
  Tech. rep., European Comission (2018).
\newline\urlprefix\url{https://www.artelys.com/app/uploads/2019/04/METIS_S8_The-role-and-potential-of-power-to-X-in-2050.pdf}

\bibitem{trinomics_study_2022}
{Trinomics}, {Artelys},
  \href{https://www.energy-community.org/dam/jcr:2db406e5-294f-4285-9209-ec90349ce5cb/Flexiblity_EnCreport_0722.pdf}{Study
  on flexibility options to support decarbonization in the {Energy}
  {Community}}, Final {Report}, Artelys, Vienna, Austria (Jul. 2022).
\newline\urlprefix\url{https://www.energy-community.org/dam/jcr:2db406e5-294f-4285-9209-ec90349ce5cb/Flexiblity_EnCreport_0722.pdf}

\bibitem{european_commission_role_2019}
{European Commission}, {Artelys},
  \href{https://data.europa.eu/doi/10.2833/639890}{The role and need of
  flexibility in 2030 focus on energy storage: study {S07}.}, Publications
  Office, LU, 2019.
\newline\urlprefix\url{https://data.europa.eu/doi/10.2833/639890}

\bibitem{artelys_artelys_2025}
{Artelys}, \href{https://www.artelys.com/crystal/super-grid/}{Artelys {Crystal}
  {Super} {Grid}} (2025).
\newline\urlprefix\url{https://www.artelys.com/crystal/super-grid/}

\bibitem{artelys_power_2020}
{Artelys}, Power system flexibility {The} {METIS} approach to identify needs
  and solutions (Jan. 2020).

\bibitem{lindner_pypsa-_2025}
M.~Lindner, J.~Geis, T.~Seibold, T.~Brown,
  \href{https://ieeexplore.ieee.org/document/11050093}{Pypsa-{De}:
  {Open}-{Source} {German} {Energy} {System} {Model} {Reveals} {Savings} {From}
  {Integrated} {Planning}}, in: 2025 21st {International} {Conference} on the
  {European} {Energy} {Market} ({EEM}), 2025, pp. 1--6, iSSN: 2165-4093.
\newblock \href {https://doi.org/10.1109/EEM64765.2025.11050093}
  {\path{doi:10.1109/EEM64765.2025.11050093}}.
\newline\urlprefix\url{https://ieeexplore.ieee.org/document/11050093}

\bibitem{horsch_pypsa-eur_2018}
J.~Hörsch, F.~Hofmann, D.~Schlachtberger, T.~Brown,
  \href{https://arxiv.org/pdf/1806.01613}{{PyPSA}-{Eur}: {An} open optimisation
  model of the {European} transmission system}, Energy Strategy Reviews 22
  (2018) 207--215.
\newline\urlprefix\url{https://arxiv.org/pdf/1806.01613}

\bibitem{brown_pypsa_2018}
T.~Brown, J.~Hörsch, D.~Schlachtberger, {PyPSA}: {Python} for {Power} {System}
  {Analysis}, Journal of Open Research Software 6~(1) (2018) 4.
\newblock \href {https://doi.org/10.5334/jors.188}
  {\path{doi:10.5334/jors.188}}.

\bibitem{plotz_modelling_2014}
P.~Plötz, T.~Gnann, M.~Wietschel,
  \href{https://www.sciencedirect.com/science/article/pii/S0921800914002912}{Modelling
  market diffusion of electric vehicles with real world driving data — {Part}
  {I}: {Model} structure and validation}, Ecological Economics 107 (2014)
  411--421.
\newblock \href {https://doi.org/10.1016/j.ecolecon.2014.09.021}
  {\path{doi:10.1016/j.ecolecon.2014.09.021}}.
\newline\urlprefix\url{https://www.sciencedirect.com/science/article/pii/S0921800914002912}

\bibitem{fraunhofer_isi_forecasteload_2024}
{Fraunhofer ISI}, {TEP Energy GmbH}, {IREES},
  \href{https://www.forecast-model.eu/forecast-en/index.php}{{FORECAST}/{eLOAD}
  – {Energy} {Demand} {Model}} (2024).
\newline\urlprefix\url{https://www.forecast-model.eu/forecast-en/index.php}

\bibitem{hofmann_atlite_2021}
F.~Hofmann, J.~Hampp, F.~Neumann, T.~Brown, J.~Hörsch,
  \href{https://joss.theoj.org/papers/10.21105/joss.03294}{atlite: {A}
  {Lightweight} {Python} {Package} for {Calculating} {Renewable} {Power}
  {Potentials} and {Time} {Series}}, Journal of Open Source Software 6~(62)
  (2021) 3294.
\newblock \href {https://doi.org/10.21105/joss.03294}
  {\path{doi:10.21105/joss.03294}}.
\newline\urlprefix\url{https://joss.theoj.org/papers/10.21105/joss.03294}

\bibitem{simon_pezzutto_and_stefano_zambotti_and_silvia_croce_and_hotmaps_2018}
{Simon Pezzutto and Stefano Zambotti and Silvia Croce and}, {Pietro Zambelli
  and Giulia Garegnani and Chiara Scaramuzzino and}, {Ramón Pascual Pascuas
  and Alyona Zubaryeva and Franziska Haas and}, {Dagmar Exner and Andreas
  Mueller and Michael Hartner and}, {Tobias Fleiter and Anna-Lena Klingler and
  Matthias Kuehnbach and}, {Pia Manz and Simon Marwitz and Matthias Rehfeldt
  and}, {Jan Steinbach and Eftim Popovski},
  \href{https://www.hotmaps-project.eu}{Hotmaps {Project}, {D2}.3 {WP2}
  {Report} -- {Open} {Data} {Set} for the {EU28}}, Tech. rep., Hotmaps Project,
  reviewed by Lukas Kranzl and Sara Fritz (2018).
\newline\urlprefix\url{https://www.hotmaps-project.eu}

\bibitem{kirstin_ganz_wie_2021}
{Kirstin Ganz}, {Andrej Guminski}, {Michael Kolb}, {Serafin von Roon},
  \href{https://www.ffe.de/wp-content/uploads/2021/11/Wie-koennen-europaeische-Branchen-Lastgaenge-die-Energiewende-im-Industriesektor-unterstuetzen.pdf}{Wie
  können europäische {Branchen}-{Lastgänge} die {Energiewende} im
  {Industriesektor} unterstützen?}, Energiewirtschaftliche {Tagesfragen},
  {Ausgabe} 1/2, FfE (Forschungsgesellschaft für Energiewirtschaft mbH)
  (2021).
\newline\urlprefix\url{https://www.ffe.de/wp-content/uploads/2021/11/Wie-koennen-europaeische-Branchen-Lastgaenge-die-Energiewende-im-Industriesektor-unterstuetzen.pdf}

\bibitem{kondziella_flexibility_2016}
H.~Kondziella, T.~Bruckner,
  \href{https://linkinghub.elsevier.com/retrieve/pii/S1364032115008643}{Flexibility
  requirements of renewable energy based electricity systems – a review of
  research results and methodologies}, Renewable and Sustainable Energy Reviews
  53 (2016) 10--22, publisher: Elsevier BV.
\newblock \href {https://doi.org/10.1016/j.rser.2015.07.199}
  {\path{doi:10.1016/j.rser.2015.07.199}}.
\newline\urlprefix\url{https://linkinghub.elsevier.com/retrieve/pii/S1364032115008643}

\bibitem{shariatzadeh_electric_2025}
M.~Shariatzadeh, M.~A. Lopes, C.~Henggeler~Antunes,
  \href{https://linkinghub.elsevier.com/retrieve/pii/S0306261925008979}{Electric
  vehicle users' charging behavior: {A} review of influential factors, methods
  and modeling approaches}, Applied Energy 396 (2025) 126167.
\newblock \href {https://doi.org/10.1016/j.apenergy.2025.126167}
  {\path{doi:10.1016/j.apenergy.2025.126167}}.
\newline\urlprefix\url{https://linkinghub.elsevier.com/retrieve/pii/S0306261925008979}

\bibitem{schill_power_2015}
W.-P. Schill, C.~Gerbaulet,
  \href{https://www.sciencedirect.com/science/article/pii/S0306261915008417}{Power
  system impacts of electric vehicles in {Germany}: {Charging} with coal or
  renewables?}, Applied Energy 156 (2015) 185--196.
\newblock \href {https://doi.org/10.1016/j.apenergy.2015.07.012}
  {\path{doi:10.1016/j.apenergy.2015.07.012}}.
\newline\urlprefix\url{https://www.sciencedirect.com/science/article/pii/S0306261915008417}

\bibitem{hanemann_effects_2017}
P.~Hanemann, M.~Behnert, T.~Bruckner,
  \href{https://www.sciencedirect.com/science/article/pii/S0306261917307924}{Effects
  of electric vehicle charging strategies on the {German} power system},
  Applied Energy 203 (2017) 608--622.
\newblock \href {https://doi.org/10.1016/j.apenergy.2017.06.039}
  {\path{doi:10.1016/j.apenergy.2017.06.039}}.
\newline\urlprefix\url{https://www.sciencedirect.com/science/article/pii/S0306261917307924}

\bibitem{lauvergne_integration_2022}
R.~Lauvergne, Y.~Perez, M.~Françon, A.~Tejeda De La~Cruz,
  \href{https://www.sciencedirect.com/science/article/pii/S0306261922012879}{Integration
  of electric vehicles into transmission grids: {A} case study on generation
  adequacy in {Europe} in 2040}, Applied Energy 326 (2022) 120030.
\newblock \href {https://doi.org/10.1016/j.apenergy.2022.120030}
  {\path{doi:10.1016/j.apenergy.2022.120030}}.
\newline\urlprefix\url{https://www.sciencedirect.com/science/article/pii/S0306261922012879}

\bibitem{muratori_impact_2018}
M.~Muratori, \href{https://www.nature.com/articles/s41560-017-0074-z}{Impact of
  uncoordinated plug-in electric vehicle charging on residential power demand},
  Nature Energy 3~(3) (2018) 193--201.
\newblock \href {https://doi.org/10.1038/s41560-017-0074-z}
  {\path{doi:10.1038/s41560-017-0074-z}}.
\newline\urlprefix\url{https://www.nature.com/articles/s41560-017-0074-z}

\bibitem{fischer_electric_2019}
D.~Fischer, A.~Harbrecht, A.~Surmann, R.~McKenna,
  \href{https://www.sciencedirect.com/science/article/pii/S0306261918315666}{Electric
  vehicles’ impacts on residential electric local profiles – {A} stochastic
  modelling approach considering socio-economic, behavioural and spatial
  factors}, Applied Energy 233-234 (2019) 644--658.
\newblock \href {https://doi.org/10.1016/j.apenergy.2018.10.010}
  {\path{doi:10.1016/j.apenergy.2018.10.010}}.
\newline\urlprefix\url{https://www.sciencedirect.com/science/article/pii/S0306261918315666}

\bibitem{taljegard_impacts_2019}
M.~Taljegard, L.~Göransson, M.~Odenberger, F.~Johnsson,
  \href{https://www.sciencedirect.com/science/article/pii/S0306261918316970}{Impacts
  of electric vehicles on the electricity generation portfolio – {A}
  {Scandinavian}-{German} case study}, Applied Energy 235 (2019) 1637--1650.
\newblock \href {https://doi.org/10.1016/j.apenergy.2018.10.133}
  {\path{doi:10.1016/j.apenergy.2018.10.133}}.
\newline\urlprefix\url{https://www.sciencedirect.com/science/article/pii/S0306261918316970}

\bibitem{ffe_regionale_2021}
{FfE},
  \href{https://www.ffe.de/wp-content/uploads/2022/01/Regionale_Lastmanagementpotenziale_DE2.pdf#page=7}{Regionale
  {Lastmanagementpotenziale} {Quantifizierung} bestehender und zukünftiger
  {Lastmanagementpotenziale} in {Deutschland}}, Tech. rep., FfE (2021).
\newline\urlprefix\url{https://www.ffe.de/wp-content/uploads/2022/01/Regionale_Lastmanagementpotenziale_DE2.pdf#page=7}

\bibitem{glaum_leveraging_2023}
P.~Glaum, F.~Hofmann,
  \href{https://linkinghub.elsevier.com/retrieve/pii/S0306261923005639}{Leveraging
  the existing {German} transmission grid with dynamic line rating}, Applied
  Energy 343 (2023) 121199.
\newblock \href {https://doi.org/10.1016/j.apenergy.2023.121199}
  {\path{doi:10.1016/j.apenergy.2023.121199}}.
\newline\urlprefix\url{https://linkinghub.elsevier.com/retrieve/pii/S0306261923005639}

\bibitem{luderer_energiewende_2025}
G.~Luderer, F.~Bartels, T.~Brown, C.~Aulich, F.~Benke, T.~Fleiter, F.~Frank,
  H.~Ganal, J.~Geis, N.~Gerhardt, T.~Gnann, A.~Gunnemann, R.~Hasse, A.~Herbst,
  S.~Herkel, J.~Hoppe, C.~Kost, M.~Krail, M.~Lindner, M.~Neuwirth, H.~Nolte,
  R.~Pietzcker, P.~Plötz, M.~Rehfeldt, F.~Schreyer, T.~Seibold, C.~Senkpiel,
  D.~Sörgel, D.~Speth, B.~Steffen, P.~C. Verpoort,
  \href{https://publications.pik-potsdam.de/pubman/item/item_32090}{Die
  {Energiewende} kosteneffizient gestalten: {Szenarien} zur {Klimaneutralität}
  2045}, Tech. rep., Potsdam Institute for Climate Impact Research, artwork
  Size: 106 pages, 33,7 MB Medium: application/pdf (Mar. 2025).
\newblock \href {https://doi.org/10.48485/PIK.2025.003}
  {\path{doi:10.48485/PIK.2025.003}}.
\newline\urlprefix\url{https://publications.pik-potsdam.de/pubman/item/item_32090}

\bibitem{helisto_impact_2023}
N.~Helistö, S.~Johanndeiter, J.~Kiviluoma,
  \href{https://ieeexplore.ieee.org/document/10161962}{The {Impact} of {Sector}
  {Coupling} and {Demand}-side {Flexibility} on {Electricity} {Prices} in a
  {Close} to 100\% {Renewable} {Power} {System}}, in: 2023 19th {International}
  {Conference} on the {European} {Energy} {Market} ({EEM}), 2023, pp. 1--6,
  iSSN: 2165-4093.
\newblock \href {https://doi.org/10.1109/EEM58374.2023.10161962}
  {\path{doi:10.1109/EEM58374.2023.10161962}}.
\newline\urlprefix\url{https://ieeexplore.ieee.org/document/10161962}

\bibitem{european_commission_joint_research_centre_flexibility_2023}
{European Commission. Joint Research Centre.},
  \href{https://data.europa.eu/doi/10.2760/384443}{Flexibility requirements and
  the role of storage in future {European} power systems.}, Tech. rep.,
  Publications Office, LU (2023).
\newline\urlprefix\url{https://data.europa.eu/doi/10.2760/384443}

\bibitem{jrc_idees}
{Joint Research Centre},
  \href{https://ec.europa.eu/jrc/en/publication/eur-scientific-and-technical-research-reports/jrc-idees}{{JRC-IDEES}:
  Integrated database of the european energy sector, 2021 edition} (2021).
\newline\urlprefix\url{https://ec.europa.eu/jrc/en/publication/eur-scientific-and-technical-research-reports/jrc-idees}

\bibitem{eurostat_energy_balances}
{Eurostat},
  \href{https://ec.europa.eu/eurostat/databrowser/view/NRG_BAL_C}{Energy
  balance sheets, april 2023 edition} (2023).
\newline\urlprefix\url{https://ec.europa.eu/eurostat/databrowser/view/NRG_BAL_C}

\bibitem{bast_traffic}
{Federal Highway Research Institute (BASt)},
  \href{https://www.bast.de/DE/Verkehrstechnik/Fachthemen/v2-verkehrszaehlung/zaehl_node.html}{Permanent
  automatic counting stations on highways and federal roads}, license: CC-BY
  4.0 (2023).
\newline\urlprefix\url{https://www.bast.de/DE/Verkehrstechnik/Fachthemen/v2-verkehrszaehlung/zaehl_node.html}

\bibitem{klaucke_advanced_2025}
F.~Klaucke, K.~Neuhoff, A.~Roth, W.-P. Schill, L.~Stolle,
  \href{http://arxiv.org/abs/2506.14664}{An advanced reliability reserve
  incentivizes flexibility investments while safeguarding the electricity
  market}, arXiv:2506.14664 [econ] (Jun. 2025).
\newblock \href {https://doi.org/10.48550/arXiv.2506.14664}
  {\path{doi:10.48550/arXiv.2506.14664}}.
\newline\urlprefix\url{http://arxiv.org/abs/2506.14664}

\bibitem{steurer_analyse_2017}
M.~Steurer,
  \href{http://nbn-resolving.de/urn:nbn:de:bsz:93-opus-ds-91980}{Analyse von
  {Demand} {Side} {Integration} im {Hinblick} auf eine effiziente und
  umweltfreundliche {Energieversorgung}} (2017).
\newline\urlprefix\url{http://nbn-resolving.de/urn:nbn:de:bsz:93-opus-ds-91980}

\bibitem{stede_demand_2016}
J.~Stede, Demand {Response} in {Germany}: {Technical} {Potential}, {Benefits}
  and {Regulatory} {Challenges}, Tech. rep., DIW Berlin – Deutsches Institut
  für Wirtschaftsforschung, Berlin (2016).

\bibitem{mckinsey_flexibilitat_2025}
{McKinsey},
  \href{https://www.mckinsey.de/~/media/mckinsey/locations/europe%20and%20middle%20east/deutschland/news/presse/2025/2025-06-16%20flexibilisierung%20stromnachfrage/mckinsey_juni%202025_flexibilisierung%20ind%20stromnachfrage.pdf}{Flexibilität
  in der industriellen {Stromnachfrage}: {Ein} {Schlüssel} zur {Energie} wende
  in {Deutschland}?}, Tech. rep. (2025).
\newline\urlprefix\url{https://www.mckinsey.de/~/media/mckinsey/locations/europe%20and%20middle%20east/deutschland/news/presse/2025/2025-06-16%20flexibilisierung%20stromnachfrage/mckinsey_juni%202025_flexibilisierung%20ind%20stromnachfrage.pdf}

\bibitem{ranaboldo_comprehensive_2024}
M.~Ranaboldo, M.~Aragüés-Peñalba, E.~Arica, A.~Bade, E.~Bullich-Massagué,
  A.~Burgio, C.~Caccamo, A.~Caprara, D.~Cimmino, B.~Domenech, I.~Donoso,
  G.~Fragapane, P.~González-Font-de Rubinat, E.~Jahnke, M.~Juanpera,
  E.~Manafi, J.~Rövekamp, R.~Tani,
  \href{https://linkinghub.elsevier.com/retrieve/pii/S1364032124005239}{A
  comprehensive overview of industrial demand response status in {Europe}},
  Renewable and Sustainable Energy Reviews 203 (2024) 114797.
\newblock \href {https://doi.org/10.1016/j.rser.2024.114797}
  {\path{doi:10.1016/j.rser.2024.114797}}.
\newline\urlprefix\url{https://linkinghub.elsevier.com/retrieve/pii/S1364032124005239}

\bibitem{kochems_lastflexibilisierungspotenziale_2020}
J.~Kochems, Lastflexibilisierungspotenziale in {Deutschland} -
  {Bestandsaufnahme} und {Entwicklungsprojektionen}, Graz, 2020.

\bibitem{buber_tim_lastmanagement_2013}
{Buber, Tim}, {Gruber, Anna}, {Klobasa, Marian}, {von Roon, Serafin},
  Lastmanagement für {Systemdienstleistungen} und zur {Reduktion} der
  {Spitzenlast} 82~(3) (2013) 89--106.
\newblock \href {https://doi.org/10.3790/vjh.82.3.89}
  {\path{doi:10.3790/vjh.82.3.89}}.
  
\bibitem{frysztacki_strong_2021}
M.~M. Frysztacki, J.~Hörsch, V.~Hagenmeyer, T.~Brown, \href{https://www.sciencedirect.com/science/article/pii/S0306261921002439}{The strong effect of network resolution on electricity system models with high shares of wind and solar}, Applied Energy 291 (2021) 116726.
\newblock \href {https://doi.org/10.1016/j.apenergy.2021.116726} {\path{doi:10.1016/j.apenergy.2021.116726}}.
\newline\urlprefix\url{https://www.sciencedirect.com/science/article/pii/S0306261921002439}

\end{thebibliography}

\onecolumn
\section{Appendix}
\begin{appendices}
\section{Additional Information}
\label{app:infos}

\subsection{Proof of Additivity of the Correlation-Based Method}
\label{app:proof_additivity}

The correlation-based method decomposes total flexibility needs into technology-specific contributions. A key property of this formulation is additivity: the sum of all technology contributions equals total flexibility needs at each temporal resolution pair $\ell|h$.

\textbf{System balance.}
At every time step, residual load equals the sum of net outputs of all dispatchable technologies \textit{i}:
\begin{equation}
RL(t) = \sum_i P_i(t) \quad \forall t.
\end{equation}

Applying the averaging operator at temporal resolution $h$ yields
\begin{equation}
\overline{RL}^h(t) = \sum_i \bar{P}_i^h(t),
\end{equation}
and analogously at resolution $\ell$,
\begin{equation}
\overline{RL}^\ell(t) = \sum_i \bar{P}_i^\ell(t),
\end{equation}
since averaging is a linear operator.

Subtracting these two equations gives
\begin{equation}
\overline{RL}^h(t) - \overline{RL}^\ell(t)
=
\sum_i \left( \bar{P}_i^h(t) - \bar{P}_i^\ell(t) \right).
\end{equation}

\textbf{Flexibility needs.}
By definition,
\begin{equation}
FlexCurve^{\ell|h}(t)
=
\overline{RL}^h(t) - \overline{RL}^\ell(t),
\end{equation}
and total flexibility needs are
\begin{equation}
FlexNeed^{\ell|h}
=
\frac{1}{2}
\sum_{t=0}^{T}
FlexCurve^{\ell|h}(t)\,
FlexSign^{\ell|h}(t).
\end{equation}

\textbf{Technology contributions.}
For each technology $i$, define the flexibility provision curve
\begin{equation}
FlexProvCurve_i^{\ell|h}(t)
=
\bar{P}_i^h(t) - \bar{P}_i^\ell(t).
\end{equation}

Its contribution to flexibility needs is
\begin{equation}
FlexProv_i^{\ell|h}
=
\frac{1}{2}
\sum_{t=0}^{T}
FlexProvCurve_i^{\ell|h}(t)\,
FlexSign^{\ell|h}(t).
\end{equation}

\textbf{Proof of additivity.}
Using the system balance identity above,
\begin{align}
FlexNeed^{\ell|h}
&=
\frac{1}{2}
\sum_{t=0}^{T}
FlexSign^{\ell|h}(t)
\left(
\overline{RL}^h(t) - \overline{RL}^\ell(t)
\right)
\nonumber \\
&=
\frac{1}{2}
\sum_{t=0}^{T}
FlexSign^{\ell|h}(t)
\sum_i
\left(
\bar{P}_i^h(t) - \bar{P}_i^\ell(t)
\right)
\nonumber \\
&=
\sum_i
\left[
\frac{1}{2}
\sum_{t=0}^{T}
FlexSign^{\ell|h}(t)
\left(
\bar{P}_i^h(t) - \bar{P}_i^\ell(t)
\right)
\right]
\nonumber \\
&=
\sum_i FlexProv_i^{\ell|h}.
\end{align}
where the exchange of summation order follows from linearity, and $FlexSign^{\ell|h}(t)$ is identical for all technologies.

Thus, total flexibility needs equal the sum of all technology contributions:
\begin{equation}
\sum_i FlexProv_i^{\ell|h} = FlexNeed^{\ell|h}.
\end{equation}

This proves exact additivity of the correlation-based method.

\subsection{Non-Additivity of the Add-Tech-Profile Method}
\label{app:non_additivity_add_tech_method}

The methodology underlying flexibility provision calculations is sparsely documented in the literature. While numerous studies report flexibility provision values \cite{artelys_mainstreaming_2017, trinomics_study_2022, artelys_optimal_2018, trinomics_power_2023, european_commission_joint_research_centre_flexibility_2023, iea_managing_2023}, only three provide methodological detail \cite{artelys_mainstreaming_2017, european_commission_joint_research_centre_flexibility_2023, iea_managing_2023}. Since the remaining studies \cite{trinomics_study_2022, artelys_optimal_2018, trinomics_power_2023} were conducted with the participation of Artelys, it is reasonable to assume they employ the methodology originally developed for the METIS project \cite{artelys_mainstreaming_2017}. To the best of the authors' knowledge, the flexibility provision formulation presented in Section~\ref{sec:methods:flexibility-correlation-method} differs from those used in the METIS project \cite{artelys_mainstreaming_2017}, by the Joint Research Centre \cite{european_commission_joint_research_centre_flexibility_2023}, and presumably in the studies cited above. The IEA \cite{iea_managing_2023} describes an approach likely related to the one presented here, but without a rigorous mathematical formulation. This methodological opacity makes cross-study comparisons difficult and underscores the need for a transparent and well-defined formulation. In METIS, the contribution of a technology is computed by recalculating total flexibility needs after removing its generation profile, with the difference interpreted as its contribution, a method we refer to here as the \textit{add-tech-profile} method. While intuitive, this subtraction-based approach does not guarantee additivity, as interaction effects between technologies lead to path-dependent and non-unique allocations.

Formally, the add-tech-profile method defines the contribution of a technology $i$ as the reduction in total flexibility needs when its profile is removed from the residual load:
\begin{equation}
\Delta_i^{add} :=
FlexNeed^{\ell|h}(RL) - FlexNeed^{\ell|h}(RL - P_i).
\end{equation}
Because $FlexNeed^{\ell|h}(\cdot)$ is nonlinear (due to the absolute value operator), this attribution rule is not additive in general.
\textbf{Counterexample.}
Consider a two-timestep system at a single resolution (so $h=\ell$ for simplicity). Let
\begin{equation}
RL = [2,4], 
\quad
P_1 = [2,8], 
\quad
P_2 = [0,-4],
\end{equation}
so that system balance holds: $RL = P_1 + P_2$. Here, $P_1$ represents a conventional generator and $P_2$ a flexible consumer that shifts load between timesteps.
\textbf{Original system.}
The mean residual load is $\overline{RL}=3$, hence
\[
FlexNeed(RL)
=
\frac12 (|2-3| + |4-3|)
=
\frac12 (1+1)
=
1.
\]
\textbf{Removing $P_1$.}
\[
RL - P_1 = [0,-4],
\quad
\overline{(RL - P_1)} = -2.
\]
Thus,
\[
FlexNeed(RL - P_1)
=
\frac12 (|0+2| + |-4+2|)
=
\frac12 (2+2)
=
2.
\]
Hence,
\[
\Delta_1^{add} = 1 - 2 = -1.
\]
\textbf{Removing $P_2$.}
\[
RL - P_2 = [2,8],
\quad
\overline{(RL - P_2)} = 5.
\]
Thus,
\[
FlexNeed(RL - P_2)
=
\frac12 (|2-5| + |8-5|)
=
\frac12 (3+3)
=
3.
\]
Hence,
\[
\Delta_2^{add} = 1 - 3 = -2.
\]
\textbf{Non-additivity.}
Summing individual contributions yields
\[
\Delta_1^{add} + \Delta_2^{add} = -3,
\]
whereas total flexibility needs equal
\[
FlexNeed(RL) = 1.
\]
Thus,
\[
\sum_i \Delta_i^{add} \neq FlexNeed(RL),
\]
demonstrating that the add-tech-profile method does not produce an additive decomposition.
The non-additivity arises because each marginal contribution is evaluated relative to the full system, and interactions between technologies are not accounted for consistently. Since $FlexNeed(\cdot)$ is nonlinear, marginal effects cannot be summed to recover the total system requirement.\newline

Note that in this example both contributions are negative, implying that removing either technology increases flexibility needs. While counterintuitive, this can occur when technologies are operationally constrained and therefore inflexible in practice, yet are nonetheless excluded from the residual load definition. The negative contributions thus reflect a definitional issue: if a technology's output is neither fully dispatchable nor accounted for in the residual load, the method may misattribute its role in flexibility provision.\newline

\textbf{Comparison with the correlation-based method.}
Applying the correlation-based method to the same example confirms exact additivity. The flexibility sign is determined by the residual load deviations from its mean $\overline{RL}=3$:
\[
FlexSign(t) = \mathrm{sign}(RL(t) - \overline{RL}) = \mathrm{sign}([2-3,\ 4-3]) = [-1,\ +1].
\]
The flexibility provision curves are computed as deviations from each technology's mean:
\[
\bar{P}_1 = 5, \quad FlexProvCurve_1 = [2-5,\ 8-5] = [-3,\ +3],
\]
\[
\bar{P}_2 = -2, \quad FlexProvCurve_2 = [0-(-2),\ -4-(-2)] = [+2,\ -2].
\]
The flexibility contributions are then:
\[
FlexProv_1 = \frac{1}{2}\left((-3)(-1) + (3)(+1)\right) = \frac{1}{2}(3+3) = 3,
\]
\[
FlexProv_2 = \frac{1}{2}\left((+2)(-1) + (-2)(+1)\right) = \frac{1}{2}(-2-2) = -2.
\]
Summing the contributions yields
\[
FlexProv_1 + FlexProv_2 = 3 + (-2) = 1 = FlexNeed(RL),
\]
confirming exact additivity. Note that $P_2$ receives a negative contribution, correctly identifying it as a flexibility-increasing consumer whose consumption pattern moves against system requirements.

\subsection{Detailed information on battery electric vehicle modelling}
\label{app:bev_modelling}

Battery electric vehicles (BEV) are modelled with an inflexible load that can be met through direct charging from the grid. If flexible charging is enabled, a battery is added with the energy capacity of the share of BEVs that participate in flexible charging. This decouples BEV electricity consumption and BEV charging temporally. If V2G is enabled, the energy stored in the BEV battery can also be fed back into the grid.\newline

\subsection*{BEV load and charging:}

The total BEV electricity demand is derived from road and rail transport energy balances from the JRC Integrated Database of the European Energy Sector (JRC-IDEES) \citep{jrc_idees}, rescaled to match Eurostat Energy Balances \citep{eurostat_energy_balances} and disaggregated to the nodal level by population distribution, then converted from fuel-equivalent kilometres 
to electricity consumption using the assumed electric drivetrain efficiency (\textit{default: 0.18 MWh/100km}). Both the load and charger availability profiles are shaped by hourly traffic counts from the Federal Highway Research Institute (BASt) \citep{bast_traffic}, averaged over 2010--2015, with availability defined as the inverse of traffic intensity, implicitly reflecting post-arrival plug-in behaviour. Therefore the load manifests during the day, with peaks around 8 am and 5 pm. \cite{schill_power_2015} find a slightly different behavior for their "plug-in and forget" strategy, with peaks at 9 am, 2 and 7 pm, but confirm low charging during the night; similar to \cite{lauvergne_integration_2022}. Other studies \cite{taljegard_impacts_2019,fischer_electric_2019,hanemann_effects_2017,muratori_impact_2018} confirm the general trend of increased charging during midday and reduced charging in the night, if an uncontrolled charging strategy is used. The availability profile is rescaled between a minimum baseline (\textit{default: }$\mathit{0.80}$) and a maximum plugged-in share (\textit{default: }$\mathit{0.95}$), and the total charging capacity is further determined by the number of EVs, their individual charger rating (\textit{default:} \SI{11}{kW}\textit{, corresponding to a standard three-phase AC charger}; in line with \cite{hanemann_effects_2017} and 10.45kW from \cite{schill_power_2015}), 
and the assumed electrification share of the passenger fleet. Charging efficiency is set to \textit{90\%} by default.

\begin{figure}[h]
    \centering
    \includegraphics[width=0.9\linewidth]{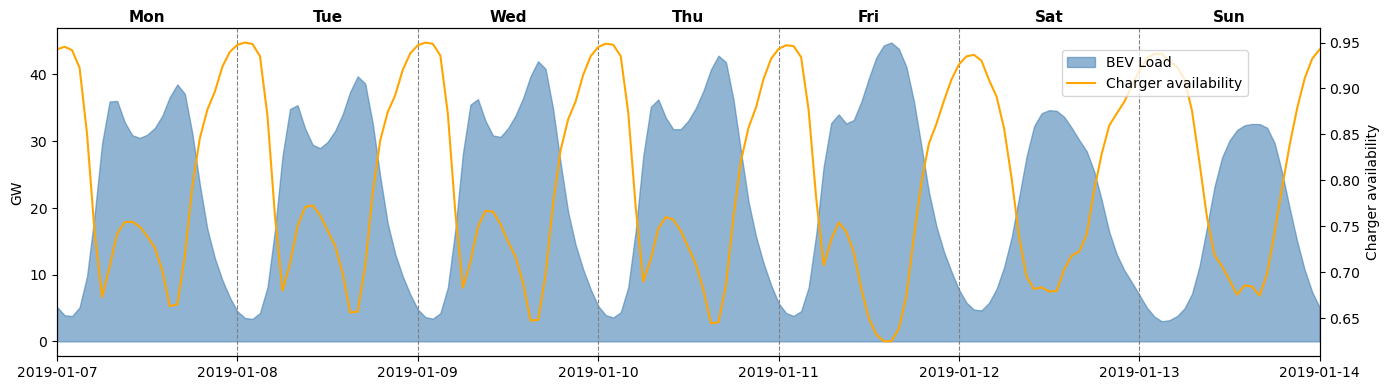}
    \caption{BEV transport energy demand and charger availability over a representative winter week (January 2019, Germany).}
    \label{fig:bev_avail}
\end{figure}

Figure~\ref{fig:bev_avail} illustrates this inverse relationship over a representative winter week: charger availability peaks during nights and weekends when most vehicles are parked, while the transport energy demand concentrates around morning and evening commute hours on weekdays.\newline

\subsection*{Demand-Side Management:}

Only a fraction of the fleet (\textit{default: 50\%}) contributes usable battery capacity for load shifting. The remaining vehicles do not add to the flexible energy buffer, though since the total transport load is not split between flexible and inflexible charging, all vehicles implicitly benefit from the available storage. The total flexible energy buffer therefore scales with the DSM participation rate, the number of participating vehicles, and their average battery capacity (\textit{default:} \SI{50}{kWh}). This is a conservative assumption taking average capacities of BEVs on the market in 2025 of about 80 kWh. 

A minimum state-of-charge (SoC) constraint is enforced each morning at \textit{07:00}, requiring participating batteries to hold at least \textit{80\%} of capacity at that time. This ensures vehicles are adequately charged for daily use while permitting the optimiser to shift grid electricity uptake freely outside this window. Energy balance is maintained over the full optimisation horizon through a periodicity constraint, enforcing equal SoC at the start and end of each optimisation period.

\begin{figure}[h]
    \centering
    \includegraphics[width=0.9\linewidth]{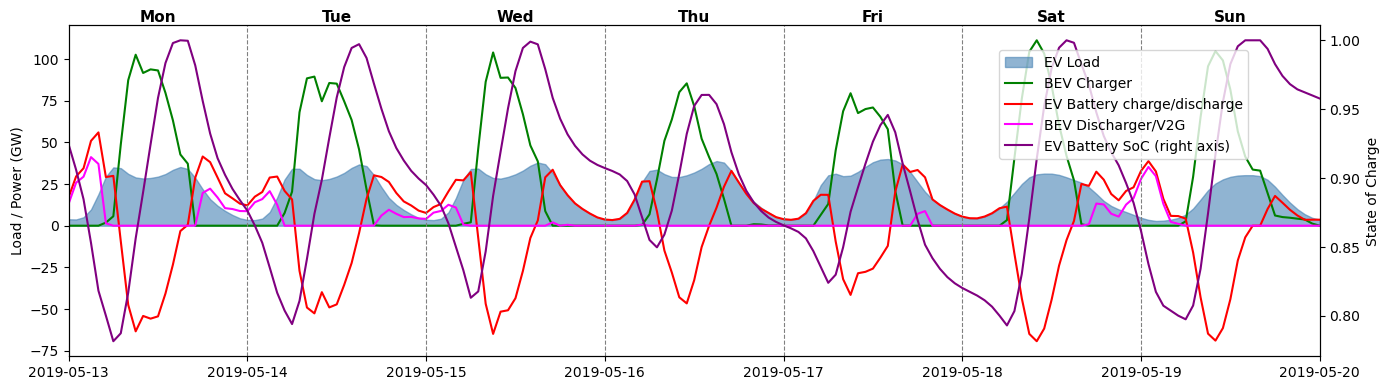}
    \caption{EV fleet dispatch over a representative spring week (May 2019, Germany) in a scenario with V2G enabled. Shown are the transport energy demand, grid charging power, net battery charge/discharge, V2G discharge to grid, and battery state of charge.}
    \label{fig:bev_dispatch}
\end{figure}

\subsection*{Vehicle-to-Grid:}

When V2G capability is enabled, DSM-participating vehicles may discharge stored energy back to the grid. Discharge is subject to the same availability profile as charging and the round-trip charging efficiency applies in both directions (\textit{default: 90\% per direction}).

Figure~\ref{fig:bev_dispatch} shows a representative dispatch week with V2G enabled. Charging of BEVs and battery storage takes place during low-price hours (midday solar peaks), building up battery SoC, and discharges back to the grid when economically favourable. The SoC consistently recovers to its minimum threshold before the morning commute. Schill and Gerbaulet \cite{schill_power_2015} confirm the shift in BEV load towards the midday and night with cost-driven charging.\newline

\subsection{Industrial Demand-Side Management: Implementation and Modelling Assumptions}
\label{app:industrial_dsm}

\subsubsection*{Taxonomy and terminology}

Industrial DSM can be decomposed into three flexibility types: (i)~\textit{load shedding} (permanent demand reduction), (ii)~\textit{load shifting} (a reduction compensated by a corresponding increase within a bounded time window), and (iii)~\textit{uncompensated load increase}. The implementation presented here models \textit{load shifting} within a daily time frame.

Flexibility-relevant processes are typically grouped into \textit{industrial processes}, \textit{general-purpose technologies} (GPT), and \textit{trade, commerce and services} (TCS) \cite{ffe_regionale_2021, klaucke_advanced_2025}. These categories are characterised by the \textit{holding duration} (maximum duration of a load adjustment) and the \textit{shifting duration} (maximum time until energy balance must be restored).

\subsubsection*{Evidence base for flexibility potentials}

The industrial load shedding potential in Germany was already 1.2 GW for one hour in 2019 \cite{ffe_regionale_2021}. Estimates of \textit{current technical potentials} in Germany consistently lie in the range of 5--8~GW for load reduction with one-hour activation, declining by roughly 30--40\% for four-hour durations \cite{steurer_analyse_2017, ffe_regionale_2021, stede_demand_2016}. For example, \cite{steurer_analyse_2017} report a current technical potential of 7.5~GW (1~h) decreasing to 5.2~GW (4~h). Other studies find similar magnitudes of several gigawatts available for at least one hour, with longer durations requiring additional investments \cite{stede_demand_2016}.

\textit{Future projections} suggest moderate expansion: \cite{ffe_regionale_2021} project 5.3~GW in 2035 and 12.6~GW in 2045 (both for at least one-hour holding duration), while \cite{mckinsey_flexibilitat_2025} estimate that 3--4~GW will be realised already by 2028.

Load shifting potentials are generally smaller than load reduction potentials and depend strongly on process assumptions. Importantly, some studies consider only \textit{industrial processes} and exclude GPT and TCS, leading to lower estimates. For instance, \cite{klaucke_advanced_2025} estimate a \textit{future} shifting potential of 1.5--1.6~GW (3--12~h holding duration) in 2030 based solely on industrial processes, assuming 10\% flexible operation enabled by product storage buffers. More broadly, \cite{ranaboldo_comprehensive_2024} report \textit{current} short-term load reduction potentials of 3--5~GW, typically exceeding load increase potentials.

Holding durations vary substantially across processes but consistently cluster in the range of \textit{1--4 hours}, with some processes extending to 5--8 hours (e.g.\ chlor-alkali electrolysis), while shorter durations below two hours dominate in many applications \cite{ffe_regionale_2021, kochems_lastflexibilisierungspotenziale_2020, buber_tim_lastmanagement_2013}. Empirical evidence shows that activation beyond one to two hours is feasible but often requires additional investments such as intermediate storage. Shifting durations are predominantly intra-day and typically limited to within 24 hours \cite{kochems_lastflexibilisierungspotenziale_2020, stede_demand_2016}.

\subsubsection*{Modelling choices and justification}

Industrial DSM is modelled as a debt-tracking virtual storage: load reductions charge a debt store that must be compensated by subsequent increases within the same day. The total DSM capacity is distributed across industrial load buses proportional to their average electricity demand.

The assumed shiftable capacity of 6~GW in 2035 and 10~GW in 2045 reflects \textit{future potentials} and lies within the range of projected total technical potentials when accounting for duration effects and increasing electrification \cite{ffe_regionale_2021, mckinsey_flexibilitat_2025}. A uniform holding duration of 4~hours is chosen as a compromise across heterogeneous processes, consistent with the commonly reported 2--4~hour range and the upper bound of typical operation without major additional investments. The shifting duration is set to 24~hours via a daily cycling constraint, reflecting the predominantly intra-day nature of industrial load shifting.

\subsubsection*{Limitations}

Several simplifications should be noted. First, zero activation costs and instantaneous ramping are assumed, overstating operational flexibility. Second, activation frequency is unconstrained despite real-world recovery times. Third, DSM capacity is spatially distributed proportional to demand rather than process-specific locations. Finally, the aggregation into a single technology with fixed parameters cannot capture the heterogeneity of industrial processes, in particular the wide range of holding durations.

\clearpage
\section{Tables and Figure}
\label{app:figures-tables}

\subsection{Tables}
\label{app:tables}

\begin{table}[htbp]
\caption{\textbf{Storage capacities for Germany in 2035 and 2045.} Storage capacities and BEV charging parameters across flexibility scenarios.}
\begin{center}
\small
\begin{tabular}{lcccccccc}
\toprule
 & \multicolumn{4}{c}{\textbf{2035}} & \multicolumn{4}{c}{\textbf{2045}} \\
\cmidrule(lr){2-5} \cmidrule(lr){6-9}
\multicolumn{9}{c}{\textit{Storage Energy Capacity [GWh]}} \\
\midrule
\textbf{Parameter} & \textbf{LF} & \textbf{LB} & \textbf{BA} & \textbf{HF} & \textbf{LF} & \textbf{LB} & \textbf{BA} & \textbf{HF} \\
\midrule
PHS                    & 41   & 41   & 41   & 41   & 41    & 41    & 41    & 41    \\
Battery                & 305  & 215  & 310  & 21   & 600   & 285   & 429   & 25    \\
BEV                    & --   & --   & 650  & 1040 & 402   & 402   & 1006  & 1609  \\
Iron-air battery       & --   & --   & --   & 1278 & --    & --    & --    & 1765  \\
Industry DSM           & --   & --   & --   & 24   & --    & --    & --    & 40   \\
Heat (central)         & 4848 & 7661 & 6426 & 5972 & 6258  & 8581  & 8366  & 6444  \\
Heat (decentral/rural) & 11   & 52   & 14   & 9    & 11    & 52    & 14    & 9     \\
H$_2$                  & 2738 & 6368 & 3646 & 3043 & 15527 & 20413 & 20689 & 21781 \\
\midrule
\multicolumn{9}{c}{\textit{Power Capacity [GW]}} \\
\midrule
\textbf{Parameter} & \textbf{LF} & \textbf{LB} & \textbf{BA} & \textbf{HF} & \textbf{LF} & \textbf{LB} & \textbf{BA} & \textbf{HF} \\
\midrule
PHS                         & 8   & 8   & 8   & 8   & 8   & 8   & 8   & 8   \\
Battery                     & 29  & 19  & 39  & 3   & 36  & 24  & 48  & 3   \\
Iron-air battery            & --  & --  & --  & 35  & --  & --  & --  & 45  \\
Industry DSM                & --  & --  & --  & 6   & --  & --  & --  & 10  \\
Heat (central)              & 34  & 52  & 44  & 41  & 43  & 58  & 57  & 44  \\
Heat (decentral/rural)      & 75  & 349 & 93  & 57  & 75  & 349 & 93  & 57  \\
BEV max.\ charge$^{a}$      & 38  & 38  & 82  & 98  & 85  & 94  & 108 & 121 \\
BEV max.\ discharge$^{a}$   & --  & --  & --  & 52  & --  & --  & --  & 52  \\
\bottomrule
\end{tabular}
\begin{tablenotes}
\item LF = LowFlex, LB = LowBattery, BA = Base, HF = HighFlex.
Storage energy capacities represent total installed capacity across all nodes. BEV storage energy and V2G discharge are only available where smart charging / V2G is enabled.
\item[$a$] For BEV, values represent the fleet-level peak charging/discharging constraint rather than an optimised infrastructure capacity.
\end{tablenotes}
\label{tab:storage-capacities}
\end{center}
\end{table}

\clearpage

\subsection{Figures}
\label{app:figures}

\begin{figure}[htbp]
    \centering
    \includegraphics[width=0.8\linewidth]{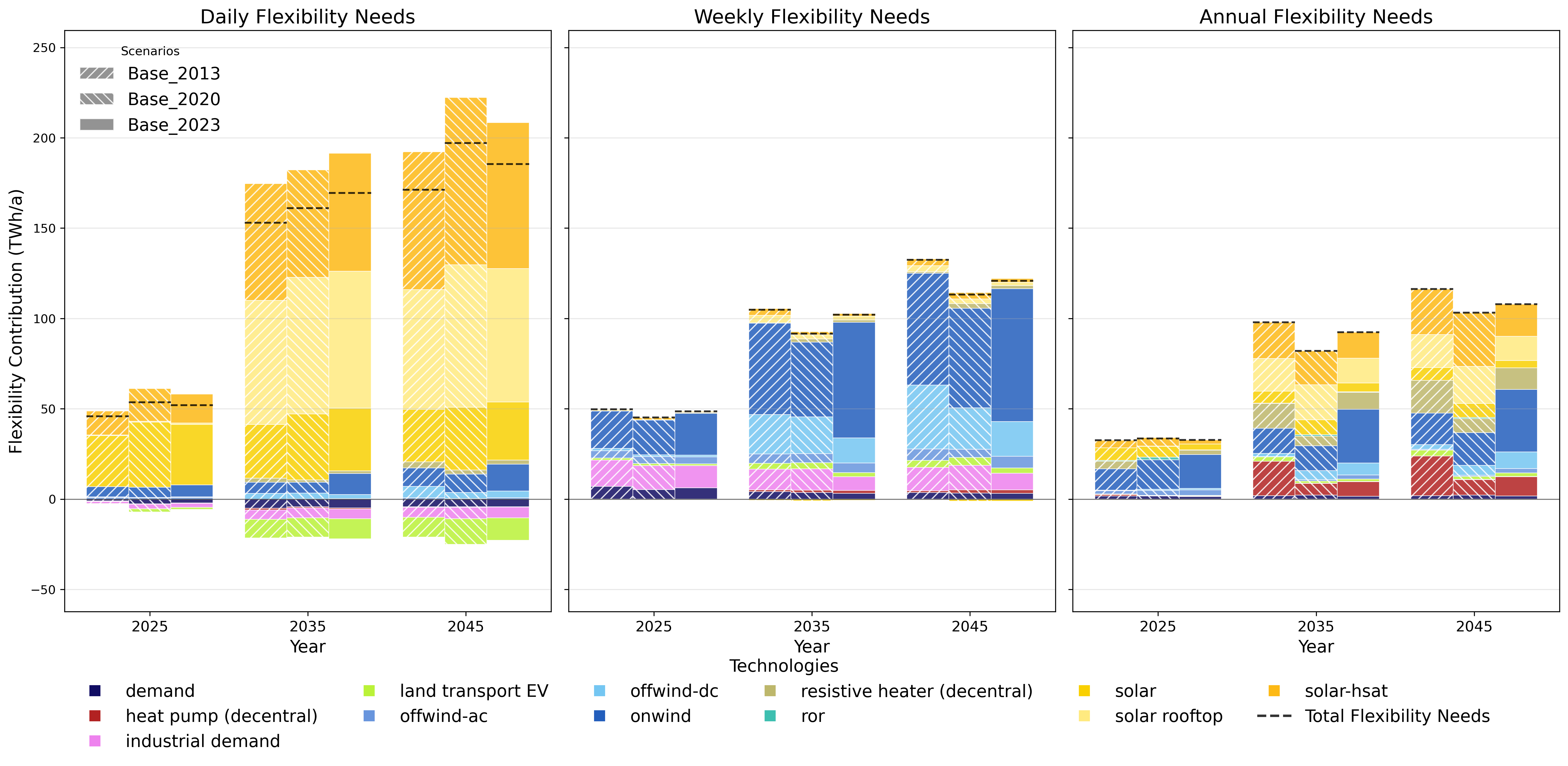}
    \caption{\textbf{Flexibility needs and causes for different weather years in the Base scenario.}
        The Figure shows the variation in flexibility needs across the weather years 2013, 2020 and 2023 in Germany for the Base scenario.}
    \label{fig:app:flex-needs-weather-years}
\end{figure}

\begin{figure}[htbp]
    \centering
    \includegraphics[width=0.9\linewidth]{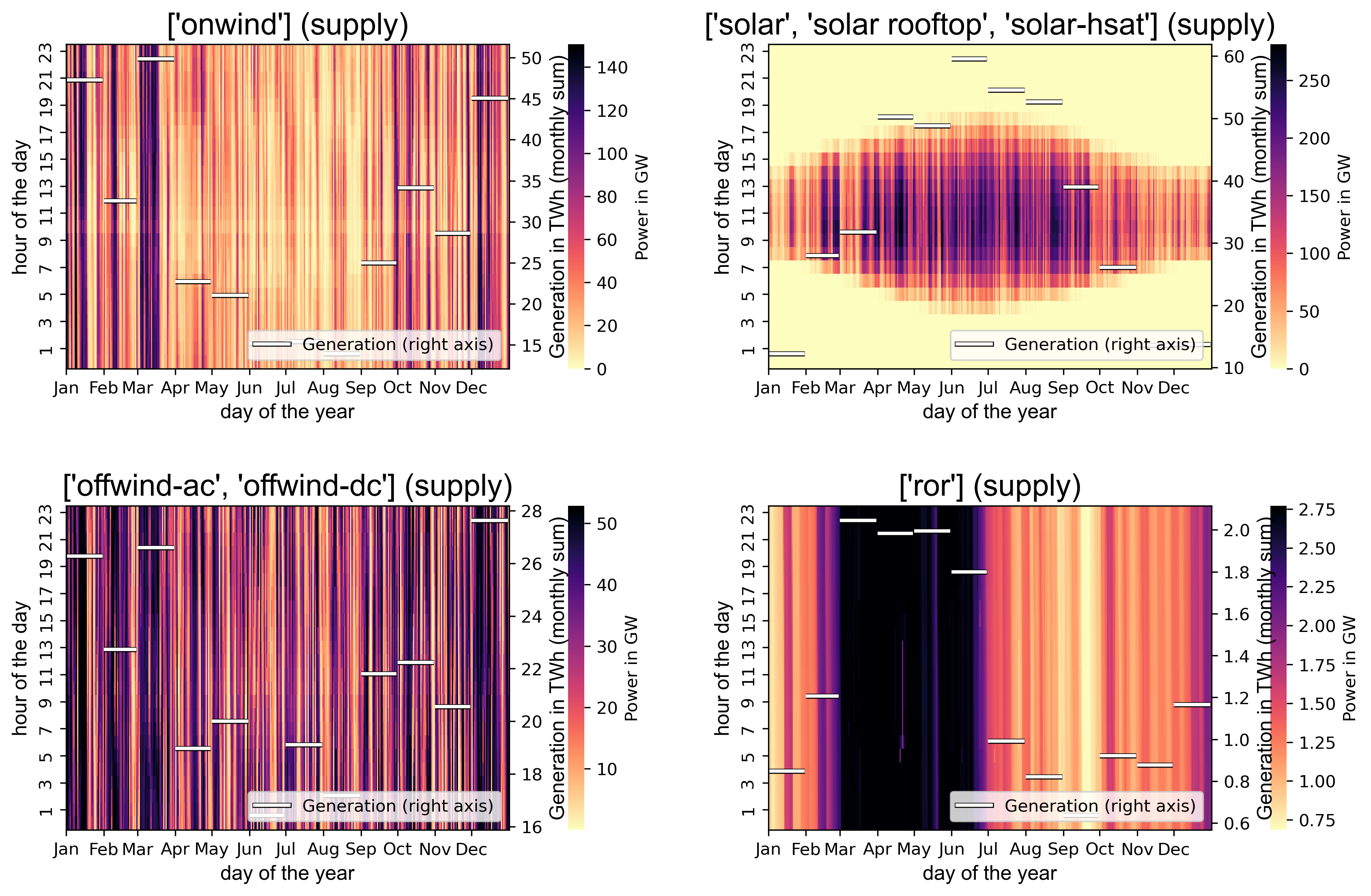}
    \caption{\textbf{Heatmap of non-dispatchable supply for the Base scenario in 2045.}
        The Figure shows the generation profiles of VRES in Germany for the year 2045.}
    \label{fig:app:heatmap_non-disp-supply}
\end{figure}

\begin{figure}[htbp]
    \centering
    \includegraphics[width=0.9\linewidth]{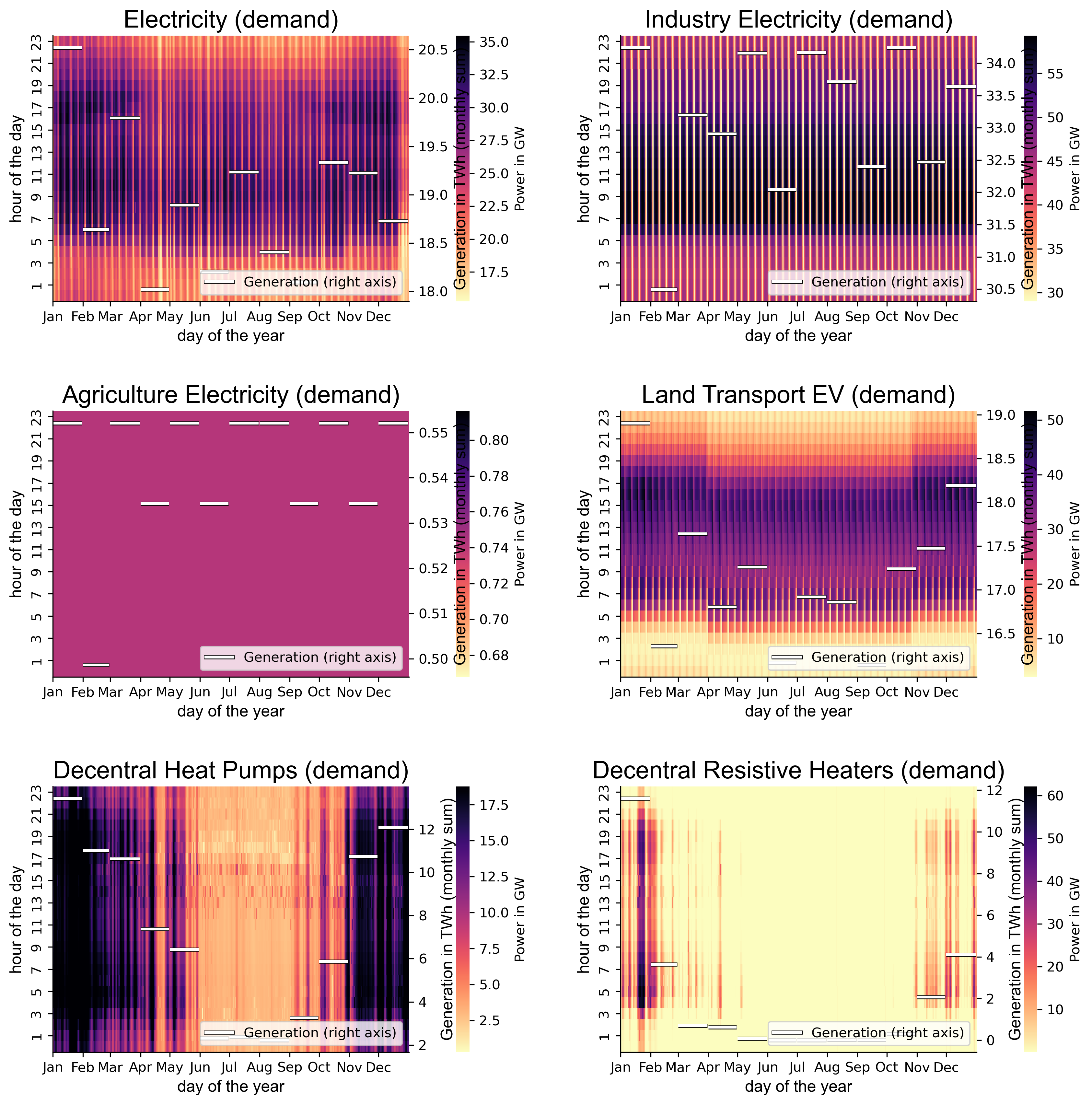}
    \caption{\textbf{Heatmap of non-dispatchable demand for the Base scenario in 2045.}
        The Figure shows the consumption profiles of the temporally resolved electricity loads in Germany for the year 2045.}
    \label{fig:app:heatmap_non-disp-demand}
\end{figure}

\begin{figure}[htbp]
    \centering
    \includegraphics[width=0.7\linewidth]{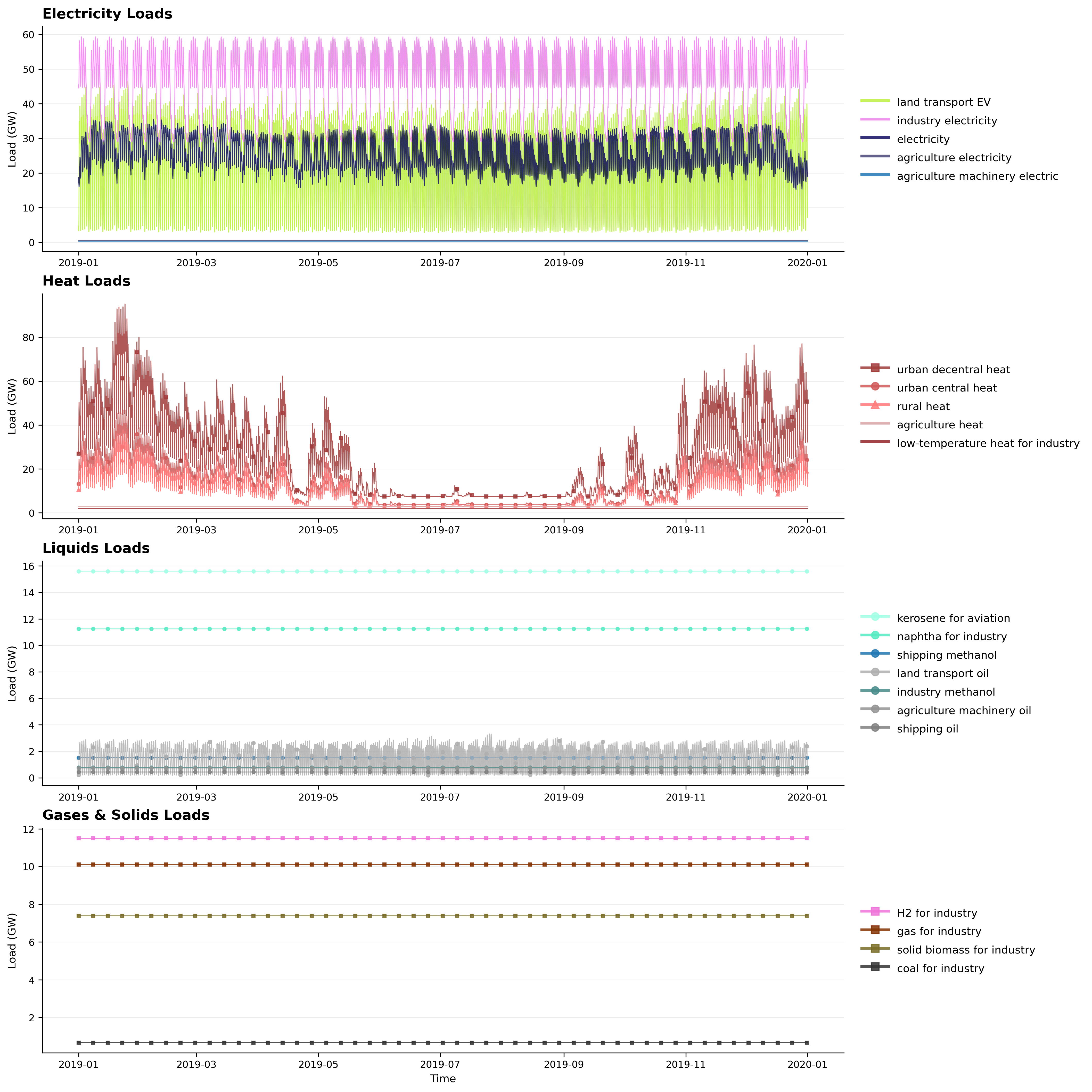}
    \caption{\textbf{Consumption profiles of loads for the Base scenario in 2045.}
        The Figure shows the consumption profiles of different load types in Germany for the year 2045.}
    \label{fig:app:loads_by_carrier}
\end{figure}

\begin{figure}[htb]
    \centering
    \footnotesize
    \begin{minipage}{0.6\columnwidth}
        \centering
        (A) Evolution of residual load and renewables penetration in February and July\\[0.5em]
        \includegraphics[width=\textwidth]{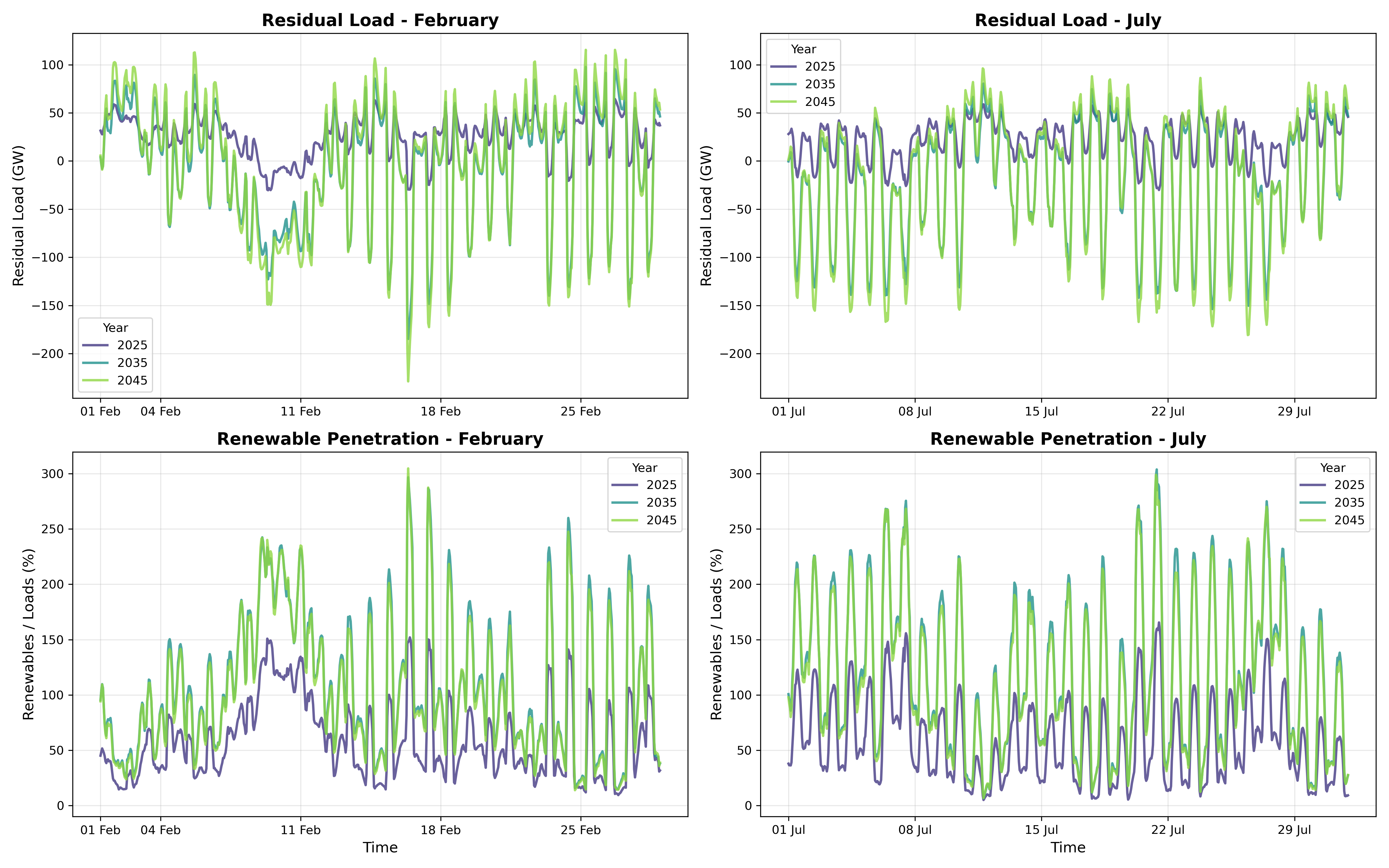}
    \end{minipage}%
    \hfill
    \begin{minipage}{0.4\columnwidth}
        \centering
        (B) Evolution of residual load duration curve\\[0.5em]
        \includegraphics[width=\textwidth]{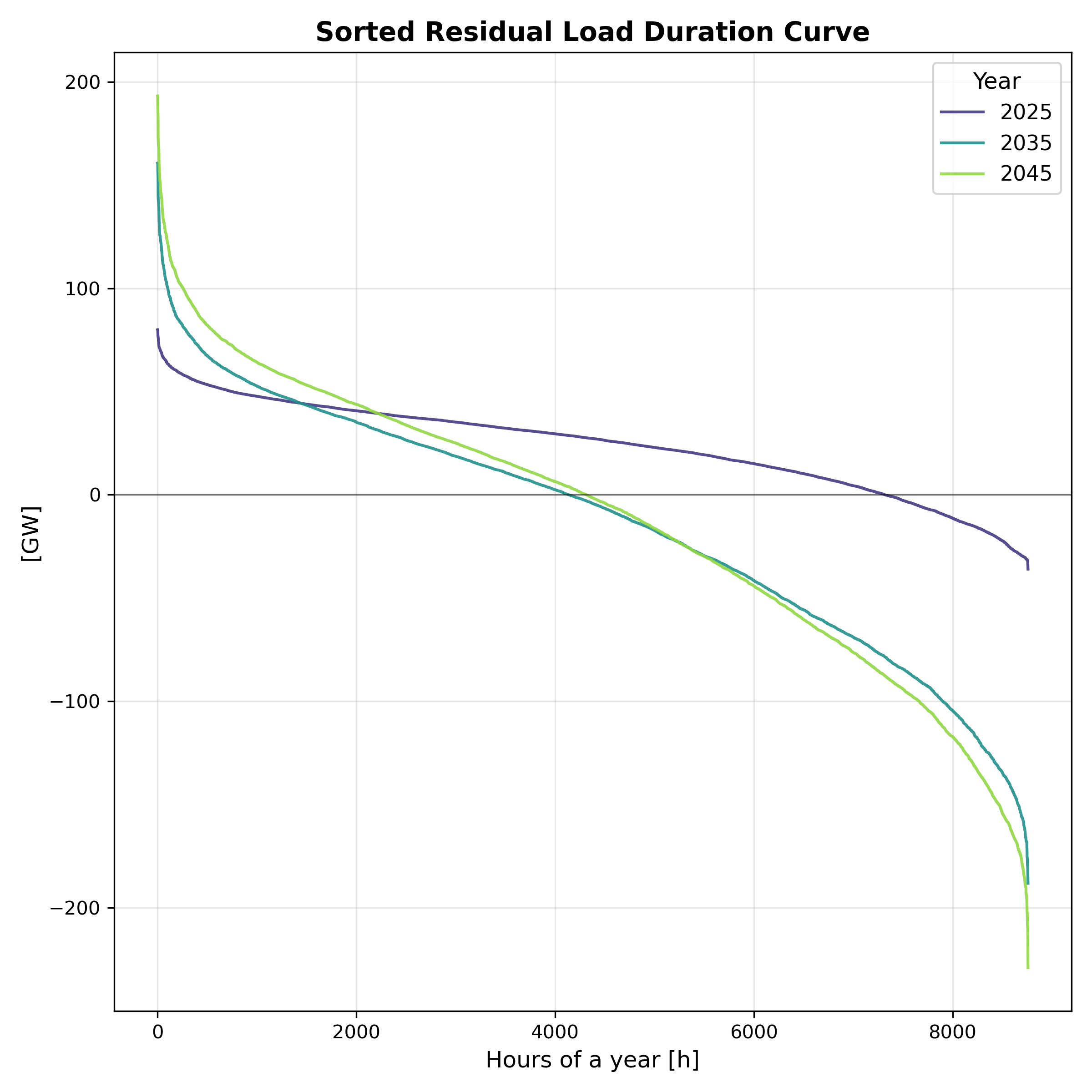}
    \end{minipage}
    \caption{\textbf{Evolution of Residual Load}
    The Figure shows the evolution of the residual load and renewables penetration for the months of February and July (left plot) as well as the evolution of the residual load duration curve (right plot)}
    \label{fig:app:residual-load}
\end{figure}

\begin{figure}[htb!]
    \centering
    \footnotesize

    \begin{subfigure}{\textwidth}
        \centering
        \includegraphics[width=\textwidth]{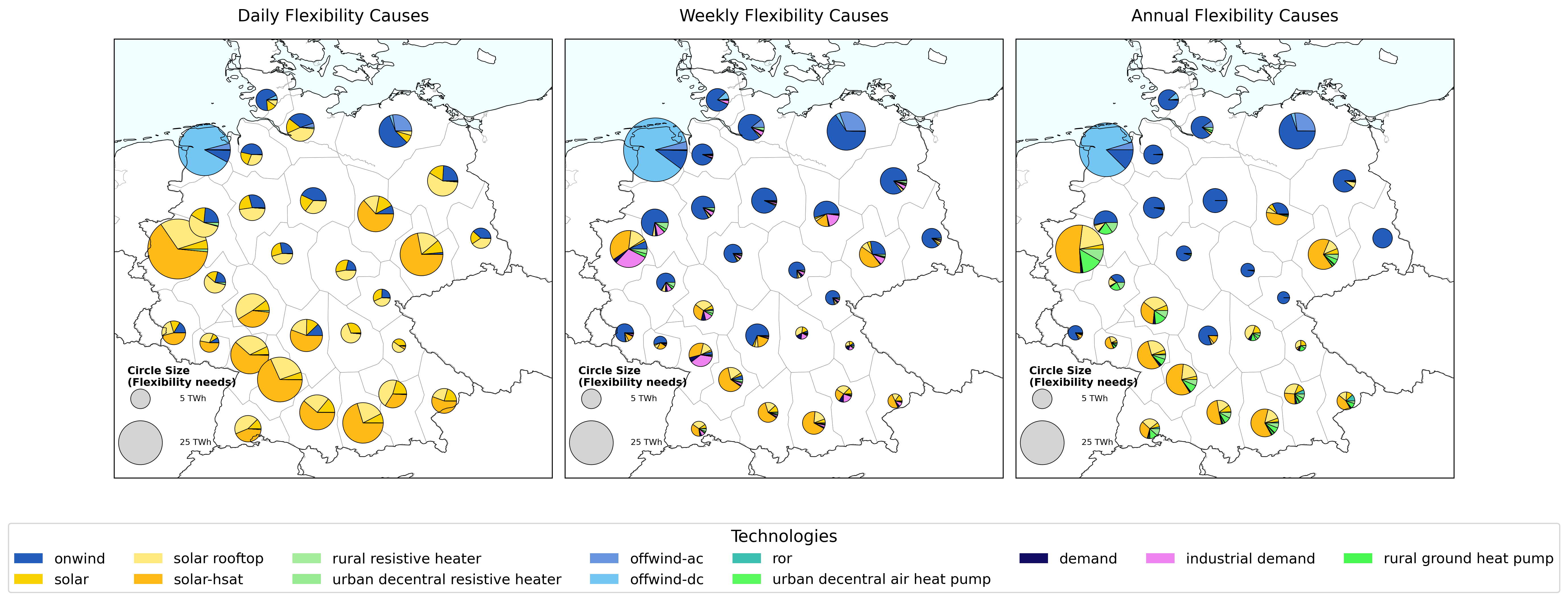}
        \caption{Flexibility causes across regions for Base scenario in 2045}
    \end{subfigure}

    \vspace{0.8em}

    \begin{subfigure}{\textwidth}
        \centering
        \includegraphics[width=\textwidth]{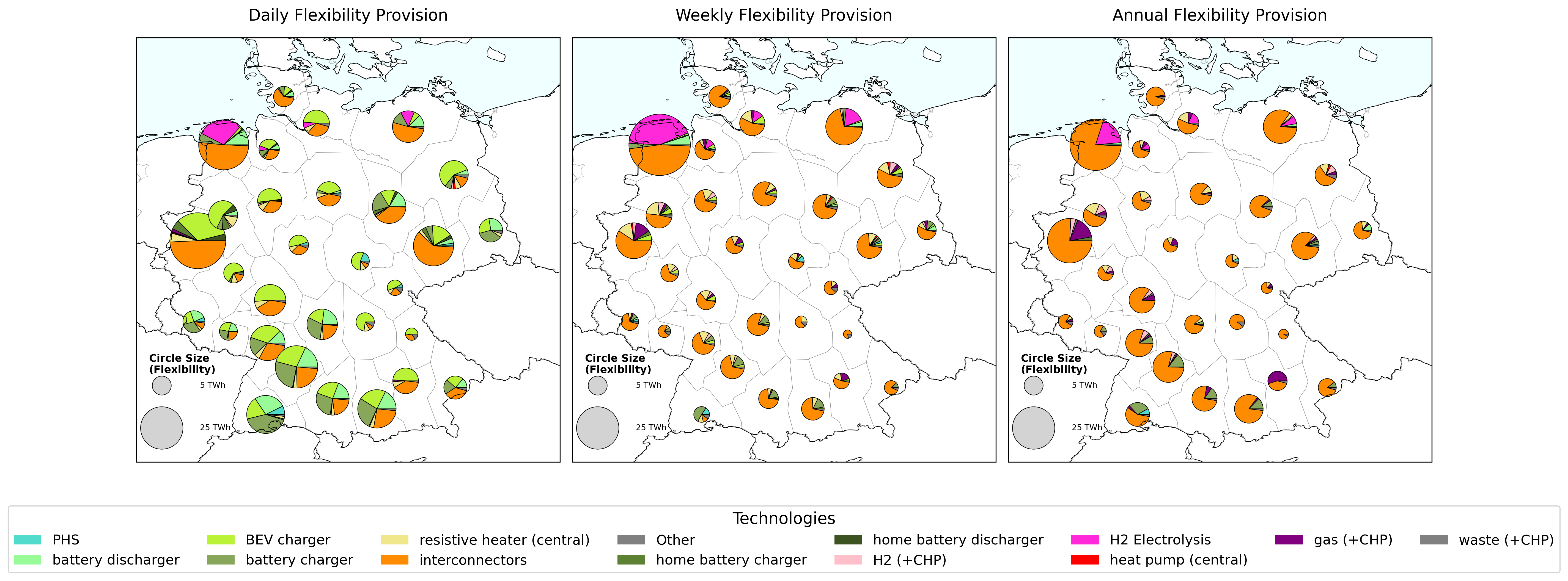}
        \caption{Flexibility provision across regions for Base scenario in 2045}
    \end{subfigure}

    \caption{\textbf{Flexibility causes and provision across different regions for the Base scenario in 2045.} 
    The Figure shows flexibility needs (A) and provision (B) for different regions across various timescales.}
    \label{fig:app:flex-spatial}
\end{figure}

\begin{figure}[htb!]
    \centering
    \footnotesize

    \begin{subfigure}{\textwidth}
        \centering
        \includegraphics[width=\textwidth]{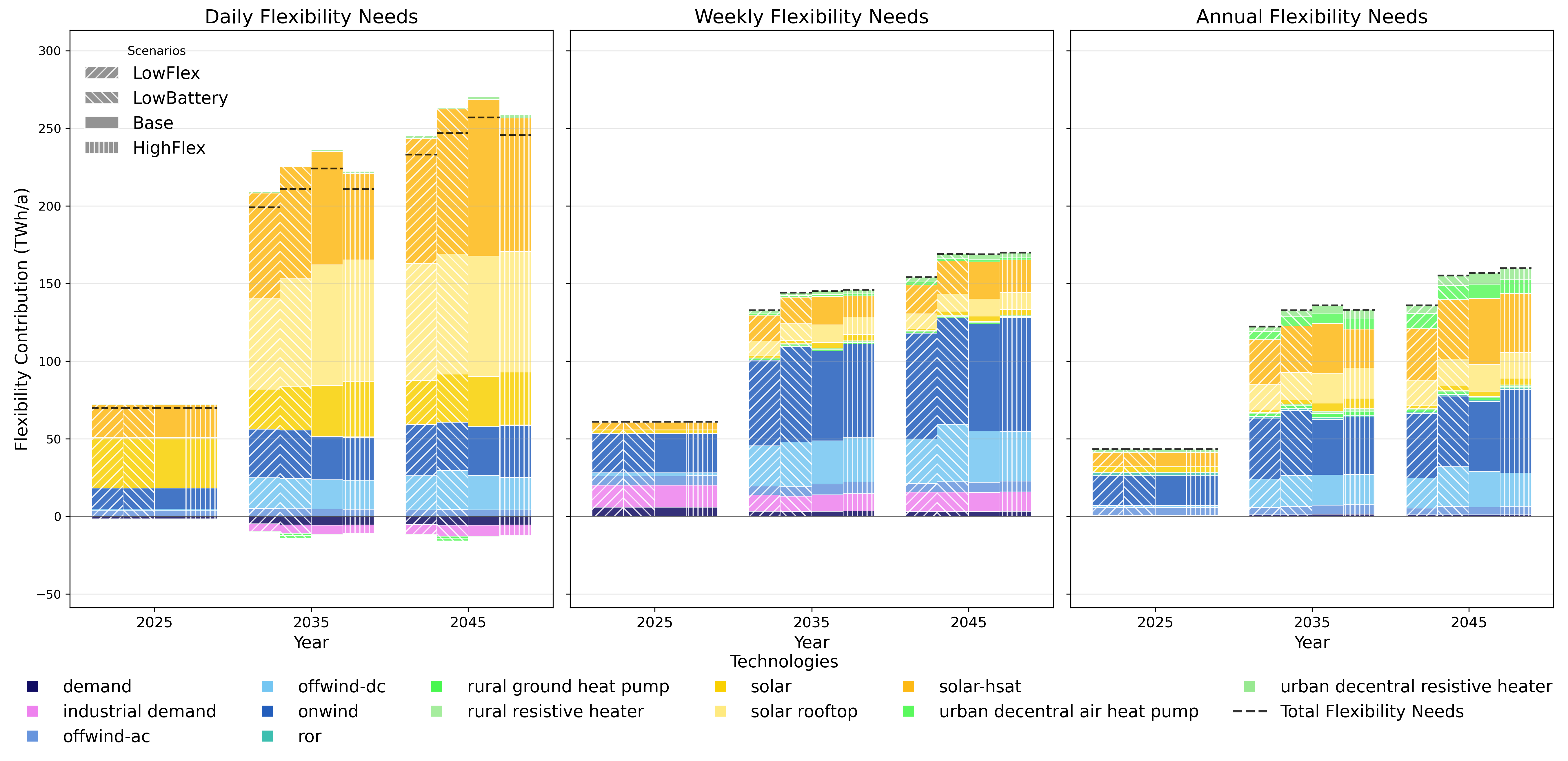}
        \caption{Flexibility needs across different scenarios (sum of regions)}
    \end{subfigure}

    \vspace{0.8em}

    \begin{subfigure}{\textwidth}
        \centering
        \includegraphics[width=\textwidth]{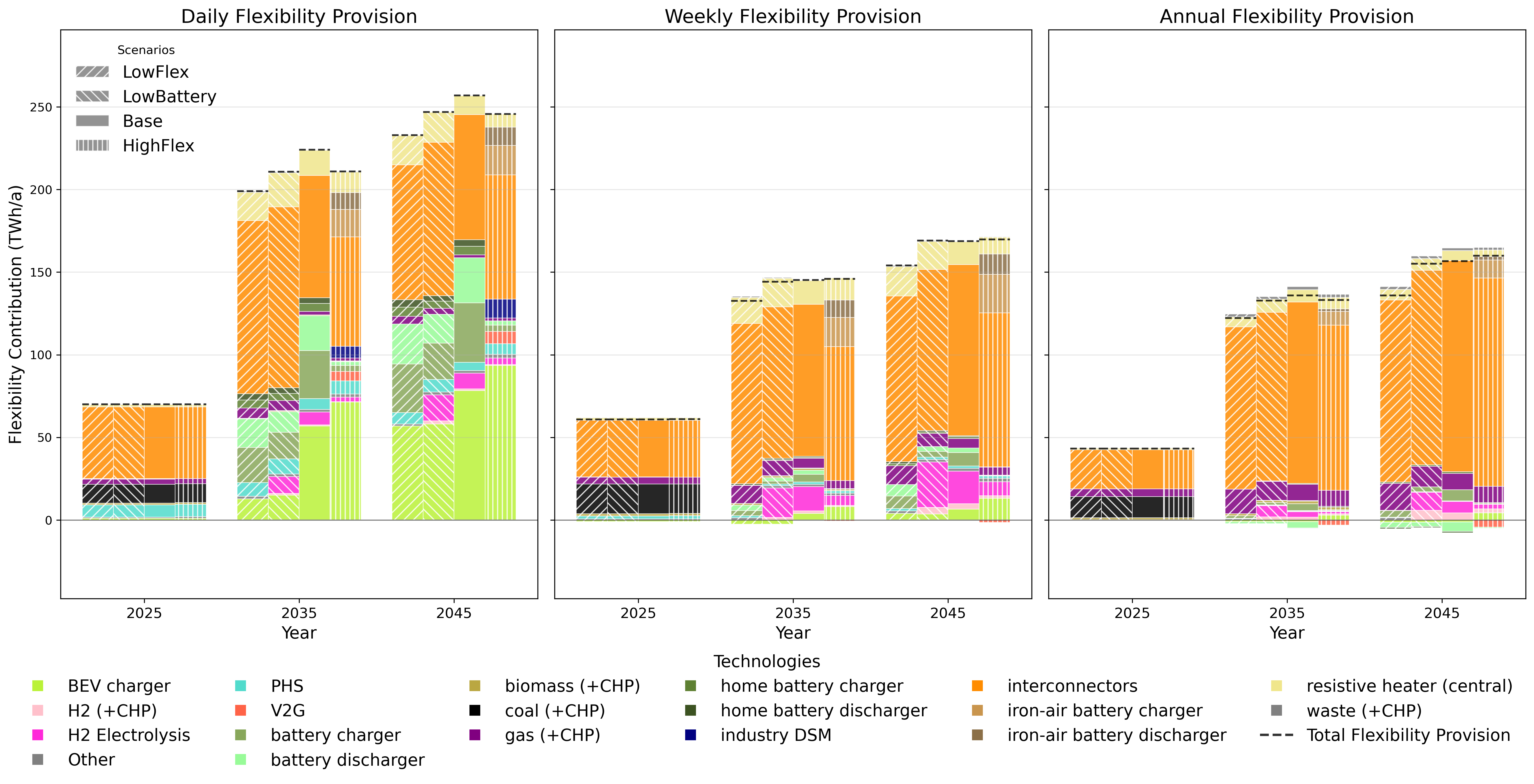}
        \caption{Flexibility provision across different scenarios (sum of regions)}
    \end{subfigure}

    \caption{\textbf{Flexibility causes and provision across different scenarios.} 
    The Figure shows flexibility needs (A) and provision (B) for different scenarios across various timescales. System characteristics are not aggregated prior to applying the method; instead, the method is applied to each individual region, and the resulting flexibility metrics are aggregated afterwards.}
    \label{fig:app:flex-per-node}
\end{figure}

\begin{figure*}[htb!]
    \centering
    \footnotesize

    \begin{subfigure}{\textwidth}
        \centering
        \includegraphics[width=\textwidth]{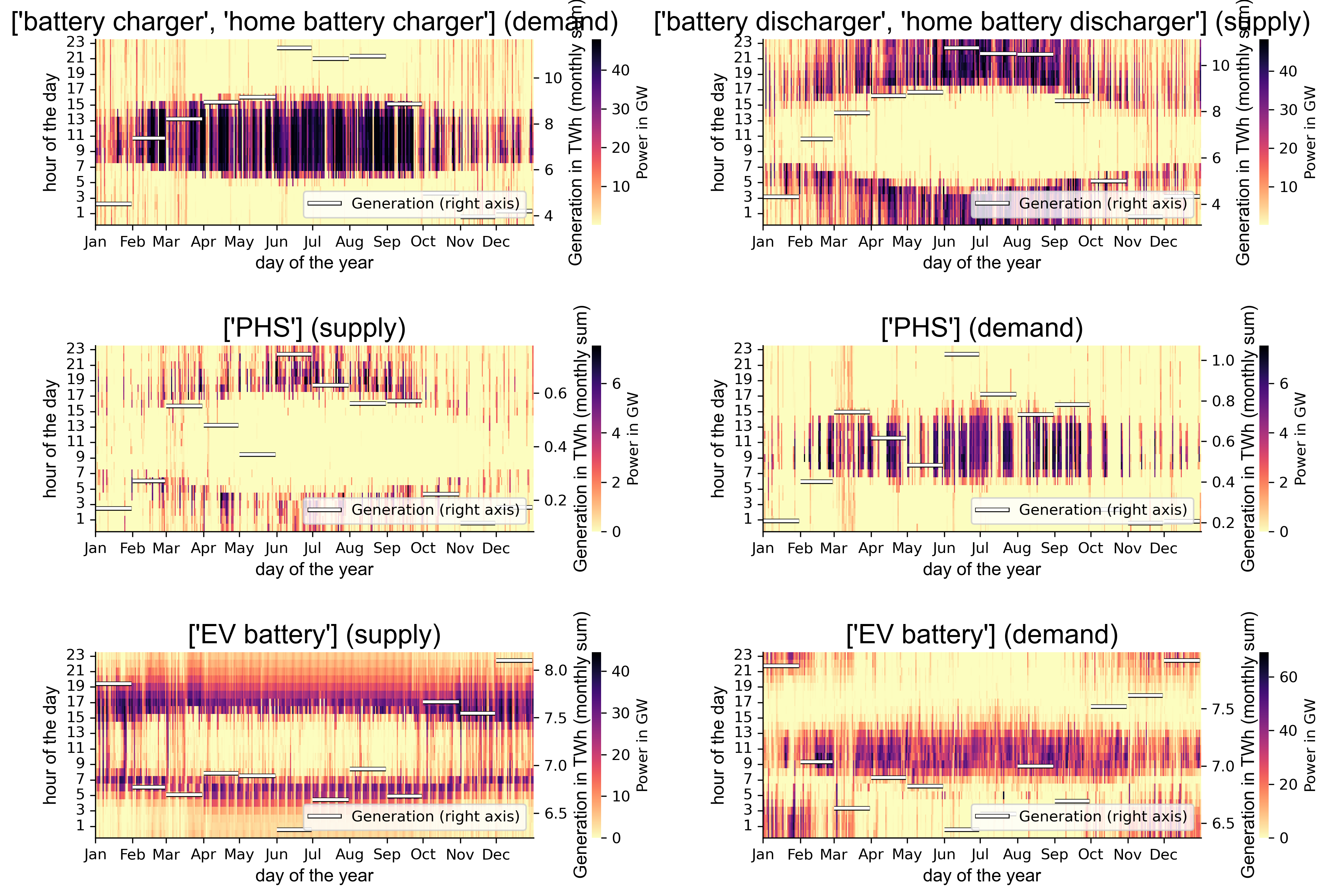}
        \caption{Storage}
    \end{subfigure}

    \vspace{0.2em}

    \begin{subfigure}{\textwidth}
        \centering
        \includegraphics[width=\textwidth]{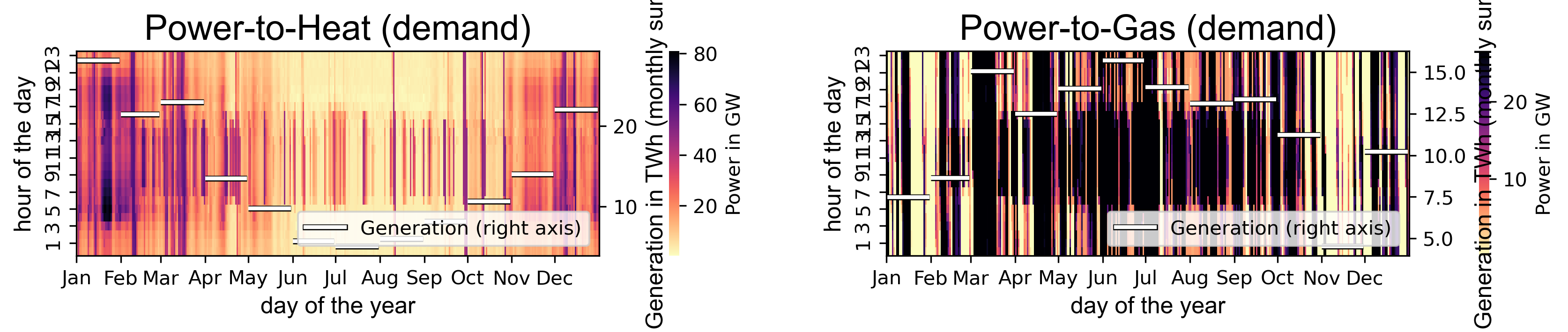}
        \caption{PtX}
    \end{subfigure}

    \vspace{0.2em}

    \begin{subfigure}{\textwidth}
        \centering
        \includegraphics[width=\textwidth]{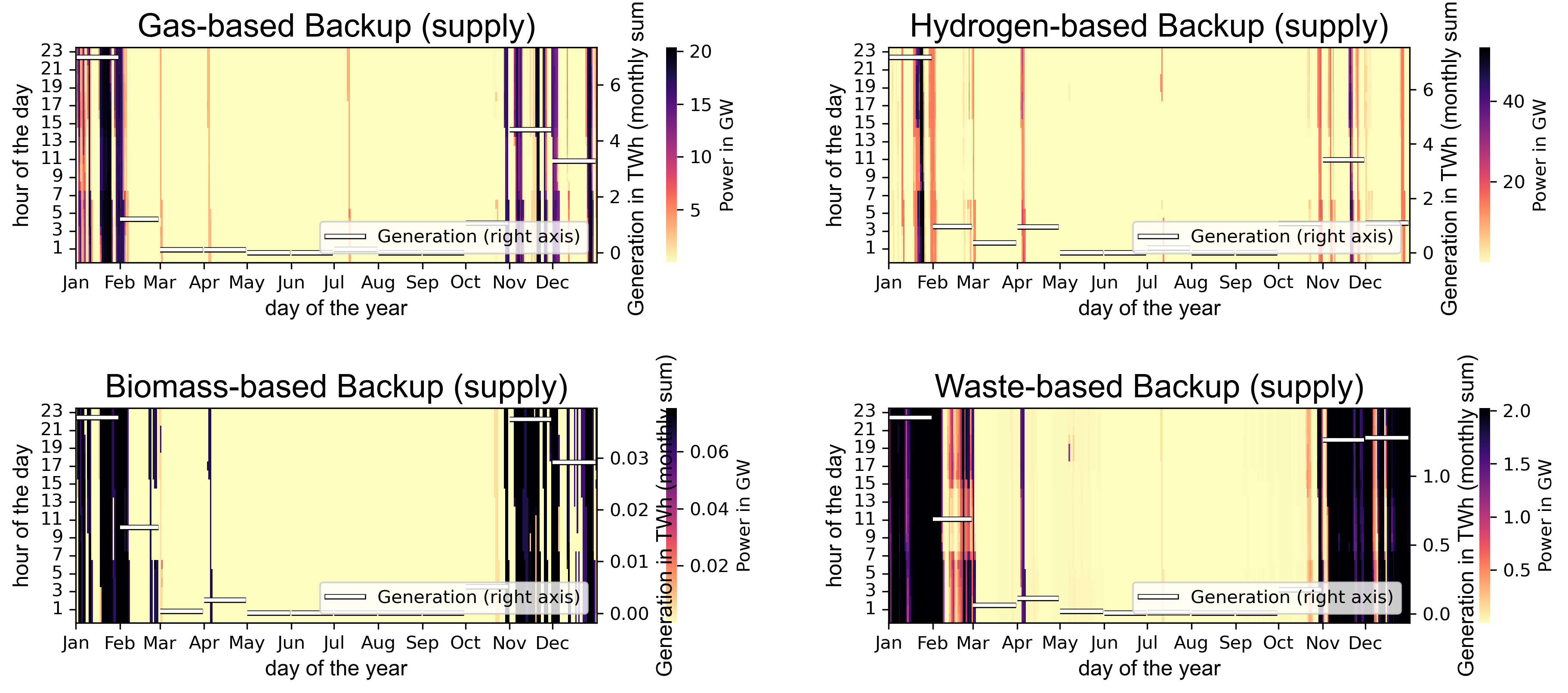}
        \caption{Backup}
    \end{subfigure}

    \caption{\textbf{Operation profiles of key flexibility technologies.} 
    The Figure shows the operation profiles for storage technologies (a), PtX (b) and Backup (c).}
    \label{fig:app:tech-profiles}
\end{figure*}

\begin{figure}[htbp]
    \centering
    \begin{subfigure}{\linewidth}
        \centering
        \includegraphics[width=0.7\linewidth]{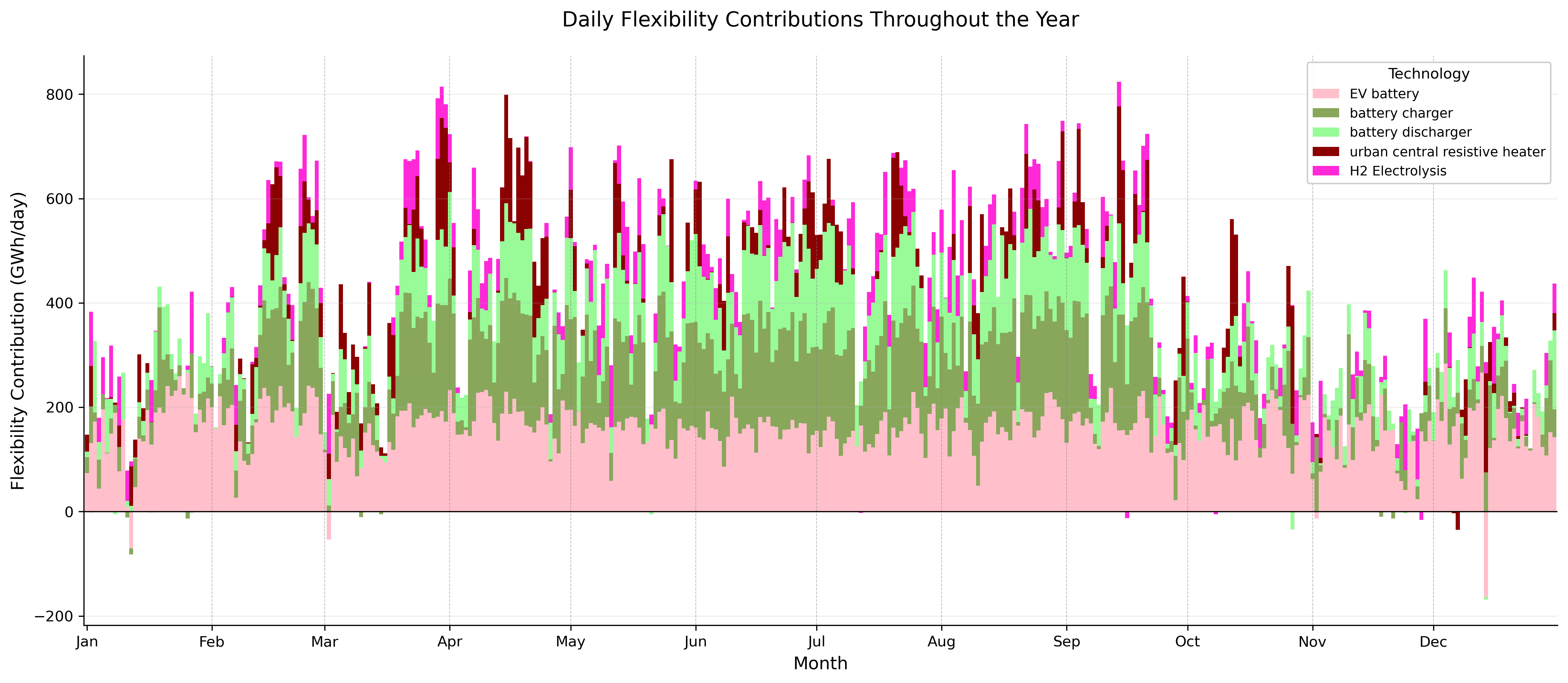}
        \caption{Top 5 Flexibility Contributors}
    \end{subfigure}\\[1em]
    \begin{subfigure}{\linewidth}
        \centering
        \includegraphics[width=0.7\linewidth]{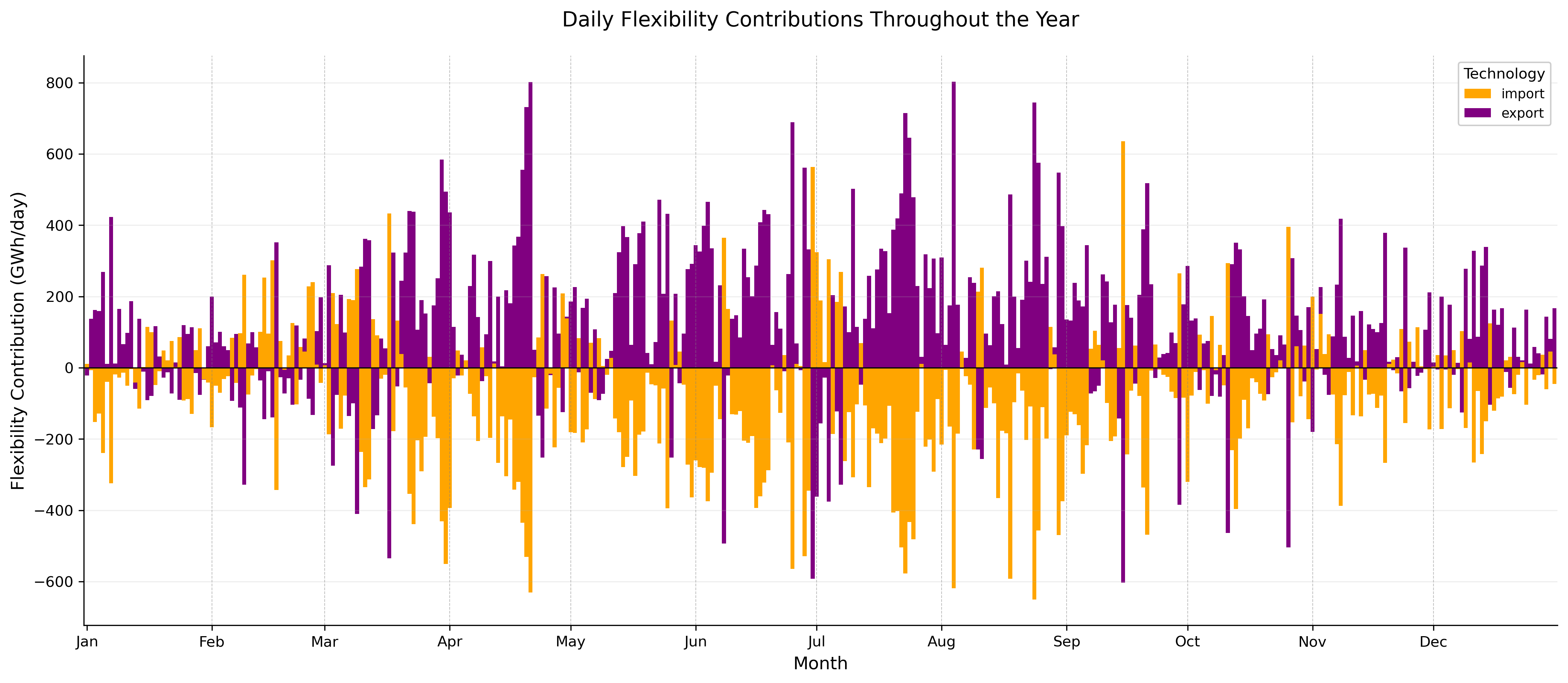}
        \caption{Import \& Export Flexibility Contributions}
    \end{subfigure}\\[1em]
    \begin{subfigure}{\linewidth}
        \centering
        \includegraphics[width=0.7\linewidth]{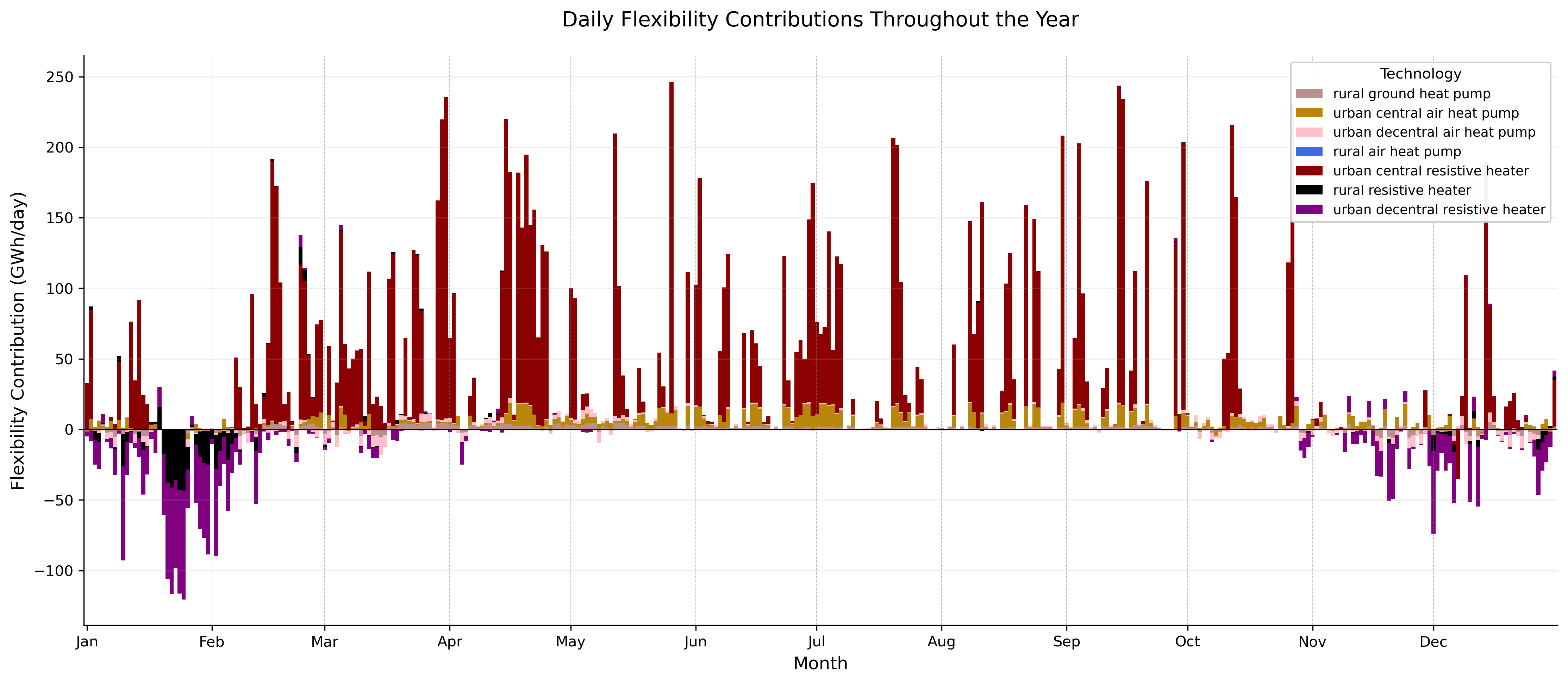}
        \caption{Heat Flexibility Contributions}
    \end{subfigure}
    \begin{subfigure}{\linewidth}
        \centering
        \includegraphics[width=0.7\linewidth]{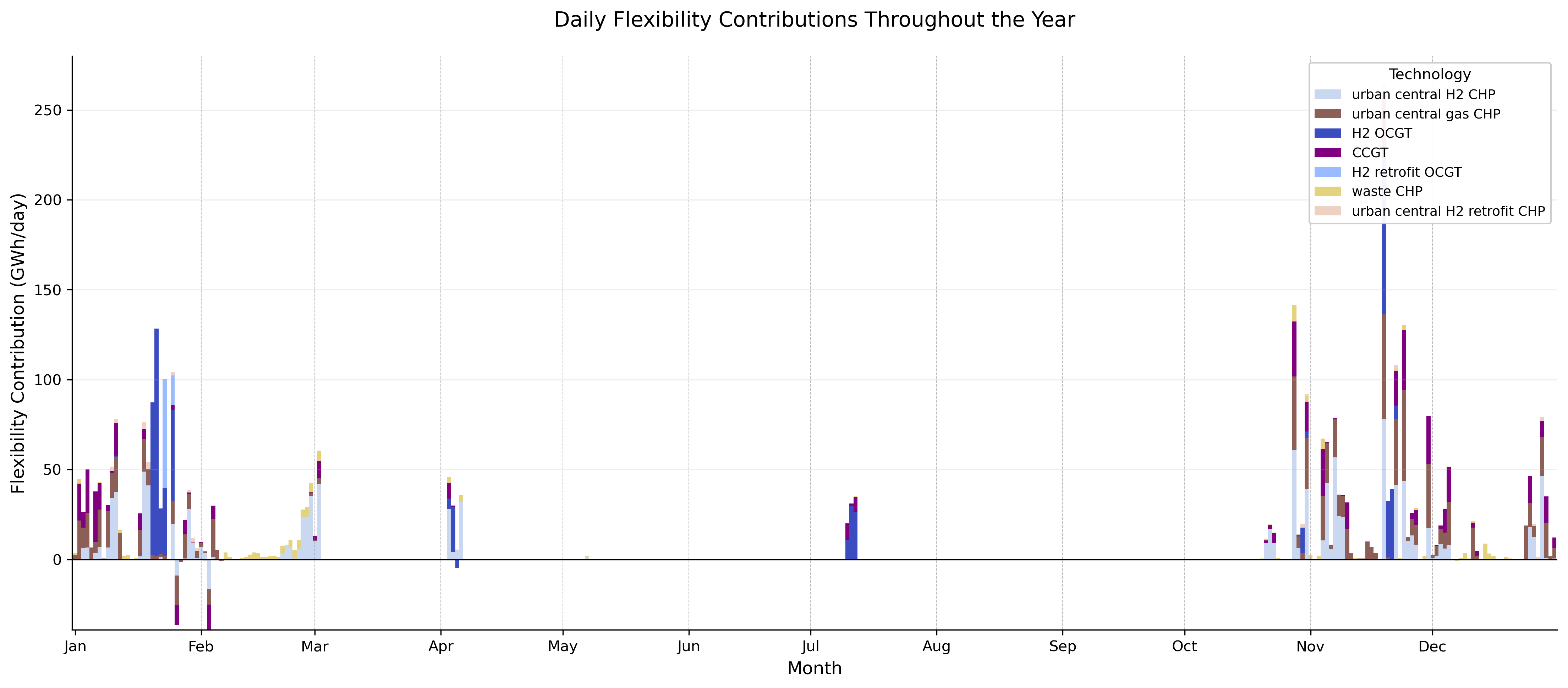}
        \caption{Backup Flexibility Contributions}
    \end{subfigure}
    \caption{\textbf{Temporal view on daily flexibility.}
            The Figure shows the daily flexibility contribution temporally resolved for A) the five technologies with the highest daily flexibility provision, B) import and export, C) additional heat technologies and D) backup technologies.}
    \label{fig:app:temporal-flex-daily}
\end{figure}

\begin{figure}[htbp]
    \centering
    \begin{subfigure}{\linewidth}
        \centering
        \includegraphics[width=0.7\linewidth]{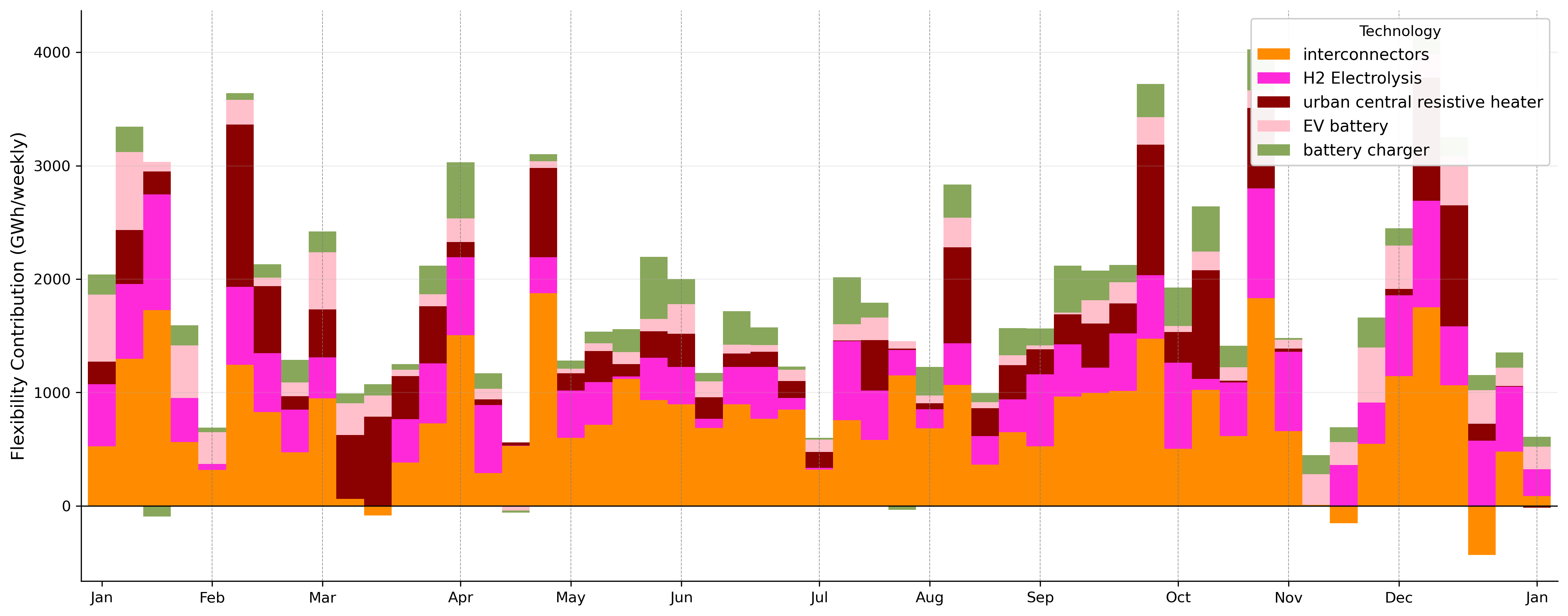}
        \caption{Top 5 Flexibility Contributors}
    \end{subfigure}
    \begin{subfigure}{\linewidth}
        \centering
        \includegraphics[width=0.7\linewidth]{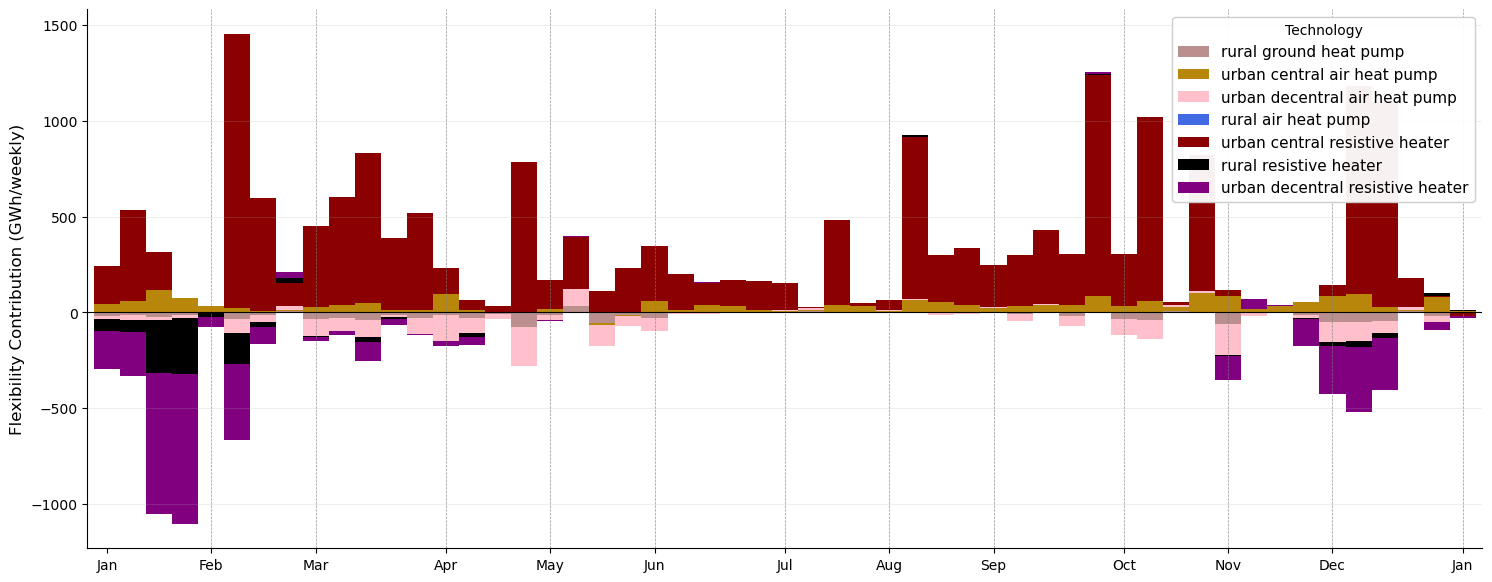}
        \caption{Heat Flexibility Contributions}
    \end{subfigure}
    \begin{subfigure}{\linewidth}
        \centering
        \includegraphics[width=0.7\linewidth]{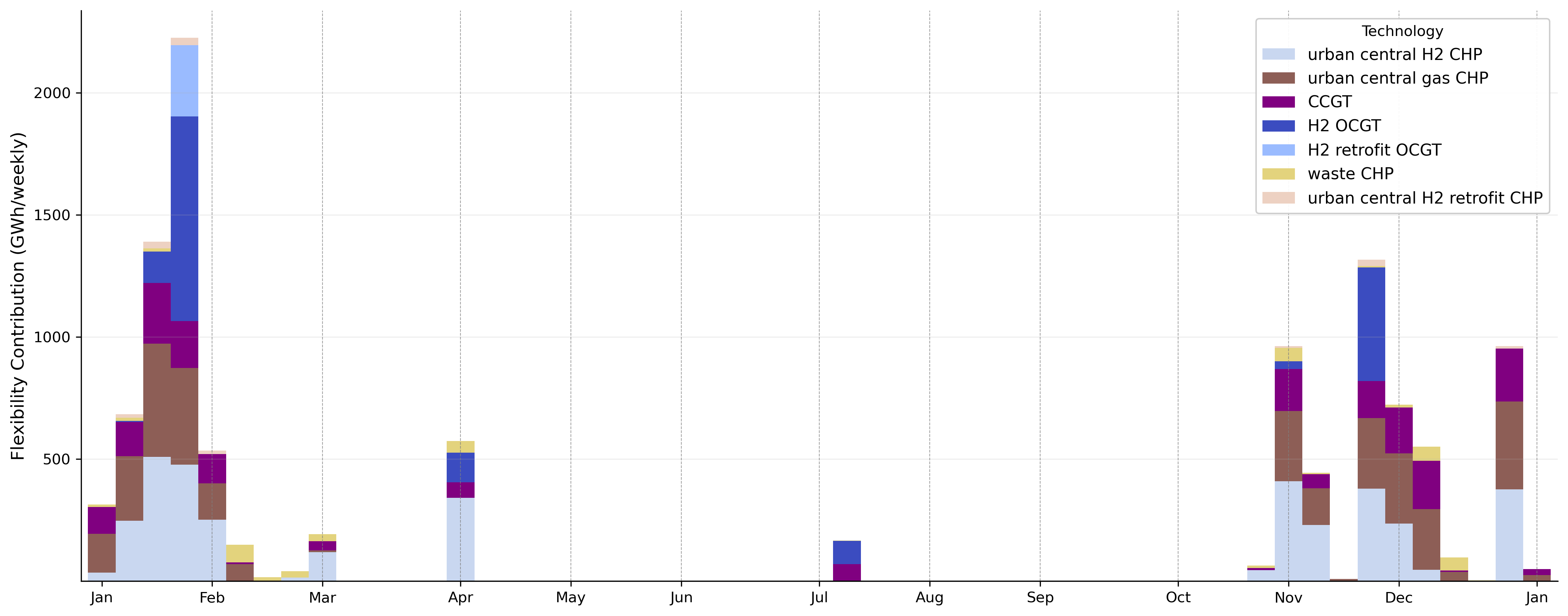}
        \caption{Backup Flexibility Contributions}
    \end{subfigure}
    \caption{\textbf{Temporal view on weekly flexibility.}
        The Figure shows the weekly flexibility contribution temporally resolved for A) the five technologies with the highest weekly flexibility provision, B) electrified heat technologies and C) backup technologies.}
    \label{fig:app:temporal-flex-weekly}
\end{figure}

\end{appendices} 

\end{document}